\documentclass[12pt]{article}
\pdfoutput=1
\usepackage{amsmath,amsfonts,graphicx,color,bbm,tikz}
\usepackage{comment}
\usepackage[nosort]{cite}
\usepackage{subfigure}
\usetikzlibrary{patterns, calc, arrows, automata, shadings}
\usetikzlibrary{calc,positioning}
\usetikzlibrary{patterns,arrows,decorations.pathreplacing, decorations.markings}
\usepackage{caption}
\usepackage{ulem}
\tikzset{>=stealth}
\usepackage{lipsum}
\usepackage{tabularx}
\usepackage{fullpage}
\usepackage{hyperref}

\definecolor{ForestGreen}{RGB}{ 0, 146, 49}
\definecolor{red}{RGB}{ 211, 0, 0}
\definecolor{blue}{RGB}{1, 62, 177}

\makeatletter\@addtoreset{equation}{section}\makeatother

\newcommand{\be}{\begin{equation}}
\newcommand{\ee}{\end{equation}}
\def\beq{\begin{equation}}
\def\eeq{\end{equation}}
\newcommand{\bea}{\begin{eqnarray}}
\newcommand{\eea}{\end{eqnarray}}
\newcommand{\Tr}{{\rm Tr\,}}

\newcommand{\bra}[1]{{\left< {#1} \right|}}
\newcommand{\ket}[1]{{\left| {#1} \right>}}

\renewcommand{\title}[1]{\vbox{\center\LARGE{#1}}\vspace{3mm}}
\renewcommand{\author}[1]{\vbox{\center{#1}}\vspace{3mm}}

\newcommand{\email}[1]{\vbox{\center\tt#1}\vspace{3mm}}

\begin{document}
\begin{titlepage}

\begin{center}
{\large {\bf Groupoid Toric Codes 
}}

\author{ 
Pramod Padmanabhan and Indrajit Jana}
{{\it School of Basic Sciences,\\ Indian Institute of Technology, Bhubaneswar, India}}

\email{pramod23phys, jana.indrajit@gmail.com}
\vskip 1cm 

\end{center}


\abstract{
\noindent 
 The toric code can be constructed as a gauge theory of finite groups on oriented two dimensional lattices. Here we construct analogous models with the gauge fields belonging to groupoids, which are categories where every morphism has an inverse. We show that a consistent system can be constructed for an arbitrary groupoid and analyze the simplest example that can be seen as the analog of the Abelian $\mathbb{Z}_2$ toric code. We find several exactly solvable models that have fracton-like features which include an extensive ground state degeneracy and excitations that are either immobile or have restricted mobility. Among the possibilities we study in detail the one where the ground state degeneracy scales as $2\times 2^{N_v}$, where $N_v$ is the number of vertices in the lattice. The origin of this degeneracy can be traced to loop operators supported on both contractible and non-contractible loops. In particular, different non-contractible loops, along the same direction on a torus, result in different ground states. This is an exponential increase in the number of logical qubits that can be encoded in this code. Moreover the face excitations in this system are deconfined, free to move without an energy cost along certain directions of the lattice, whereas in certain other directions their movement incurs an energy cost. This places a restriction on the types of loop operators that contribute to the ground state degeneracy. The vertex excitations are immobile. The results are also extended to the groupoid analogs of Abelian $\mathbb{Z}_N$ toric codes.

}

\end{titlepage}
\tableofcontents 

\section{Introduction}
\label{sec:Introduction}

Groups are synonymous with symmetry in the mathematician's and physicist's toolkit. For the latter they are in one-to-one correspondence with the set of automorphisms of geometrical structures whereas for the former they are the guiding principle in constructing any consistent quantum field theory. However groups are in some sense far too ideal as they only capture the transformations of homogeneous objects, like that of the sphere or the infinite plane. When we think of the set of symmetries of a finite space, like a tiled room for example, we observe that there are symmetries for these structures but they cannot be completely described by non-trivial automorphisms. In such situations it is wise to enlarge our notion of symmetries and introduce a structure such as {\it groupoids} \cite{https://doi.org/10.48550/arxiv.math/9602220, Brown1987FromGT, Higgins1971NotesOC}. In physics the notion of symmetries assume more general structures at the Planck scale where spacetime is thought to be noncommutative radically changing the way we study quantum fields on such `quantum' spacetimes \cite{Tureanu_2007,Fiore_2007,connes2014noncommutative,Huggett_2021,Balachandran:2005ew,Balachandran_2009,Doplicher_1995,Doplicher1994SpacetimeQI,Madore_1998,https://doi.org/10.48550/arxiv.2203.16052}. In this context the spacetime symmetries are deformed to Hopf algebras and quantum groups \cite{Majid1995FoundationsOQ,Aschieri2007LecturesOH,Chari:1994pz}. More examples of exotic symmetries include {\it inverse semigroups} that are thought of as partial symmetries of aperiodic structures and quasicrystals \cite{Lawson1998InverseST, Kellendonk1999TilingS, Senechal1995QuasicrystalsAG,landsman2012mathematical}. Semigroups also appear in the construction of supersymmetric spin chains \cite{Padmanabhan:2017ekk} and in building analogs of highly entangled spin chains such as the Motzkin and Fredkin spin chains \cite{Sugino:2018fkw, Sugino2018AreaLV, Padmanabhan:2018dae,2018EPJST.227..269S}. Groupoids also make a more fundamental appearance in the structure of quantum mechanics \cite{Ibort_2013,Ciaglia2019SchwingersPO,Ciaglia2021SchwingersPO,Ciaglia2022QuantumTA,Cosmo2019GroupoidsAC}. More recently categorical symmetries have become important in understanding quantum phases of matter, gravity and holographic aspects and more generally in understanding higher dimensional topological theories. They appear in the form of {\it higher group} symmetries and have been crucial in obtaining a broader understanding of topological order among other important aspects of theoretical physics \cite{Wen2019EmergentAH,Bullivant_2017,Bullivant_2019,https://doi.org/10.48550/arxiv.1711.04186,Baez_2010,https://doi.org/10.48550/arxiv.math/0511710,Parzygnat_2019,Delcamp_2018,https://doi.org/10.48550/arxiv.1802.07747,C_rdova_2019,Hofman_2019,Bouzid_2018,Benini_2019, Nussinov_2009, Girelli_2008,Hastings_2005, Gaiotto_2015, https://doi.org/10.48550/arxiv.1309.4721, https://doi.org/10.48550/arxiv.1511.02929, IbietaJimenez2020TopologicalEE, Zucchini2011AKSZMO,Zucchini2017AlgebraicFO,Zucchini20214dCT, Ritter2014Lie2M, Ritter2015LalgebraMA, Soncini20144DSH, McGreevy2022GeneralizedSI, Espiro2022ClassificationOT}.

In this vein we study the role of groupoids in the quest for new quantum phases by constructing the quantum double models of Kitaev \cite{Kitaev_2003, Ferreira20142DQD, Buerschaper_2013, Cowtan2022QuantumDA, Cowtan2022AlgebraicAO,Jia2022BoundaryAD} using groupoid algebras\footnote{The Levin-Wen string-net models \cite{Levin2005StringnetCA} can be obtained as a Kitaev model of unitary quantum groupoids \cite{Chang2014KitaevMB}. The groupoids we consider in this paper are not the same as these. }. As Kitaev's models are particular fixed point of finite group lattice gauge theories \cite{PhysRevD.11.395, Zohar_2015}, we will construct the lattice gauge theory of groupoid algebras. We study in detail the groupoid analog of the simplest quantum double model, namely the $\mathbb{Z}_2$ toric code. Groupoid algebras are non-Abelian in general and hence we expect the corresponding quantum double model to be similar to the non-Abelian toric codes as well. Even for the simplest non-trivial groupoid that we consider in this paper, the models are quite hard to analyze and solve in general and has far richer structure than the Abelian versions. However we get around this problem by rewriting the non-Abelian model in an equivalent way on qubit spaces which simplifies the model to one which is similar to the $\mathbb{Z}_2$ toric code. The groupoid toric code is significantly different from the group version. To begin with we have a larger set of orthogonal operators for a given vertex and face, unlike the toric codes where this set is determined by the representations of the quantum double algebra. As a result of this we can write down several kinds of exactly solvable Hamiltonians on this Hilbert space. Among these we study in detail one set of models that change two of the features of the $\mathbb{Z}_2$ toric code in a significant way. The new features are summarized as : 
\begin{enumerate}
    \item The topological ground state degeneracy of the $\mathbb{Z}_2$ toric code, $2^{2g}$ is exponentially enlarged to an extensive quantity in the groupoid analog. The new ground state degeneracy is $2\times 2^{N_v}$. 
    \item The enlarged degeneracy can be traced to symmetry operators supported on loops, which can be both contractible and non-contractible. Thus, unlike the $\mathbb{Z}_2$ toric code which obtains a topological degeneracy from non-contractible loops, the groupoid analog obtains a degeneracy from both contractible and non-contractible loops. As a consequence different non-contractible loops, along a chosen direction on the genus $g$  surface, result in inequivalent ground states. Moreover the groupoid toric code also has a degeneracy on the 2-sphere, $S^2$ due to the symmetries supported on contractible loops.
    \item By cutting open these loop symmetry operators we obtain excited states corresponding to the face operators, more commonly known as the flux excitations. They are deconfined and free to move along the paths that form these loops. However if they deviate from these paths they gain energy by exciting adjacent vertex and face operators and hence become confined. To summarize, the flux excitations of the groupoid toric code have restricted mobility, that is they are deconfined along certain directions of the lattice whereas they are confined along certain other directions.
    \item Two adjacent vertex operators in the groupoid toric code do not share qubit space or have no common support unlike the group case. Due to this the vertex excitations, and subsequently the dyonic excitations, are localized at the vertices and are immobile. Nevertheless the flux part of the dyonic excitation can still be moved in the restricted directions of the lattice. 
\end{enumerate}
We see similar features in the other models built out of the operators acting on this Hilbert space and we describe them as well.
The extensive ground state degeneracy and the presence of low energy excitations which are either immobile or have restricted motion on the lattice make these models similar to fracton models \cite{Nandkishore2019Fractons, Gromov2022FractonM, Vijay2016FractonTO, IbietaJimenez2020FractonlikePF, Haah2013LatticeQC, Bravyi2011TopologicalOI, Wen2020SystematicCO, Bulmash2019GaugingFI, Radzihovsky2020FractonsFV, Qi2020FractonPV, Lozano2020VorticesIF, Hirono2022ASP, Pretko2020FractonPO, Gromov2019TowardsCO, Shenoy2020knM, Pretko2018TheFG, Banerjee2022HamiltonianFO, Banerjee2022NoetherTF, Bennett2022FractonsGS}. The logical operators for such systems in three dimensions \cite{Ma2022GroundSD, PhysRevX.9.021010} come in different types, with  strings or membranes being some examples. Whereas in two dimensions we may expect the logical operators to be of the string type or of lower weights that act locally. The two dimensional models in this paper have the latter features and we can view the groupoid approach as yet another source for fracton models in two dimensions\footnote{Fracton-like models also appear on graphs which are complexes consisting of just edges and vertices \cite{Padmanabhan2021NovelQP, Resende2022NonAbelianFR}.}. 

With these motivations we organize the rest of the article as follows. The setup of the system starting with the construction of the Hilbert space using groupoid algebras is described in Sec. \ref{sec:groupoidbasics}, followed by the operators acting on this space in Sec. \ref{sec:operatorsgd}. Here we also construct the vertex and face operators that form the Hamiltonian of these models. In both these sections we use the language of morphisms which obey non-Abelian relations as mentioned earlier. This is however inconvenient in the analysis of these systems which simplify significantly when the system is rewritten on equivalent qubit spaces in Sec. \ref{sec:equivalentqubitsystem}. The complete set of orthogonal face and vertex operators are constructed here and the algebra of these operators become much clearer in this new language. In Sec. \ref{sec:hamiltonian1} we construct the Hamiltonian out of these sets of operators and that has the interesting features as mentioned above. We compute the ground state degeneracy on an arbitrary square lattice in Sec. \ref{subsec:gsd} , study the operators creating these states, thought of as the logical Pauli $X$ operators in Sec. \ref{subsec:symmetries} and the operators measuring these states, thought of as the logical Pauli $Z$ operators in Sec. \ref{subsec:organizegs}. The excitations consisting of the fluxes with restricted motion and the immobile charges can be found in Sec. \ref{subsec:excitations}. In Sec. \ref{sec:hamiltonian2} we look at other possible exactly solvable Hamiltonians on this space and briefly look at their features. The generalizations of the above model to the groupoid analogs of the $\mathbb{Z}_N$ toric code is done in Sec. \ref{sec:zngdtc}. Sec. \ref{sec:discussion} contains a few remarks and some future directions. We have included three appendices that go over some of the basics of category theory and groupoids (Appendix \ref{app:groupoidreview}), operator relations for a general groupoid (Appendix \ref{app:gengroupoid}), and some interesting groupoid toric code models on manifolds with boundary (Appendix \ref{app:finitegdtc}).

\section{Groupoid algebras and the Hilbert space}
\label{sec:groupoidbasics}
We begin the construction by setting up the Hilbert space on which our models live. To do this we first recall the example of Kitaev's toric code or quantum double models based on a finite group $G$. They are defined on oriented two dimensional lattices whose local Hilbert space, on every edge, is spanned by the basis of the group algebra, $\mathbb{C}G$. The dimension is given by the order of the finite group $|G|$. This setup can naturally be translated to the {\it groupoid} case once we define them. A {\it groupoid} $\mathcal{G}$ is an example of a category\footnote{See Appendix \ref{app:groupoidreview} for a short review on category theory with groupoids as an example.} made up of {\it objects}, belonging to the set $Ob(\mathcal{G})$, and {\it morphisms} between them such that every morphism is invertible. We reserve the axiomatic definition to an appendix (See Appendix \ref{app:groupoidreview}) and use a particular example that we will be interested in to illustrate the ideas behind the category here. 

\begin{figure}[h!]
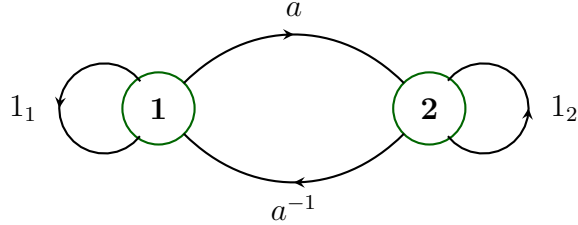

    \centering
    \tikz[scale = 1.2, baseline = 2.5ex, decoration={
	markings,
	mark=at position 0.5 with {\arrow{>}}}]{
    \draw [thick, black!60!green] (0.4,0)  arc (360:0:0.4);
    \node at (0, 0) {$\mathbf{1}$};
    \draw [thick, black!60!green] (3.4,0)  arc (360:0:0.4);
    \node at (3, 0) {$\mathbf{2}$};
    \draw [thick,domain=37:322, postaction={decorate}] plot ({-0.6+0.5*cos(\x)}, {0.5*sin(\x)});
    \node at (-1.5, 0) {$1_{1}$};
    \draw [thick,domain=-142:142, postaction={decorate}] plot ({3.6+0.5*cos(\x)}, {0.5*sin(\x)});
    \node at (4.5, 0) {$1_{2}$};
    \draw [thick, postaction={decorate}] (0.28,0.28) .. controls (1,1) and (2,1) .. (2.72,0.28);
	\node at (1.5, 1.1) {$a$};
    \draw [thick, postaction={decorate}] (2.72,-0.28) .. controls (2,-1) and (1,-1) .. (0.28, -0.28);
    \node at (1.5, -1.1) {$a^{-1}$};
}
    \caption{A groupoid with 2 objects and 1 morphism between them.}
    \label{fig:2obj1morph}
\end{figure}

Consider a groupoid with two objects and one morphism between them as shown in Fig. \ref{fig:2obj1morph}.

Every morphism has a {\it source} and {\it target} that belong to $Ob(\mathbb{G})$. For the groupoid in Fig. \ref{fig:2obj1morph} we have  
\begin{eqnarray}
    s(a)=1,~t(a)=2;&~~&s(a^{-1})=2,~t(a^{-1})=1,\nonumber \\
    s(1_1)=t(1_1)=1;&~~&s(1_2)=t(1_2)=2.
\end{eqnarray}
Two morphisims, $g_1$ and $g_2$ can be composed only when $t(g_1)=s(g_2)$. Using this rule we find the following composition rules for the morphisms of the groupoid in Fig. \ref{fig:2obj1morph}.
\begin{eqnarray}\label{eq:s21gdrelations}
    aa^{-1} = 1_1,& ~~ & a^{-1}a = 1_2, \nonumber \\
    1_1a=a1_2=a, & ~~ & a^{-1}1_1=1_2a^{-1} = a^{-1}, \nonumber \\
    1_11_1=1_1, & ~~ & 1_21_2=1_2.
\end{eqnarray}
All the other compositions are zero as the targets and sources do not coincide. From these composition rules it is clear that the morphisms $1_1$ and $1_2$ are the identity morphims on the two objects respectively. More precisely the morphisms $1_1$ and $1_2$ are the left and right identities of the morphism $a$ and they are the right and left identities of the morphism $a^{-1}$. Alternatively they can be thought of as partial identities. 

The relations in Eq. \ref{eq:s21gdrelations} are similar to the relations of the {\it symmetric inverse semigroup}(SIS), $S^2_1$ \cite{} once we make the identification, 
\begin{eqnarray}
    1_1 \rightarrow x_{11}, &~~& 1_2\rightarrow x_{22}, \nonumber \\
    a \rightarrow x_{12}, &~~& a^{-1}\rightarrow x_{21}.
\end{eqnarray}
We can think of the indices $i$, $j$ in $x_{ij}$ as being the source and target, then the multiplication rule in the SIS is given by 
\begin{equation}\label{eq:SISmultiplication}
    x_{i_1i_2}x_{j_1j_2}=\delta(i_2,j_1)~x_{i_1j_2}.
\end{equation}
It is now easy to see that the morphism relations in Eq. \ref{eq:s21gdrelations} follows from the multiplication rule in Eq. \ref{eq:SISmultiplication} once we make the identification between the two structures. The identification of the groupoid in Fig. \ref{fig:2obj1morph} to the SIS, $S^2_1$ makes it convenient to generalize the above groupoid to the case where there are $n$ objects and exactly one morphism (and its inverse) between any two objects. We shall address such cases in a later section. Henceforth we will denote the groupoid in Fig. \ref{fig:2obj1morph} as the {\it $S^2_1$-groupoid}.

With these definitions in place we are now ready to build the local Hilbert space for this system. To this end we choose an oriented edge, $e$ of the lattice supporting the local Hilbert space, $\mathcal{H}_e$. For the $S^2_1$-groupoid this space is spanned by the elements 
$$ \{\phi_{1_1}, \phi_{1_2}, \phi_a, \phi_{a^{-1}}\}\simeq \mathbb{C}^4=\mathcal{H}_e.$$
We take the states corresponding to each of these basis elements as being mutually orthogonal, that is we choose the canonical inner product. To simplify the notation we will drop the $\phi$ symbol on the basis elements and hope that no confusion arises to the reader. Hence the Hilbert space on a single edge is spanned by 
$$ \{\ket{1_1},~\ket{1_2},~\ket{a},~\ket{a^{-1}}\},$$
and the total Hilbert space is given by 
$$ \mathcal{H}_T= \bigotimes_{e=1}^{N_e}~\mathbb{C}^4_e,$$
where $N_e$ is the total number of edges in the lattice.

\section{Local operators on groupoid algebras}
\label{sec:operatorsgd}
We now define the operators acting on the local Hilbert space, $\mathcal{H}_e\simeq\mathbb{C}^4$, in a manner analogous to how it is done for the group toric codes. We define the left and right actions of the morphisms on $\mathcal{H}_e$ as follows,
\begin{equation}
    g^l~\ket{h} = \ket{gh},~~g^r~\ket{h}=\ket{hg}.
\end{equation}
Using this definition we write down the actions of the morphisms $1_1^l$ and $a^l$ as,
\begin{eqnarray}
    1_1^l~\ket{1_1} = \ket{1_1}, & ~~ & a^l~\ket{1_1} = 0, \nonumber \\
    1_1^l~\ket{1_2} = 0, & ~~ & a^l~\ket{1_2} = \ket{a}, \nonumber \\
    1_1^l~\ket{a} = \ket{a}, & ~~ & a^l~\ket{a} = 0, \nonumber \\
    1_1^l~\ket{a^{-1}} = 0, & ~~ & a^l~\ket{a^{-1}} = \ket{1_1},
\end{eqnarray}
and similarly for the morphisms $1_2^l$ and $\left(a^{-1}\right)^l$. The right actions for morphisms $1_2^r$ and $a^r$ are similarly evaluated as,
\begin{eqnarray}
    1_2^r~\ket{1_1} = 0, & ~~ & a^r~\ket{1_1} = \ket{a}, \nonumber \\
    1_2^r~\ket{1_2} = \ket{1_2}, & ~~ & a^r~\ket{1_2} = 0, \nonumber \\
    1_2^r~\ket{a} = \ket{a}, & ~~ & a^r~\ket{a} = 0, \nonumber \\
    1_2^r~\ket{a^{-1}} = 0, & ~~ & a^r~\ket{a^{-1}} = \ket{1_2},
\end{eqnarray}
and similarly for the morphisms $1_1^r$ and $\left(a^{-1}\right)^r$.

The operators for the left and right action of the morphisms obey the relations of the $S^2_1$-groupoid in Eq. \ref{eq:s21gdrelations} respectively. As in the group case the operators corresponding to the left and right action commute with each other. Also note that the identity operator on each edge is now given by $1_1^l+1_2^l$ for the left action and $1_1^r+1_2^r$ for the right action. In light of this we look at $1_1^l$ and $1_2^l$ as partial identities individually.   

\begin{figure}[h]
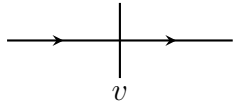

    \centering
    \tikz[scale = 1, baseline = 2.5ex, decoration={
	markings,
	mark=at position 0.5 with {\arrow{>}}}] {
    \def\a{0}
    \def\b{0}
    \draw[thick, postaction={decorate}] (-1.5+\a, \b) -- (\a, \b);
    \draw[thick, postaction={decorate}] (\a, \b) -- (1.5+\a, \b);
    \draw[thick] (\a, -0.5+\b) -- (\a, 0.5+\b);
    \node at (\a, -0.7+\b) {$v$};
    }
    \caption{A vertex $v$ with valency two. The orientation on the edges determine whether the morphism acts from the left or from the right.}
    \label{fig:2valencevertex}
\end{figure}

\paragraph{Vertex Operators - } Using the operators corresponding to the left and right action of the morphisms we can define the vertex operators acting on the edges surrounding a given vertex $v$ as shown in Fig. \ref{fig:2valencevertex}.


The vertex operators for the vertex in Fig. \ref{fig:2valencevertex} can be written as 
\begin{eqnarray}\label{eq:vertexopmorphversion}
    A_v^{(0)} & = & \frac{1}{2}\left[1_1^r1_1^l + 1_2^r1_2^l + \left(a^{-1}\right)^ra^l + a^r\left(a^{-1}\right)^l \right], \nonumber \\
    A_v^{(1)} & = & \frac{1}{2}\left[1_1^r1_1^l + 1_2^r1_2^l - \left(a^{-1}\right)^ra^l - a^r\left(a^{-1}\right)^l \right], \nonumber \\
    A_v^{(2)} & = & \frac{1}{2}\left[1_2^r1_1^l + 1_1^r1_2^l + a^ra^l + \left(a^{-1}\right)^r\left(a^{-1}\right)^l \right], \nonumber \\
    A_v^{(3)} & = & \frac{1}{2}\left[1_2^r1_1^l + 1_1^r1_2^l - a^ra^l - \left(a^{-1}\right)^r\left(a^{-1}\right)^l \right],
\end{eqnarray}
where we have omitted the edge indices on each operator to keep the notation simple. It is understood that the first(second) operator acts on the edge going into(out of) the vertex $v$ in Fig. \ref{fig:2valencevertex}. The operator acts as identity on all the other edges of the lattice. The vertex operators for vertices with higher valency can be written down in a similar manner taking into account the orientations of these edges. 

\begin{figure}[h]
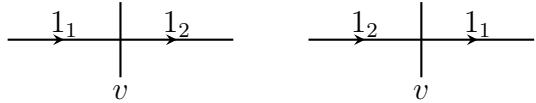

		     \centering
		     \tikz[scale = 1, baseline = 2.5ex, decoration={
			 	markings,
			 	mark=at position 0.5 with {\arrow{>}}}] {
			     \def\a{0}
			     \def\b{0}
			     \draw[thick, postaction={decorate}] (-1.5+\a, \b) -- (\a, \b);
			     \node at (-0.75+\a, \b+0.2) {$1_{1}$};
			     \draw[thick, postaction={decorate}] (\a, \b) -- (1.5+\a, \b);
			     \node at (0.75+\a, \b+0.2) {$1_{2}$};
			     \draw[thick] (\a, -0.5+\b) -- (\a, 0.5+\b);
			     \node at (\a, -0.7+\b) {$v$};
			     \def\a{4}
			     \def\b{0}
			     \draw[thick, postaction={decorate}] (-1.5+\a, \b) -- (\a, \b);
			     \node at (-0.75+\a, \b+0.2) {$1_{2}$};
			     \draw[thick, postaction={decorate}] (\a, \b) -- (1.5+\a, \b);
			     \node at (0.75+\a, \b+0.2) {$1_{1}$};
			     \draw[thick] (\a, -0.5+\b) -- (\a, 0.5+\b);
			     \node at (\a, -0.7+\b) {$v$};
			     }
		     \caption{Configurations that are not killed by the vertex operators, $A_v^{(2)}$ and $A_v^{(3)}$ in Eq. \ref{eq:vertexopmorphversion}.}
		     \label{fig:configsforvertexops}
		 \end{figure}

It is easy to check using the algebra of these local operators that each of the vertex ooperators are projectors and they are orthogonal to each other. Note that unlike the $\mathbb{Z}_2$ toric code there are more orthogonal vertex operators in the analogous groupoid case. This can be seen due to the presence of the vertex operators, $A_v^{(2)}$ and $A_v^{(3)}$ in Eq. \ref{eq:vertexopmorphversion}. These operators act non-trivially on the states that are killed by the operators, $A_v^{(0)}$ and $A_v^{(1)}$. This is better illustrated in Fig. \ref{fig:configsforvertexops} showing the configurations on which the vertex operators act non-trivially.

This is a direct consequence of the groupoid structure in this system. With these operators in place we also have,
\begin{equation}
    \sum\limits_{j=0}^3~A_v^{(j)} = 1,
\end{equation}
where 1 is the identity operator on the total Hilbert space, $\mathcal{H}_T$.
Note that for vertices with valency greater than two, the number of orthogonal operators to $A_v^{(0)}$,\footnote{The $A^{(0)}_v$ operator can be thought of as the groupoid average analogous to the vertex operators in group toric code.}
increases. 

\begin{figure}[h]
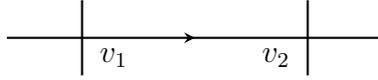

    \centering
    \tikz[scale = 1, baseline = 2.5ex, decoration={
			 	markings,
			 	mark=at position 0.5 with {\arrow{>}}}] {
			     \def\a{0}
			     \def\b{0}
			     \draw[thick, postaction={decorate}] (-2.5+\a, \b) -- (2.5+\a, \b);
            \draw[thick] (-1.5+\a, \b-0.5) -- (-1.5+\a, \b+0.5);
            \draw[thick] (1.5+\a, \b-0.5) -- (1.5+\a, \b+0.5);
			     \node[anchor = north east] at (-0.75+\a, \b) {$v_{1}$};
                \node[anchor = north west] at (0.75+\a, \b) {$v_{2}$};
        }
    \caption{Two adjacent vertices $v_1$ and $v_2$ sharing an edge. The corresponding vertex operators commute.}
    \label{fig:twoadjacentvertices}
\end{figure}

Finally it is easy to check that the vertex operators for two adjacent vertices, $v_1$ and $v_2$ sharing an edge commute with each other (See Fig. \ref{fig:twoadjacentvertices}) as one of them acts with the left action while the other acts with the right action on the Hilbert space of the shared edge.

\begin{figure}[h!]
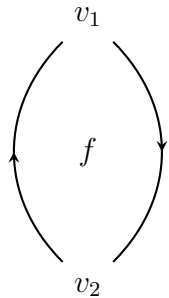

    \centering
    \tikz[scale = 1.2, baseline = 2.5ex, decoration={
	markings,
	mark=at position 0.5 with {\arrow{>}}}]{
    \draw [thick, postaction={decorate}] (0.28, 2.72) .. controls (1,2) and (1,1) .. (0.28, 0.28);
    \draw [thick, postaction={decorate}] (-0.28, 0.28) .. controls (-1, 1) and (-1,2) .. (-0.28, 2.72);
    \node at (0, 1.5) {$f$};
    \node at (0, 0) {$v_{2}$};
    \node at (0, 3) {$v_{1}$};
}
    \caption{The convention for measuring the groupoid holonomy across a face. The holonomy is measured clockwise from $v_1$. For simplicity we have chosen a face with just two edges.}
    \label{fig:holonomyconvention}
\end{figure}

\paragraph{Face Operators - } With the vertex operators in place we move on to the face operators that measure the holonomy of the groupoid gauge fields around a plaquette or face. Before we write down these operators we define more operators on the local Hilbert space, $\mathbb{C}^4$ on an edge of the lattice. These are the projectors to the different morphisms given by,
\begin{eqnarray}
    P^{1_1} = \ket{1_1}\bra{1_1}, &~~& P^{1_2} = \ket{1_2}\bra{1_2}, \nonumber \\
    P^{a} = \ket{a}\bra{a}, &~~& P^{a^{-1}} = \ket{a^{-1}}\bra{a^{-1}}.
\end{eqnarray}
We will use these operators to specify the different face operators. We begin by choosing a convention to measure the holonomy of the faces as illustrated in Fig. \ref{fig:holonomyconvention}.

With the conventions in place we are now in a position to write down the different face operators in this system as,
\begin{eqnarray}\label{eq:faceopsmorphversion}
    B_f^{1_1} &=& P^{1_1}P^{1_1} + P^{a}P^{a^{-1}}, \nonumber \\
    B_f^{1_2} &=& P^{1_2}P^{1_2} + P^{a^{-1}}P^{a}, \nonumber \\
    B_f^{a} &=& P^{1_1}P^{a} + P^{a}P^{1_2}, \nonumber \\
    B_f^{a^{-1}} &=& P^{1_2}P^{a^{-1}} + P^{a^{-1}}P^{1_1}, 
\end{eqnarray}
where we have omitted the edge indices on the local projectors to keep the notation simple. Clearly each of these operators are projectors and they are orthogonal to each other. However these operators do not add up to the identity as we have another face operator that projects to the zero,
\begin{eqnarray}\label{eq:bfzeromorphversion}
    B_f^0 & = & P^{1_1}P^{a^{-1}} + P^{a^{-1}}P^{1_2} + P^{a^{-1}}P^{a^{-1}} + P^{1_1}P^{1_2} \nonumber \\ & + & P^{a}P^{1_1} + P^{1_2}P^{1_1} + P^{a}P^{a} + P^{1_2}P^{a}. 
\end{eqnarray}
It is now easily checked that,
\begin{equation}
    B_f^{1_1} + B_f^{1_2} + B_f^{a} + B_f^{a^{-1}} + B_f^0 = 1,
\end{equation}
is the identity operator on $\mathcal{H}_T$.

It is easily seen that since each of the face operators are diagonal in the chosen basis of the Hilbert space, they all commute with each other for any face. The generalization of Eqs. \ref{eq:faceopsmorphversion} and \ref{eq:bfzeromorphversion} for faces of arbitrary shapes is straightforward even though the expressions get more complicated due to larger number of possibilities in obtaining a given morphism. 

With some algebra we can explicitly check that the vertex operators in Eq. \ref{eq:vertexopmorphversion} commute with the face operators in Eqs. \ref{eq:faceopsmorphversion} and \ref{eq:bfzeromorphversion}. However this computation will become far simpler when we rewrite this system using qubit degrees of freedom in the next section\footnote{A proof for the commutativity of the face and vertex operators for a general groupoid toric code can be found in Appendix B.}.

\section{Translating to a qubit system}
\label{sec:equivalentqubitsystem}
\begin{figure}[h]
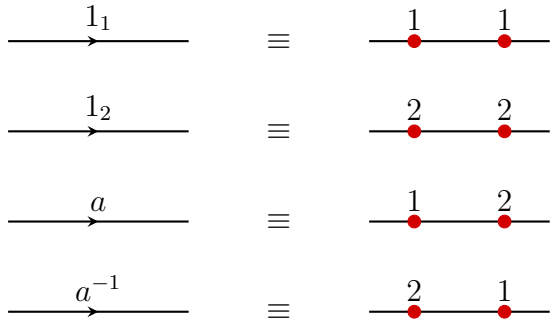

    \centering
    \tikz[scale = 1.2, baseline = 2.5ex, decoration={
	markings,
	mark=at position 0.5 with {\arrow{>}}}]{
    \def\a{0}
    \def\b{0}
    \draw[thick, postaction={decorate}] (-3+\a, \b) -- (-1+\a, \b);
    \node[anchor=south] at (-2+\a, \b) {$1_1$};
    \node at (\a, \b) {$\equiv$};
    \draw[thick] (1+\a, \b) -- (3+\a, \b);
    \filldraw[red] (1.5+\a, \b) circle (2pt) node[anchor=south, black] {$1$};
    \filldraw[red] (2.5+\a, \b) circle (2pt) node[anchor=south, black] {$1$};
    \def\a{0}
    \def\b{-1}
    \draw[thick, postaction={decorate}] (-3+\a, \b) -- (-1+\a, \b);
    \node[anchor=south] at (-2+\a, \b) {$1_2$};
    \node at (\a, \b) {$\equiv$};
    \draw[thick] (1+\a, \b) -- (3+\a, \b);
    \filldraw[red] (1.5+\a, \b) circle (2pt) node[anchor=south, black] {$2$};
    \filldraw[red] (2.5+\a, \b) circle (2pt) node[anchor=south, black] {$2$};
    \def\a{0}
    \def\b{-2}
    \draw[thick, postaction={decorate}] (-3+\a, \b) -- (-1+\a, \b);
    \node[anchor=south] at (-2+\a, \b) {$a$};
    \node at (\a, \b) {$\equiv$};
    \draw[thick] (1+\a, \b) -- (3+\a, \b);
    \filldraw[red] (1.5+\a, \b) circle (2pt) node[anchor=south, black] {$1$};
    \filldraw[red] (2.5+\a, \b) circle (2pt) node[anchor=south, black] {$2$};
    \def\a{0}
    \def\b{-3}
    \draw[thick, postaction={decorate}] (-3+\a, \b) -- (-1+\a, \b);
    \node[anchor=south] at (-2+\a, \b) {$a^{-1}$};
    \node at (\a, \b) {$\equiv$};
    \draw[thick] (1+\a, \b) -- (3+\a, \b);
    \filldraw[red] (1.5+\a, \b) circle (2pt) node[anchor=south, black] {$2$};
    \filldraw[red] (2.5+\a, \b) circle (2pt) node[anchor=south, black] {$1$};
    }
    \caption{The one-to-one map between the space of morphisms on the edges to the space of object indices on the endpoints of the edges, $\mathbb{C}^4\rightarrow\mathbb{C}^2\otimes\mathbb{C}^2$.}
    \label{fig:qubitmapping}
\end{figure}

Until now we have worked with the Hilbert space where locally, that is on every edge, we considered a $\mathbb{C}^4$. However we observe that on every edge, once we choose an orientation , we can uniquely map the four basis elements of $\mathbb{C}^4$ to the space spanned by the two object indices of the $S^2_1$-groupoid on the two endpoints of the edge, $\mathbb{C}^2\otimes\mathbb{C}^2$ as shown in Fig. \ref{fig:qubitmapping}.

\begin{figure}[h]
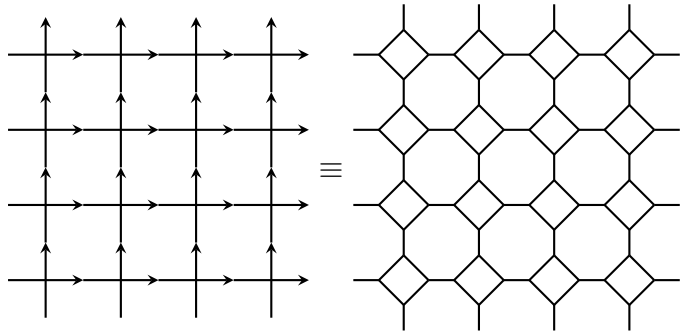

	\centering
	\begin{align*}
		\tikz[scale = 1, baseline = 7.5ex, decoration={
			markings,
			mark=at position 0.5 with {\arrow{>}}}]{
			\foreach \b in {0, 1, 2, 3}{
				\foreach \a in {0, 1, 2, 3}{
					\draw[thick, ->] (\a, \b) -- (\a+1, \b);
					\draw[thick, ->] (\a+0.5, \b-0.5) -- (\a+0.5, \b+0.5);
				}
			}
		}
	\equiv
	\tikz[scale = 1, baseline = 7.5ex]{
		\foreach \a in {0, 1, 2, 3}{
			\foreach \b in {-1, 0, 1, 2} {
				\draw[thick] (\a+1, \b+0.33) -- (\a+1, 0.67+\b) -- (\a+0.67, \b+1) -- (\a+0.33, 1+\b);
				\draw[thick] (\a+1, 0.67+\b) -- (\a+1.33, \b+1);
				\draw[thick] (\a+0.67, 1+\b) -- (\a+1, \b+1.33);
				\draw[thick] (\a+1, \b+1.33) -- (\a+1.33, \b+1);
			}
			\draw[thick] (\a+1, 3+0.33) -- (\a+1, 3+0.67);
		}
		\foreach \b in  {0, 1, 2, 3}{
			\draw[thick] (4.33, \b) -- (4.67, \b);
		}
	}	
	\end{align*}
	\caption{The Hilbert space of morphisms, $\mathcal{H}_T=\bigotimes_{e=1}^{N_e}\mathbb{C}^4_e$ on the oriented square lattice becomes the Hilbert space of qubits, $\mathcal{H}_T=\bigotimes_{v=1}^{4N_v}\mathbb{C}^2_v$ located on the vertices of the square-octagon lattice.}
	\label{fig:squaretosquareoctagon}
\end{figure}

With this identification the degrees of freedom on the edges of say, a square lattice, gets shifted to the vertices of a {\it square-octagon lattice} as shown in Fig. \ref{fig:squaretosquareoctagon}.  

With these identifications we can express the left and right actions of the morphisms in terms of Pauli $X$ and $Z$ matrices acting on the qubit system. The mapping of the left and right action operators are shown in Figs. \ref{fig:leftactiontoqubits} and \ref{fig:rightactiontoqubits} respectively.
\begin{figure}[h]
	\centering
	\begin{align*}
		1_{1}^{l}&\equiv\frac{1}{2}\left[1+\tikz[scale=1, baseline=-0.5ex]{
			\draw[thick] (0, 0) -- (2, 0);
			\filldraw[red] (0.5, 0) circle (2pt) node[anchor=south, black]{$Z$};
			\filldraw[red] (1.5, 0) circle (2pt) node[anchor=south, black]{};
		}\right]\\
		a^{l}&\equiv\left[\tikz[scale=1, baseline=-0.5ex]{
			\draw[thick] (0, 0) -- (2, 0);
			\filldraw[red] (0.5, 0) circle (2pt) node[anchor=south, black]{$X$};
			\filldraw[red] (1.5, 0) circle (2pt) node[anchor=south, black]{};
		}\right]
		\frac{1}{2}\left[1-\tikz[scale=1, baseline=-0.5ex]{
			\draw[thick] (0, 0) -- (2, 0);
			\filldraw[red] (0.5, 0) circle (2pt) node[anchor=south, black]{$Z$};
			\filldraw[red] (1.5, 0) circle (2pt) node[anchor=south, black]{};
		}\right]\\
		(a^{-1})^{l}&\equiv\left[\tikz[scale=1, baseline=-0.5ex]{
			\draw[thick] (0, 0) -- (2, 0);
			\filldraw[red] (0.5, 0) circle (2pt) node[anchor=south, black]{$X$};
			\filldraw[red] (1.5, 0) circle (2pt) node[anchor=south, black]{};
		}\right]
		\frac{1}{2}\left[1+\tikz[scale=1, baseline=-0.5ex]{
			\draw[thick] (0, 0) -- (2, 0);
			\filldraw[red] (0.5, 0) circle (2pt) node[anchor=south, black]{$Z$};
			\filldraw[red] (1.5, 0) circle (2pt) node[anchor=south, black]{};
		}\right]\\
		1_{2}^{l}&\equiv\frac{1}{2}\left[1-\tikz[scale=1, baseline=-0.5ex]{
			\draw[thick] (0, 0) -- (2, 0);
			\filldraw[red] (0.5, 0) circle (2pt) node[anchor=south, black]{$Z$};
			\filldraw[red] (1.5, 0) circle (2pt) node[anchor=south, black]{};
		}\right]
	\end{align*}
	\caption{The left action morphisms on the qubit space.}
	\label{fig:leftactiontoqubits}
\end{figure}


\begin{figure}[h]
	\centering
	\begin{align*}
		1_{1}^{r}&\equiv\frac{1}{2}\left[1+\tikz[scale=1, baseline=-0.5ex]{
			\draw[thick] (0, 0) -- (2, 0);
			\filldraw[red] (0.5, 0) circle (2pt) node[anchor=south, black]{};
			\filldraw[red] (1.5, 0) circle (2pt) node[anchor=south, black]{$Z$};
		}\right]\\
		a^{r}&\equiv\left[\tikz[scale=1, baseline=-0.5ex]{
			\draw[thick] (0, 0) -- (2, 0);
			\filldraw[red] (0.5, 0) circle (2pt) node[anchor=south, black]{};
			\filldraw[red] (1.5, 0) circle (2pt) node[anchor=south, black]{$X$};
		}\right]
		\frac{1}{2}\left[1+\tikz[scale=1, baseline=-0.5ex]{
			\draw[thick] (0, 0) -- (2, 0);
			\filldraw[red] (0.5, 0) circle (2pt) node[anchor=south, black]{};
			\filldraw[red] (1.5, 0) circle (2pt) node[anchor=south, black]{$Z$};
		}\right]\\
		(a^{-1})^{r}&\equiv\left[\tikz[scale=1, baseline=-0.5ex]{
			\draw[thick] (0, 0) -- (2, 0);
			\filldraw[red] (0.5, 0) circle (2pt) node[anchor=south, black]{};
			\filldraw[red] (1.5, 0) circle (2pt) node[anchor=south, black]{$X$};
		}\right]
		\frac{1}{2}\left[1-\tikz[scale=1, baseline=-0.5ex]{
			\draw[thick] (0, 0) -- (2, 0);
			\filldraw[red] (0.5, 0) circle (2pt) node[anchor=south, black]{};
			\filldraw[red] (1.5, 0) circle (2pt) node[anchor=south, black]{$Z$};
		}\right]\\
		1_{2}^{r}&\equiv\frac{1}{2}\left[1-\tikz[scale=1, baseline=-0.5ex]{
			\draw[thick] (0, 0) -- (2, 0);
			\filldraw[red] (0.5, 0) circle (2pt) node[anchor=south, black]{};
			\filldraw[red] (1.5, 0) circle (2pt) node[anchor=south, black]{$Z$};
		}\right]
	\end{align*}
	\caption{The right action morphisms on the qubit space.}
	\label{fig:rightactiontoqubits}
\end{figure}


These can be used to write down the vertex operators in Eq. \ref{eq:vertexopmorphversion} as 

\begin{align}\label{eq:vertexopqubitversionvalence2}
	A_{v}^{(0)}&=\frac{1}{2}\left[1+\tikz[baseline=-0.5ex]{
			\draw (-1, 0) -- (1, 0);
			\draw (0, -1/6) -- (0, 1/6);
			\node at (0, -1/3) {$v$};
			\filldraw[red] (-0.5,0) circle (2pt) node[anchor=south, black]{$X$};
			\filldraw[red] (0.5,0) circle (2pt) node[anchor=south, black]{$X$};
	}
	\right]
	\frac{1}{2}\left[1+\tikz[baseline=-0.5ex]{
		\draw (-1, 0) -- (1, 0);
		\draw (0, -1/6) -- (0, 1/6);
		\node at (0, -1/3) {$v$};
		\filldraw[red] (-0.5,0) circle (2pt) node[anchor=south, black]{$Z$};
		\filldraw[red] (0.5,0) circle (2pt) node[anchor=south, black]{$Z$};
	}
	\right]\nonumber\\
	A_{v}^{(1)}&=\frac{1}{2}\left[1-\tikz[baseline=-0.5ex]{
		\draw (-1, 0) -- (1, 0);
		\draw (0, -1/6) -- (0, 1/6);
		\node at (0, -1/3) {$v$};
		\filldraw[red] (-0.5,0) circle (2pt) node[anchor=south, black]{$X$};
		\filldraw[red] (0.5,0) circle (2pt) node[anchor=south, black]{$X$};
	}
	\right]
	\frac{1}{2}\left[1+\tikz[baseline=-0.5ex]{
		\draw (-1, 0) -- (1, 0);
		\draw (0, -1/6) -- (0, 1/6);
		\node at (0, -1/3) {$v$};
		\filldraw[red] (-0.5,0) circle (2pt) node[anchor=south, black]{$Z$};
		\filldraw[red] (0.5,0) circle (2pt) node[anchor=south, black]{$Z$};
	}
	\right]\nonumber\\
	A_{v}^{(2)}&=\frac{1}{2}\left[1+\tikz[baseline=-0.5ex]{
		\draw (-1, 0) -- (1, 0);
		\draw (0, -1/6) -- (0, 1/6);
		\node at (0, -1/3) {$v$};
		\filldraw[red] (-0.5,0) circle (2pt) node[anchor=south, black]{$X$};
		\filldraw[red] (0.5,0) circle (2pt) node[anchor=south, black]{$X$};
	}
	\right]
	\frac{1}{2}\left[1-\tikz[baseline=-0.5ex]{
		\draw (-1, 0) -- (1, 0);
		\draw (0, -1/6) -- (0, 1/6);
		\node at (0, -1/3) {$v$};
		\filldraw[red] (-0.5,0) circle (2pt) node[anchor=south, black]{$Z$};
		\filldraw[red] (0.5,0) circle (2pt) node[anchor=south, black]{$Z$};
	}
	\right]\nonumber\\
	A_{v}^{(3)}&=\frac{1}{2}\left[1-\tikz[baseline=-0.5ex]{
		\draw (-1, 0) -- (1, 0);
		\draw (0, -1/6) -- (0, 1/6);
		\node at (0, -1/3) {$v$};
		\filldraw[red] (-0.5,0) circle (2pt) node[anchor=south, black]{$X$};
		\filldraw[red] (0.5,0) circle (2pt) node[anchor=south, black]{$X$};
	}
	\right]
	\frac{1}{2}\left[1-\tikz[baseline=-0.5ex]{
		\draw (-1, 0) -- (1, 0);
		\draw (0, -1/6) -- (0, 1/6);
		\node at (0, -1/3) {$v$};
		\filldraw[red] (-0.5,0) circle (2pt) node[anchor=south, black]{$Z$};
		\filldraw[red] (0.5,0) circle (2pt) node[anchor=south, black]{$Z$};
	}
	\right]
\end{align}

The Pauli matrices in the expressions of Eq. \ref{eq:vertexopqubitversionvalence2} act on the dots surrounding the vertex $v$. Note that these operators are almost similar to that of the $\mathbb{Z}_2$-toric code except that now they also check if the qubits surrounding the vertex are either matched, 
\tikz[baseline=-0.5ex, scale=0.7]{
	\draw (-1, 0) -- (1, 0);
	\draw (0, -1/6) -- (0, 1/6);
	\filldraw[red] (-0.5,0) circle (2pt) node[anchor=south, black]{$\textbf{\scriptsize{1}}$};
	\filldraw[red] (0.5,0) circle (2pt) node[anchor=south, black]{$\textbf{\scriptsize{1}}$};
	\def\x{2.3};
	\draw (-1+\x, 0) -- (1+\x, 0);
	\draw (\x, -1/6) -- (\x, 1/6);
	\filldraw[red] (-0.5+\x,0) circle (2pt) node[anchor=south, black]{$\textbf{\scriptsize{2}}$};
	\filldraw[red] (0.5+\x,0) circle (2pt) node[anchor=south, black]{$\textbf{\scriptsize{2}}$};
}  
or unmatched
\tikz[baseline=-0.5ex, scale=0.7]{
	\draw (-1, 0) -- (1, 0);
	\draw (0, -1/6) -- (0, 1/6);
	\filldraw[red] (-0.5,0) circle (2pt) node[anchor=south, black]{$\textbf{\scriptsize{1}}$};
	\filldraw[red] (0.5,0) circle (2pt) node[anchor=south, black]{$\textbf{\scriptsize{2}}$};
	\def\x{2.3};
	\draw (-1+\x, 0) -- (1+\x, 0);
	\draw (\x, -1/6) -- (\x, 1/6);
	\filldraw[red] (-0.5+\x,0) circle (2pt) node[anchor=south, black]{$\textbf{\scriptsize{2}}$};
	\filldraw[red] (0.5+\x,0) circle (2pt) node[anchor=south, black]{$\textbf{\scriptsize{1}}$};
}.
These are all the mutually orthogonal vertex operators that we can write down for a vertex with valency two and we have,
\begin{equation}
    \sum\limits_{j=0}^3~A_v^{(j)} = 1.
\end{equation}

It is clear that the expressions for the vertex operator depend on the valency of the vertices. As we analyze our models on a square lattice let us write down the vertex operators on vertices with valency four. The expressions for the corresponding operators on valence 2 vertices in Eq. \ref{eq:vertexopqubitversionvalence2} give us a clue as to how we should go about obtaining these operators. Consider a vertex of valency 4,
	\tikz[baseline=-0.5ex]{
		\draw[thick] (-1, 0) -- (1, 0);
		\draw[thick] (0, -1) -- (0, 1);
		\filldraw[red] (-0.5,0) circle (2pt);
		\filldraw[red] (0.5,0) circle (2pt);
		\filldraw[red] (0, -0.5) circle (2pt);
		\filldraw[red] (0, 0.5) circle (2pt);
	}
with four corners. Let us denote these corners as north-west (NW), north-east (NE), south-west (SW) and south-east (SE). We can determine the configuration on this vertex by specifying three of the corners. In each corner we either have a match
	\tikz[baseline = -0.5ex, scale=0.5]{
		\def\a{-1.25}
		\draw[thick] (-1+\a, 0) -- (1+\a, 0);
		\draw[thick] (0+\a, -1) -- (0+\a, 1);
		\filldraw[red] (-0.5+\a,0) circle (2pt) node[anchor=north, black] {$\textbf{\scriptsize{1}}$};
		\filldraw[red] (0.5+\a,0) circle (2pt);
		\filldraw[red] (0+\a, -0.5) circle (2pt);
		\filldraw[red] (0+\a, 0.5) circle (2pt) node[anchor=west, black] {$\textbf{\scriptsize{1}}$};
		\def\a{1.25}
		\draw[thick] (-1+\a, 0) -- (1+\a, 0);
		\draw[thick] (0+\a, -1) -- (0+\a, 1);
		\filldraw[red] (-0.5+\a,0) circle (2pt) node[anchor=north, black] {$\textbf{\scriptsize{2}}$};
		\filldraw[red] (0.5+\a,0) circle (2pt);
		\filldraw[red] (0+\a, -0.5) circle (2pt);
		\filldraw[red] (0+\a, 0.5) circle (2pt) node[anchor=west, black] {$\textbf{\scriptsize{2}}$};
	}
or a mismatch
\tikz[baseline = -0.5ex, scale=0.5]{
    \def\a{-1.25}
    \def\b{0}
    \draw[thick] (-1+\a, 0+\b) -- (1+\a, 0+\b);
    \draw[thick] (0+\a, -1+\b) -- (0+\a, 1+\b);
    \filldraw[red] (-0.5+\a,0+\b) circle (2pt) node[anchor=north, black] {$\textbf{\scriptsize{2}}$};
    \filldraw[red] (0.5+\a,0+\b) circle (2pt);
    \filldraw[red] (0+\a, -0.5+\b) circle (2pt);
    \filldraw[red] (0+\a, 0.5+\b) circle (2pt) node[anchor=west, black] {$\textbf{\scriptsize{1}}$};
    \def\a{1.25}
    \def\b{0}
    \draw[thick] (-1+\a, 0+\b) -- (1+\a, 0+\b);
    \draw[thick] (0+\a, -1+\b) -- (0+\a, 1+\b);
    \filldraw[red] (-0.5+\a,0+\b) circle (2pt) node[anchor=north, black] {$\textbf{\scriptsize{1}}$};
    \filldraw[red] (0.5+\a,0+\b) circle (2pt);
    \filldraw[red] (0+\a, -0.5+\b) circle (2pt);
    \filldraw[red] (0+\a, 0.5+\b) circle (2pt) node[anchor=west, black] {$\textbf{\scriptsize{2}}$};
}. Thus we have two possibilities in each corner, making it eight for the three independent corners and this specifies the configurations on the entire vertex. Thus on a vertex with valency four, we expect to find eight mutually orthogonal vertex operators and these are given by,

\begin{align}\label{eq:vertexopsqubitversionvalence4}
	A_{v}^{(0)}&=\frac{1}{2}\left[1+\tikz[baseline=-0.5ex, scale=0.8]{
			\def\a{0}
			\def\b{0}
			\draw[thick] (-1+\a, 0+\b) -- (1+\a, 0+\b);
			\draw[thick] (0+\a, -1+\b) -- (0+\a, 1+\b);
			\node[anchor=north west] at (-0.1, 0.1) {$v$};
			\filldraw[red] (-0.5+\a,0+\b) circle (2pt) node[anchor=north, black] {$\textbf{\scriptsize{X}}$};
			\filldraw[red] (0.5+\a,0+\b) circle (2pt) node[anchor=south, black] {$\textbf{\scriptsize{X}}$};
			\filldraw[red] (0+\a, -0.5+\b) circle (2pt) node[anchor=north west, black] {$\textbf{\scriptsize{X}}$};
			\filldraw[red] (0+\a, 0.5+\b) circle (2pt) node[anchor=south east, black] {$\textbf{\scriptsize{X}}$};
	}\right]
	\frac{1}{2}\left[1+\tikz[baseline=-0.5ex, scale=0.8]{
		\def\a{0}
		\def\b{0}
		\draw[thick] (-1+\a, 0+\b) -- (1+\a, 0+\b);
		\draw[thick] (0+\a, -1+\b) -- (0+\a, 1+\b);
		\node[anchor=north west] at (-0.1, 0.1) {$v$};
		\filldraw[red] (-0.5+\a,0+\b) circle (2pt) node[anchor=north, black] {$\textbf{\scriptsize{Z}}$};
		\filldraw[red] (0+\a, -0.5+\b) circle (2pt) node[anchor=north west, black] {$\textbf{\scriptsize{}}$};
		\filldraw[red] (0.5+\a,0+\b) circle (2pt) node[anchor=south, black] {$\textbf{\scriptsize{}}$};
		\filldraw[red] (0+\a, 0.5+\b) circle (2pt) node[anchor=south east, black] {$\textbf{\scriptsize{Z}}$};
	}\right]
	\frac{1}{2}\left[1+\tikz[baseline=-0.5ex, scale=0.8]{
		\def\a{0}
		\def\b{0}
		\draw[thick] (-1+\a, 0+\b) -- (1+\a, 0+\b);
		\draw[thick] (0+\a, -1+\b) -- (0+\a, 1+\b);
		\node[anchor=north west] at (-0.1, 0.1) {$v$};
		\filldraw[red] (-0.5+\a,0+\b) circle (2pt) node[anchor=north, black] {$\textbf{\scriptsize{Z}}$};
		\filldraw[red] (0+\a, -0.5+\b) circle (2pt) node[anchor=north west, black] {$\textbf{\scriptsize{Z}}$};
		\filldraw[red] (0.5+\a,0+\b) circle (2pt) node[anchor=south, black] {$\textbf{\scriptsize{}}$};
		\filldraw[red] (0+\a, 0.5+\b) circle (2pt) node[anchor=south east, black] {$\textbf{\scriptsize{}}$};
	}\right]
	\frac{1}{2}\left[1+\tikz[baseline=-0.5ex, scale=0.8]{
		\def\a{0}
		\def\b{0}
		\draw[thick] (-1+\a, 0+\b) -- (1+\a, 0+\b);
		\draw[thick] (0+\a, -1+\b) -- (0+\a, 1+\b);
		\node[anchor=north west] at (-0.1, 0.1) {$v$};
		\filldraw[red] (-0.5+\a,0+\b) circle (2pt) node[anchor=north, black] {$\textbf{\scriptsize{}}$};
		\filldraw[red] (0+\a, -0.5+\b) circle (2pt) node[anchor=north west, black] {$\textbf{\scriptsize{Z}}$};
		\filldraw[red] (0.5+\a,0+\b) circle (2pt) node[anchor=south, black] {$\textbf{\scriptsize{Z}}$};
		\filldraw[red] (0+\a, 0.5+\b) circle (2pt) node[anchor=south east, black] {$\textbf{\scriptsize{}}$};
	}\right]\nonumber\\
A_{v}^{(1)}&=\frac{1}{2}\left[1+\tikz[baseline=-0.5ex, scale=0.8]{
	\def\a{0}
	\def\b{0}
	\draw[thick] (-1+\a, 0+\b) -- (1+\a, 0+\b);
	\draw[thick] (0+\a, -1+\b) -- (0+\a, 1+\b);
	\node[anchor=north west] at (-0.1, 0.1) {$v$};
	\filldraw[red] (-0.5+\a,0+\b) circle (2pt) node[anchor=north, black] {$\textbf{\scriptsize{X}}$};
	\filldraw[red] (0.5+\a,0+\b) circle (2pt) node[anchor=south, black] {$\textbf{\scriptsize{X}}$};
	\filldraw[red] (0+\a, -0.5+\b) circle (2pt) node[anchor=north west, black] {$\textbf{\scriptsize{X}}$};
	\filldraw[red] (0+\a, 0.5+\b) circle (2pt) node[anchor=south east, black] {$\textbf{\scriptsize{X}}$};
}\right]
\frac{1}{2}\left[1-\tikz[baseline=-0.5ex, scale=0.8]{
	\def\a{0}
	\def\b{0}
	\draw[thick] (-1+\a, 0+\b) -- (1+\a, 0+\b);
	\draw[thick] (0+\a, -1+\b) -- (0+\a, 1+\b);
	\node[anchor=north west] at (-0.1, 0.1) {$v$};
	\filldraw[red] (-0.5+\a,0+\b) circle (2pt) node[anchor=north, black] {$\textbf{\scriptsize{Z}}$};
	\filldraw[red] (0+\a, -0.5+\b) circle (2pt) node[anchor=north west, black] {$\textbf{\scriptsize{}}$};
	\filldraw[red] (0.5+\a,0+\b) circle (2pt) node[anchor=south, black] {$\textbf{\scriptsize{}}$};
	\filldraw[red] (0+\a, 0.5+\b) circle (2pt) node[anchor=south east, black] {$\textbf{\scriptsize{Z}}$};
}\right]
\frac{1}{2}\left[1+\tikz[baseline=-0.5ex, scale=0.8]{
	\def\a{0}
	\def\b{0}
	\draw[thick] (-1+\a, 0+\b) -- (1+\a, 0+\b);
	\draw[thick] (0+\a, -1+\b) -- (0+\a, 1+\b);
	\node[anchor=north west] at (-0.1, 0.1) {$v$};
	\filldraw[red] (-0.5+\a,0+\b) circle (2pt) node[anchor=north, black] {$\textbf{\scriptsize{Z}}$};
	\filldraw[red] (0+\a, -0.5+\b) circle (2pt) node[anchor=north west, black] {$\textbf{\scriptsize{Z}}$};
	\filldraw[red] (0.5+\a,0+\b) circle (2pt) node[anchor=south, black] {$\textbf{\scriptsize{}}$};
	\filldraw[red] (0+\a, 0.5+\b) circle (2pt) node[anchor=south east, black] {$\textbf{\scriptsize{}}$};
}\right]
\frac{1}{2}\left[1+\tikz[baseline=-0.5ex, scale=0.8]{
	\def\a{0}
	\def\b{0}
	\draw[thick] (-1+\a, 0+\b) -- (1+\a, 0+\b);
	\draw[thick] (0+\a, -1+\b) -- (0+\a, 1+\b);
	\node[anchor=north west] at (-0.1, 0.1) {$v$};
	\filldraw[red] (-0.5+\a,0+\b) circle (2pt) node[anchor=north, black] {$\textbf{\scriptsize{}}$};
	\filldraw[red] (0+\a, -0.5+\b) circle (2pt) node[anchor=north west, black] {$\textbf{\scriptsize{Z}}$};
	\filldraw[red] (0.5+\a,0+\b) circle (2pt) node[anchor=south, black] {$\textbf{\scriptsize{Z}}$};
	\filldraw[red] (0+\a, 0.5+\b) circle (2pt) node[anchor=south east, black] {$\textbf{\scriptsize{}}$};
}\right]\nonumber\\
A_{v}^{(2)}&=\frac{1}{2}\left[1+\tikz[baseline=-0.5ex, scale=0.8]{
	\def\a{0}
	\def\b{0}
	\draw[thick] (-1+\a, 0+\b) -- (1+\a, 0+\b);
	\draw[thick] (0+\a, -1+\b) -- (0+\a, 1+\b);
	\node[anchor=north west] at (-0.1, 0.1) {$v$};
	\filldraw[red] (-0.5+\a,0+\b) circle (2pt) node[anchor=north, black] {$\textbf{\scriptsize{X}}$};
	\filldraw[red] (0.5+\a,0+\b) circle (2pt) node[anchor=south, black] {$\textbf{\scriptsize{X}}$};
	\filldraw[red] (0+\a, -0.5+\b) circle (2pt) node[anchor=north west, black] {$\textbf{\scriptsize{X}}$};
	\filldraw[red] (0+\a, 0.5+\b) circle (2pt) node[anchor=south east, black] {$\textbf{\scriptsize{X}}$};
}\right]
\frac{1}{2}\left[1+\tikz[baseline=-0.5ex, scale=0.8]{
	\def\a{0}
	\def\b{0}
	\draw[thick] (-1+\a, 0+\b) -- (1+\a, 0+\b);
	\draw[thick] (0+\a, -1+\b) -- (0+\a, 1+\b);
	\node[anchor=north west] at (-0.1, 0.1) {$v$};
	\filldraw[red] (-0.5+\a,0+\b) circle (2pt) node[anchor=north, black] {$\textbf{\scriptsize{Z}}$};
	\filldraw[red] (0+\a, -0.5+\b) circle (2pt) node[anchor=north west, black] {$\textbf{\scriptsize{}}$};
	\filldraw[red] (0.5+\a,0+\b) circle (2pt) node[anchor=south, black] {$\textbf{\scriptsize{}}$};
	\filldraw[red] (0+\a, 0.5+\b) circle (2pt) node[anchor=south east, black] {$\textbf{\scriptsize{Z}}$};
}\right]
\frac{1}{2}\left[1-\tikz[baseline=-0.5ex, scale=0.8]{
	\def\a{0}
	\def\b{0}
	\draw[thick] (-1+\a, 0+\b) -- (1+\a, 0+\b);
	\draw[thick] (0+\a, -1+\b) -- (0+\a, 1+\b);
	\node[anchor=north west] at (-0.1, 0.1) {$v$};
	\filldraw[red] (-0.5+\a,0+\b) circle (2pt) node[anchor=north, black] {$\textbf{\scriptsize{Z}}$};
	\filldraw[red] (0+\a, -0.5+\b) circle (2pt) node[anchor=north west, black] {$\textbf{\scriptsize{Z}}$};
	\filldraw[red] (0.5+\a,0+\b) circle (2pt) node[anchor=south, black] {$\textbf{\scriptsize{}}$};
	\filldraw[red] (0+\a, 0.5+\b) circle (2pt) node[anchor=south east, black] {$\textbf{\scriptsize{}}$};
}\right]
\frac{1}{2}\left[1+\tikz[baseline=-0.5ex, scale=0.8]{
	\def\a{0}
	\def\b{0}
	\draw[thick] (-1+\a, 0+\b) -- (1+\a, 0+\b);
	\draw[thick] (0+\a, -1+\b) -- (0+\a, 1+\b);
	\node[anchor=north west] at (-0.1, 0.1) {$v$};
	\filldraw[red] (-0.5+\a,0+\b) circle (2pt) node[anchor=north, black] {$\textbf{\scriptsize{}}$};
	\filldraw[red] (0+\a, -0.5+\b) circle (2pt) node[anchor=north west, black] {$\textbf{\scriptsize{Z}}$};
	\filldraw[red] (0.5+\a,0+\b) circle (2pt) node[anchor=south, black] {$\textbf{\scriptsize{Z}}$};
	\filldraw[red] (0+\a, 0.5+\b) circle (2pt) node[anchor=south east, black] {$\textbf{\scriptsize{}}$};
}\right]\nonumber\\
A_{v}^{(3)}&=\frac{1}{2}\left[1+\tikz[baseline=-0.5ex, scale=0.8]{
	\def\a{0}
	\def\b{0}
	\draw[thick] (-1+\a, 0+\b) -- (1+\a, 0+\b);
	\draw[thick] (0+\a, -1+\b) -- (0+\a, 1+\b);
	\node[anchor=north west] at (-0.1, 0.1) {$v$};
	\filldraw[red] (-0.5+\a,0+\b) circle (2pt) node[anchor=north, black] {$\textbf{\scriptsize{X}}$};
	\filldraw[red] (0.5+\a,0+\b) circle (2pt) node[anchor=south, black] {$\textbf{\scriptsize{X}}$};
	\filldraw[red] (0+\a, -0.5+\b) circle (2pt) node[anchor=north west, black] {$\textbf{\scriptsize{X}}$};
	\filldraw[red] (0+\a, 0.5+\b) circle (2pt) node[anchor=south east, black] {$\textbf{\scriptsize{X}}$};
}\right]
\frac{1}{2}\left[1+\tikz[baseline=-0.5ex, scale=0.8]{
	\def\a{0}
	\def\b{0}
	\draw[thick] (-1+\a, 0+\b) -- (1+\a, 0+\b);
	\draw[thick] (0+\a, -1+\b) -- (0+\a, 1+\b);
	\node[anchor=north west] at (-0.1, 0.1) {$v$};
	\filldraw[red] (-0.5+\a,0+\b) circle (2pt) node[anchor=north, black] {$\textbf{\scriptsize{Z}}$};
	\filldraw[red] (0+\a, -0.5+\b) circle (2pt) node[anchor=north west, black] {$\textbf{\scriptsize{}}$};
	\filldraw[red] (0.5+\a,0+\b) circle (2pt) node[anchor=south, black] {$\textbf{\scriptsize{}}$};
	\filldraw[red] (0+\a, 0.5+\b) circle (2pt) node[anchor=south east, black] {$\textbf{\scriptsize{Z}}$};
}\right]
\frac{1}{2}\left[1+\tikz[baseline=-0.5ex, scale=0.8]{
	\def\a{0}
	\def\b{0}
	\draw[thick] (-1+\a, 0+\b) -- (1+\a, 0+\b);
	\draw[thick] (0+\a, -1+\b) -- (0+\a, 1+\b);
	\node[anchor=north west] at (-0.1, 0.1) {$v$};
	\filldraw[red] (-0.5+\a,0+\b) circle (2pt) node[anchor=north, black] {$\textbf{\scriptsize{Z}}$};
	\filldraw[red] (0+\a, -0.5+\b) circle (2pt) node[anchor=north west, black] {$\textbf{\scriptsize{Z}}$};
	\filldraw[red] (0.5+\a,0+\b) circle (2pt) node[anchor=south, black] {$\textbf{\scriptsize{}}$};
	\filldraw[red] (0+\a, 0.5+\b) circle (2pt) node[anchor=south east, black] {$\textbf{\scriptsize{}}$};
}\right]
\frac{1}{2}\left[1-\tikz[baseline=-0.5ex, scale=0.8]{
	\def\a{0}
	\def\b{0}
	\draw[thick] (-1+\a, 0+\b) -- (1+\a, 0+\b);
	\draw[thick] (0+\a, -1+\b) -- (0+\a, 1+\b);
	\node[anchor=north west] at (-0.1, 0.1) {$v$};
	\filldraw[red] (-0.5+\a,0+\b) circle (2pt) node[anchor=north, black] {$\textbf{\scriptsize{}}$};
	\filldraw[red] (0+\a, -0.5+\b) circle (2pt) node[anchor=north west, black] {$\textbf{\scriptsize{Z}}$};
	\filldraw[red] (0.5+\a,0+\b) circle (2pt) node[anchor=south, black] {$\textbf{\scriptsize{Z}}$};
	\filldraw[red] (0+\a, 0.5+\b) circle (2pt) node[anchor=south east, black] {$\textbf{\scriptsize{}}$};
}\right]\nonumber\\
A_{v}^{(4)}&=\frac{1}{2}\left[1+\tikz[baseline=-0.5ex, scale=0.8]{
	\def\a{0}
	\def\b{0}
	\draw[thick] (-1+\a, 0+\b) -- (1+\a, 0+\b);
	\draw[thick] (0+\a, -1+\b) -- (0+\a, 1+\b);
	\node[anchor=north west] at (-0.1, 0.1) {$v$};
	\filldraw[red] (-0.5+\a,0+\b) circle (2pt) node[anchor=north, black] {$\textbf{\scriptsize{X}}$};
	\filldraw[red] (0.5+\a,0+\b) circle (2pt) node[anchor=south, black] {$\textbf{\scriptsize{X}}$};
	\filldraw[red] (0+\a, -0.5+\b) circle (2pt) node[anchor=north west, black] {$\textbf{\scriptsize{X}}$};
	\filldraw[red] (0+\a, 0.5+\b) circle (2pt) node[anchor=south east, black] {$\textbf{\scriptsize{X}}$};
}\right]
\frac{1}{2}\left[1-\tikz[baseline=-0.5ex, scale=0.8]{
	\def\a{0}
	\def\b{0}
	\draw[thick] (-1+\a, 0+\b) -- (1+\a, 0+\b);
	\draw[thick] (0+\a, -1+\b) -- (0+\a, 1+\b);
	\node[anchor=north west] at (-0.1, 0.1) {$v$};
	\filldraw[red] (-0.5+\a,0+\b) circle (2pt) node[anchor=north, black] {$\textbf{\scriptsize{Z}}$};
	\filldraw[red] (0+\a, -0.5+\b) circle (2pt) node[anchor=north west, black] {$\textbf{\scriptsize{}}$};
	\filldraw[red] (0.5+\a,0+\b) circle (2pt) node[anchor=south, black] {$\textbf{\scriptsize{}}$};
	\filldraw[red] (0+\a, 0.5+\b) circle (2pt) node[anchor=south east, black] {$\textbf{\scriptsize{Z}}$};
}\right]
\frac{1}{2}\left[1-\tikz[baseline=-0.5ex, scale=0.8]{
	\def\a{0}
	\def\b{0}
	\draw[thick] (-1+\a, 0+\b) -- (1+\a, 0+\b);
	\draw[thick] (0+\a, -1+\b) -- (0+\a, 1+\b);
	\node[anchor=north west] at (-0.1, 0.1) {$v$};
	\filldraw[red] (-0.5+\a,0+\b) circle (2pt) node[anchor=north, black] {$\textbf{\scriptsize{Z}}$};
	\filldraw[red] (0+\a, -0.5+\b) circle (2pt) node[anchor=north west, black] {$\textbf{\scriptsize{Z}}$};
	\filldraw[red] (0.5+\a,0+\b) circle (2pt) node[anchor=south, black] {$\textbf{\scriptsize{}}$};
	\filldraw[red] (0+\a, 0.5+\b) circle (2pt) node[anchor=south east, black] {$\textbf{\scriptsize{}}$};
}\right]
\frac{1}{2}\left[1+\tikz[baseline=-0.5ex, scale=0.8]{
	\def\a{0}
	\def\b{0}
	\draw[thick] (-1+\a, 0+\b) -- (1+\a, 0+\b);
	\draw[thick] (0+\a, -1+\b) -- (0+\a, 1+\b);
	\node[anchor=north west] at (-0.1, 0.1) {$v$};
	\filldraw[red] (-0.5+\a,0+\b) circle (2pt) node[anchor=north, black] {$\textbf{\scriptsize{}}$};
	\filldraw[red] (0+\a, -0.5+\b) circle (2pt) node[anchor=north west, black] {$\textbf{\scriptsize{Z}}$};
	\filldraw[red] (0.5+\a,0+\b) circle (2pt) node[anchor=south, black] {$\textbf{\scriptsize{Z}}$};
	\filldraw[red] (0+\a, 0.5+\b) circle (2pt) node[anchor=south east, black] {$\textbf{\scriptsize{}}$};
}\right]\nonumber\\
A_{v}^{(5)}&=\frac{1}{2}\left[1+\tikz[baseline=-0.5ex, scale=0.8]{
	\def\a{0}
	\def\b{0}
	\draw[thick] (-1+\a, 0+\b) -- (1+\a, 0+\b);
	\draw[thick] (0+\a, -1+\b) -- (0+\a, 1+\b);
	\node[anchor=north west] at (-0.1, 0.1) {$v$};
	\filldraw[red] (-0.5+\a,0+\b) circle (2pt) node[anchor=north, black] {$\textbf{\scriptsize{X}}$};
	\filldraw[red] (0.5+\a,0+\b) circle (2pt) node[anchor=south, black] {$\textbf{\scriptsize{X}}$};
	\filldraw[red] (0+\a, -0.5+\b) circle (2pt) node[anchor=north west, black] {$\textbf{\scriptsize{X}}$};
	\filldraw[red] (0+\a, 0.5+\b) circle (2pt) node[anchor=south east, black] {$\textbf{\scriptsize{X}}$};
}\right]
\frac{1}{2}\left[1-\tikz[baseline=-0.5ex, scale=0.8]{
	\def\a{0}
	\def\b{0}
	\draw[thick] (-1+\a, 0+\b) -- (1+\a, 0+\b);
	\draw[thick] (0+\a, -1+\b) -- (0+\a, 1+\b);
	\node[anchor=north west] at (-0.1, 0.1) {$v$};
	\filldraw[red] (-0.5+\a,0+\b) circle (2pt) node[anchor=north, black] {$\textbf{\scriptsize{Z}}$};
	\filldraw[red] (0+\a, -0.5+\b) circle (2pt) node[anchor=north west, black] {$\textbf{\scriptsize{}}$};
	\filldraw[red] (0.5+\a,0+\b) circle (2pt) node[anchor=south, black] {$\textbf{\scriptsize{}}$};
	\filldraw[red] (0+\a, 0.5+\b) circle (2pt) node[anchor=south east, black] {$\textbf{\scriptsize{Z}}$};
}\right]
\frac{1}{2}\left[1+\tikz[baseline=-0.5ex, scale=0.8]{
	\def\a{0}
	\def\b{0}
	\draw[thick] (-1+\a, 0+\b) -- (1+\a, 0+\b);
	\draw[thick] (0+\a, -1+\b) -- (0+\a, 1+\b);
	\node[anchor=north west] at (-0.1, 0.1) {$v$};
	\filldraw[red] (-0.5+\a,0+\b) circle (2pt) node[anchor=north, black] {$\textbf{\scriptsize{Z}}$};
	\filldraw[red] (0+\a, -0.5+\b) circle (2pt) node[anchor=north west, black] {$\textbf{\scriptsize{Z}}$};
	\filldraw[red] (0.5+\a,0+\b) circle (2pt) node[anchor=south, black] {$\textbf{\scriptsize{}}$};
	\filldraw[red] (0+\a, 0.5+\b) circle (2pt) node[anchor=south east, black] {$\textbf{\scriptsize{}}$};
}\right]
\frac{1}{2}\left[1-\tikz[baseline=-0.5ex, scale=0.8]{
	\def\a{0}
	\def\b{0}
	\draw[thick] (-1+\a, 0+\b) -- (1+\a, 0+\b);
	\draw[thick] (0+\a, -1+\b) -- (0+\a, 1+\b);
	\node[anchor=north west] at (-0.1, 0.1) {$v$};
	\filldraw[red] (-0.5+\a,0+\b) circle (2pt) node[anchor=north, black] {$\textbf{\scriptsize{}}$};
	\filldraw[red] (0+\a, -0.5+\b) circle (2pt) node[anchor=north west, black] {$\textbf{\scriptsize{Z}}$};
	\filldraw[red] (0.5+\a,0+\b) circle (2pt) node[anchor=south, black] {$\textbf{\scriptsize{Z}}$};
	\filldraw[red] (0+\a, 0.5+\b) circle (2pt) node[anchor=south east, black] {$\textbf{\scriptsize{}}$};
}\right]\nonumber\\
A_{v}^{(6)}&=\frac{1}{2}\left[1+\tikz[baseline=-0.5ex, scale=0.8]{
	\def\a{0}
	\def\b{0}
	\draw[thick] (-1+\a, 0+\b) -- (1+\a, 0+\b);
	\draw[thick] (0+\a, -1+\b) -- (0+\a, 1+\b);
	\node[anchor=north west] at (-0.1, 0.1) {$v$};
	\filldraw[red] (-0.5+\a,0+\b) circle (2pt) node[anchor=north, black] {$\textbf{\scriptsize{X}}$};
	\filldraw[red] (0.5+\a,0+\b) circle (2pt) node[anchor=south, black] {$\textbf{\scriptsize{X}}$};
	\filldraw[red] (0+\a, -0.5+\b) circle (2pt) node[anchor=north west, black] {$\textbf{\scriptsize{X}}$};
	\filldraw[red] (0+\a, 0.5+\b) circle (2pt) node[anchor=south east, black] {$\textbf{\scriptsize{X}}$};
}\right]
\frac{1}{2}\left[1+\tikz[baseline=-0.5ex, scale=0.8]{
	\def\a{0}
	\def\b{0}
	\draw[thick] (-1+\a, 0+\b) -- (1+\a, 0+\b);
	\draw[thick] (0+\a, -1+\b) -- (0+\a, 1+\b);
	\node[anchor=north west] at (-0.1, 0.1) {$v$};
	\filldraw[red] (-0.5+\a,0+\b) circle (2pt) node[anchor=north, black] {$\textbf{\scriptsize{Z}}$};
	\filldraw[red] (0+\a, -0.5+\b) circle (2pt) node[anchor=north west, black] {$\textbf{\scriptsize{}}$};
	\filldraw[red] (0.5+\a,0+\b) circle (2pt) node[anchor=south, black] {$\textbf{\scriptsize{}}$};
	\filldraw[red] (0+\a, 0.5+\b) circle (2pt) node[anchor=south east, black] {$\textbf{\scriptsize{Z}}$};
}\right]
\frac{1}{2}\left[1-\tikz[baseline=-0.5ex, scale=0.8]{
	\def\a{0}
	\def\b{0}
	\draw[thick] (-1+\a, 0+\b) -- (1+\a, 0+\b);
	\draw[thick] (0+\a, -1+\b) -- (0+\a, 1+\b);
	\node[anchor=north west] at (-0.1, 0.1) {$v$};
	\filldraw[red] (-0.5+\a,0+\b) circle (2pt) node[anchor=north, black] {$\textbf{\scriptsize{Z}}$};
	\filldraw[red] (0+\a, -0.5+\b) circle (2pt) node[anchor=north west, black] {$\textbf{\scriptsize{Z}}$};
	\filldraw[red] (0.5+\a,0+\b) circle (2pt) node[anchor=south, black] {$\textbf{\scriptsize{}}$};
	\filldraw[red] (0+\a, 0.5+\b) circle (2pt) node[anchor=south east, black] {$\textbf{\scriptsize{}}$};
}\right]
\frac{1}{2}\left[1-\tikz[baseline=-0.5ex, scale=0.8]{
	\def\a{0}
	\def\b{0}
	\draw[thick] (-1+\a, 0+\b) -- (1+\a, 0+\b);
	\draw[thick] (0+\a, -1+\b) -- (0+\a, 1+\b);
	\node[anchor=north west] at (-0.1, 0.1) {$v$};
	\filldraw[red] (-0.5+\a,0+\b) circle (2pt) node[anchor=north, black] {$\textbf{\scriptsize{}}$};
	\filldraw[red] (0+\a, -0.5+\b) circle (2pt) node[anchor=north west, black] {$\textbf{\scriptsize{Z}}$};
	\filldraw[red] (0.5+\a,0+\b) circle (2pt) node[anchor=south, black] {$\textbf{\scriptsize{Z}}$};
	\filldraw[red] (0+\a, 0.5+\b) circle (2pt) node[anchor=south east, black] {$\textbf{\scriptsize{}}$};
}\right]\nonumber\\
A_{v}^{(7)}&=\frac{1}{2}\left[1+\tikz[baseline=-0.5ex, scale=0.8]{
	\def\a{0}
	\def\b{0}
	\draw[thick] (-1+\a, 0+\b) -- (1+\a, 0+\b);
	\draw[thick] (0+\a, -1+\b) -- (0+\a, 1+\b);
	\node[anchor=north west] at (-0.1, 0.1) {$v$};
	\filldraw[red] (-0.5+\a,0+\b) circle (2pt) node[anchor=north, black] {$\textbf{\scriptsize{X}}$};
	\filldraw[red] (0.5+\a,0+\b) circle (2pt) node[anchor=south, black] {$\textbf{\scriptsize{X}}$};
	\filldraw[red] (0+\a, -0.5+\b) circle (2pt) node[anchor=north west, black] {$\textbf{\scriptsize{X}}$};
	\filldraw[red] (0+\a, 0.5+\b) circle (2pt) node[anchor=south east, black] {$\textbf{\scriptsize{X}}$};
}\right]
\frac{1}{2}\left[1-\tikz[baseline=-0.5ex, scale=0.8]{
	\def\a{0}
	\def\b{0}
	\draw[thick] (-1+\a, 0+\b) -- (1+\a, 0+\b);
	\draw[thick] (0+\a, -1+\b) -- (0+\a, 1+\b);
	\node[anchor=north west] at (-0.1, 0.1) {$v$};
	\filldraw[red] (-0.5+\a,0+\b) circle (2pt) node[anchor=north, black] {$\textbf{\scriptsize{Z}}$};
	\filldraw[red] (0+\a, -0.5+\b) circle (2pt) node[anchor=north west, black] {$\textbf{\scriptsize{}}$};
	\filldraw[red] (0.5+\a,0+\b) circle (2pt) node[anchor=south, black] {$\textbf{\scriptsize{}}$};
	\filldraw[red] (0+\a, 0.5+\b) circle (2pt) node[anchor=south east, black] {$\textbf{\scriptsize{Z}}$};
}\right]
\frac{1}{2}\left[1-\tikz[baseline=-0.5ex, scale=0.8]{
	\def\a{0}
	\def\b{0}
	\draw[thick] (-1+\a, 0+\b) -- (1+\a, 0+\b);
	\draw[thick] (0+\a, -1+\b) -- (0+\a, 1+\b);
	\node[anchor=north west] at (-0.1, 0.1) {$v$};
	\filldraw[red] (-0.5+\a,0+\b) circle (2pt) node[anchor=north, black] {$\textbf{\scriptsize{Z}}$};
	\filldraw[red] (0+\a, -0.5+\b) circle (2pt) node[anchor=north west, black] {$\textbf{\scriptsize{Z}}$};
	\filldraw[red] (0.5+\a,0+\b) circle (2pt) node[anchor=south, black] {$\textbf{\scriptsize{}}$};
	\filldraw[red] (0+\a, 0.5+\b) circle (2pt) node[anchor=south east, black] {$\textbf{\scriptsize{}}$};
}\right]
\frac{1}{2}\left[1-\tikz[baseline=-0.5ex, scale=0.8]{
	\def\a{0}
	\def\b{0}
	\draw[thick] (-1+\a, 0+\b) -- (1+\a, 0+\b);
	\draw[thick] (0+\a, -1+\b) -- (0+\a, 1+\b);
	\node[anchor=north west] at (-0.1, 0.1) {$v$};
	\filldraw[red] (-0.5+\a,0+\b) circle (2pt) node[anchor=north, black] {$\textbf{\scriptsize{}}$};
	\filldraw[red] (0+\a, -0.5+\b) circle (2pt) node[anchor=north west, black] {$\textbf{\scriptsize{Z}}$};
	\filldraw[red] (0.5+\a,0+\b) circle (2pt) node[anchor=south, black] {$\textbf{\scriptsize{Z}}$};
	\filldraw[red] (0+\a, 0.5+\b) circle (2pt) node[anchor=south east, black] {$\textbf{\scriptsize{}}$};
}\right]
\end{align}
However note that these are not the only orthogonal operators that we expect to find. The other set of eight operators is obtained by replacing
	$\frac{1}{2}\left[1+\tikz[baseline=-0.5ex, scale=0.5]{
		\def\a{0}
		\def\b{0}
		\draw[thick] (-1+\a, 0+\b) -- (1+\a, 0+\b);
		\draw[thick] (0+\a, -1+\b) -- (0+\a, 1+\b);
		\node[anchor=north west] at (-0.1, 0.1) {$v$};
		\filldraw[red] (-0.5+\a,0+\b) circle (2pt) node[anchor=north, black] {$\textbf{\scriptsize{X}}$};
		\filldraw[red] (0+\a, -0.5+\b) circle (2pt) node[anchor=north west, black] {$\textbf{\scriptsize{X}}$};
		\filldraw[red] (0.5+\a,0+\b) circle (2pt) node[anchor=south, black] {$\textbf{\scriptsize{X}}$};
		\filldraw[red] (0+\a, 0.5+\b) circle (2pt) node[anchor=south east, black] {$\textbf{\scriptsize{X}}$};
	}\right]$
with
	$\frac{1}{2}\left[1-\tikz[baseline=-0.5ex, scale=0.5]{
		\def\a{0}
		\def\b{0}
		\draw[thick] (-1+\a, 0+\b) -- (1+\a, 0+\b);
		\draw[thick] (0+\a, -1+\b) -- (0+\a, 1+\b);
		\node[anchor=north west] at (-0.1, 0.1) {$v$};
		\filldraw[red] (-0.5+\a,0+\b) circle (2pt) node[anchor=north, black] {$\textbf{\scriptsize{X}}$};
		\filldraw[red] (0+\a, -0.5+\b) circle (2pt) node[anchor=north west, black] {$\textbf{\scriptsize{X}}$};
		\filldraw[red] (0.5+\a,0+\b) circle (2pt) node[anchor=south, black] {$\textbf{\scriptsize{X}}$};
		\filldraw[red] (0+\a, 0.5+\b) circle (2pt) node[anchor=south east, black] {$\textbf{\scriptsize{X}}$};
	}\right].$
Thus these 16 mutually orthogonal operators add up to the identity operator and exhaust all possible vertex operators on every vertex of the square lattice.

\begin{figure}
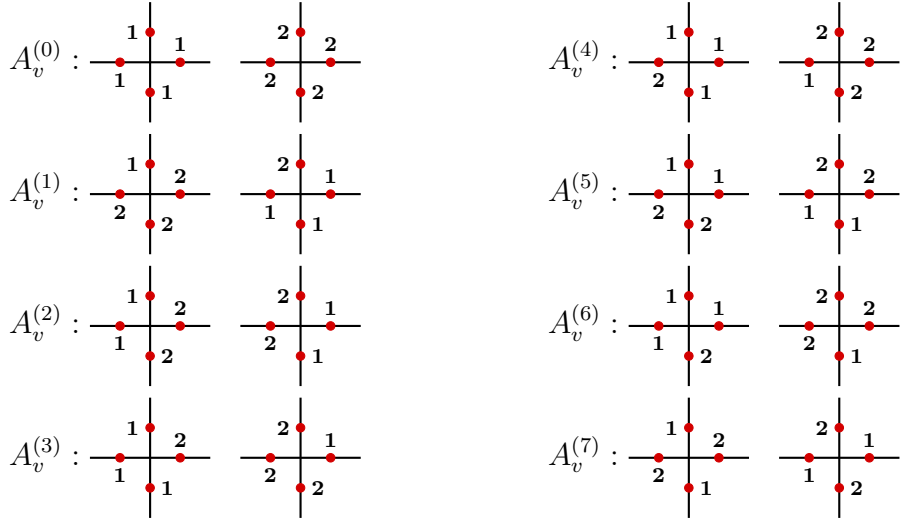

    \begin{align*}
	A_{v}^{(0)}&:
	\tikz[baseline=-0.5ex, scale=0.8]{
		\def\a{-1.25}
		\def\b{0}
		\draw[thick] (-1+\a, 0+\b) -- (1+\a, 0+\b);
		\draw[thick] (0+\a, -1+\b) -- (0+\a, 1+\b);
		\filldraw[red] (-0.5+\a,0+\b) circle (2pt) node[anchor=north, black] {$\textbf{\scriptsize{1}}$};
		\filldraw[red] (0+\a, -0.5+\b) circle (2pt) node[anchor=west, black] {$\textbf{\scriptsize{1}}$};
		\filldraw[red] (0.5+\a,0+\b) circle (2pt) node[anchor=south, black] {$\textbf{\scriptsize{1}}$};
		\filldraw[red] (0+\a, 0.5+\b) circle (2pt) node[anchor=east, black] {$\textbf{\scriptsize{1}}$};
		\def\a{1.25}
		\def\b{0}
		\draw[thick] (-1+\a, 0+\b) -- (1+\a, 0+\b);
		\draw[thick] (0+\a, -1+\b) -- (0+\a, 1+\b);
		\filldraw[red] (-0.5+\a,0+\b) circle (2pt) node[anchor=north, black] {$\textbf{\scriptsize{2}}$};
		\filldraw[red] (0+\a, -0.5+\b) circle (2pt) node[anchor=west, black] {$\textbf{\scriptsize{2}}$};
		\filldraw[red] (0.5+\a,0+\b) circle (2pt) node[anchor=south, black] {$\textbf{\scriptsize{2}}$};
		\filldraw[red] (0+\a, 0.5+\b) circle (2pt) node[anchor=east, black] {$\textbf{\scriptsize{2}}$};
	}
	 &A_{v}^{(4)}&:
	 \tikz[baseline=-0.5ex, scale=0.8]{
	 	\def\a{-1.25}
	 	\def\b{0}
	 	\draw[thick] (-1+\a, 0+\b) -- (1+\a, 0+\b);
	 	\draw[thick] (0+\a, -1+\b) -- (0+\a, 1+\b);
	 	\filldraw[red] (-0.5+\a,0+\b) circle (2pt) node[anchor=north, black] {$\textbf{\scriptsize{2}}$};
	 	\filldraw[red] (0+\a, -0.5+\b) circle (2pt) node[anchor=west, black] {$\textbf{\scriptsize{1}}$};
	 	\filldraw[red] (0.5+\a,0+\b) circle (2pt) node[anchor=south, black] {$\textbf{\scriptsize{1}}$};
	 	\filldraw[red] (0+\a, 0.5+\b) circle (2pt) node[anchor=east, black] {$\textbf{\scriptsize{1}}$};
	 	\def\a{1.25}
	 	\def\b{0}
	 	\draw[thick] (-1+\a, 0+\b) -- (1+\a, 0+\b);
	 	\draw[thick] (0+\a, -1+\b) -- (0+\a, 1+\b);
	 	\filldraw[red] (-0.5+\a,0+\b) circle (2pt) node[anchor=north, black] {$\textbf{\scriptsize{1}}$};
	 	\filldraw[red] (0+\a, -0.5+\b) circle (2pt) node[anchor=west, black] {$\textbf{\scriptsize{2}}$};
	 	\filldraw[red] (0.5+\a,0+\b) circle (2pt) node[anchor=south, black] {$\textbf{\scriptsize{2}}$};
	 	\filldraw[red] (0+\a, 0.5+\b) circle (2pt) node[anchor=east, black] {$\textbf{\scriptsize{2}}$};
	 }\\
 	A_{v}^{(1)}&:
 \tikz[baseline=-0.5ex, scale=0.8]{
 	\def\a{-1.25}
 	\def\b{0}
 	\draw[thick] (-1+\a, 0+\b) -- (1+\a, 0+\b);
 	\draw[thick] (0+\a, -1+\b) -- (0+\a, 1+\b);
 	\filldraw[red] (-0.5+\a,0+\b) circle (2pt) node[anchor=north, black] {$\textbf{\scriptsize{2}}$};
 	\filldraw[red] (0+\a, -0.5+\b) circle (2pt) node[anchor=west, black] {$\textbf{\scriptsize{2}}$};
 	\filldraw[red] (0.5+\a,0+\b) circle (2pt) node[anchor=south, black] {$\textbf{\scriptsize{2}}$};
 	\filldraw[red] (0+\a, 0.5+\b) circle (2pt) node[anchor=east, black] {$\textbf{\scriptsize{1}}$};
 	\def\a{1.25}
 	\def\b{0}
 	\draw[thick] (-1+\a, 0+\b) -- (1+\a, 0+\b);
 	\draw[thick] (0+\a, -1+\b) -- (0+\a, 1+\b);
 	\filldraw[red] (-0.5+\a,0+\b) circle (2pt) node[anchor=north, black] {$\textbf{\scriptsize{1}}$};
 	\filldraw[red] (0+\a, -0.5+\b) circle (2pt) node[anchor=west, black] {$\textbf{\scriptsize{1}}$};
 	\filldraw[red] (0.5+\a,0+\b) circle (2pt) node[anchor=south, black] {$\textbf{\scriptsize{1}}$};
 	\filldraw[red] (0+\a, 0.5+\b) circle (2pt) node[anchor=east, black] {$\textbf{\scriptsize{2}}$};
 }
 &A_{v}^{(5)}&:
 \tikz[baseline=-0.5ex, scale=0.8]{
 	\def\a{-1.25}
 	\def\b{0}
 	\draw[thick] (-1+\a, 0+\b) -- (1+\a, 0+\b);
 	\draw[thick] (0+\a, -1+\b) -- (0+\a, 1+\b);
 	\filldraw[red] (-0.5+\a,0+\b) circle (2pt) node[anchor=north, black] {$\textbf{\scriptsize{2}}$};
 	\filldraw[red] (0+\a, -0.5+\b) circle (2pt) node[anchor=west, black] {$\textbf{\scriptsize{2}}$};
 	\filldraw[red] (0.5+\a,0+\b) circle (2pt) node[anchor=south, black] {$\textbf{\scriptsize{1}}$};
 	\filldraw[red] (0+\a, 0.5+\b) circle (2pt) node[anchor=east, black] {$\textbf{\scriptsize{1}}$};
 	\def\a{1.25}
 	\def\b{0}
 	\draw[thick] (-1+\a, 0+\b) -- (1+\a, 0+\b);
 	\draw[thick] (0+\a, -1+\b) -- (0+\a, 1+\b);
 	\filldraw[red] (-0.5+\a,0+\b) circle (2pt) node[anchor=north, black] {$\textbf{\scriptsize{1}}$};
 	\filldraw[red] (0+\a, -0.5+\b) circle (2pt) node[anchor=west, black] {$\textbf{\scriptsize{1}}$};
 	\filldraw[red] (0.5+\a,0+\b) circle (2pt) node[anchor=south, black] {$\textbf{\scriptsize{2}}$};
 	\filldraw[red] (0+\a, 0.5+\b) circle (2pt) node[anchor=east, black] {$\textbf{\scriptsize{2}}$};
 }\\
	A_{v}^{(2)}&:
\tikz[baseline=-0.5ex, scale=0.8]{
	\def\a{-1.25}
	\def\b{0}
	\draw[thick] (-1+\a, 0+\b) -- (1+\a, 0+\b);
	\draw[thick] (0+\a, -1+\b) -- (0+\a, 1+\b);
	\filldraw[red] (-0.5+\a,0+\b) circle (2pt) node[anchor=north, black] {$\textbf{\scriptsize{1}}$};
	\filldraw[red] (0+\a, -0.5+\b) circle (2pt) node[anchor=west, black] {$\textbf{\scriptsize{2}}$};
	\filldraw[red] (0.5+\a,0+\b) circle (2pt) node[anchor=south, black] {$\textbf{\scriptsize{2}}$};
	\filldraw[red] (0+\a, 0.5+\b) circle (2pt) node[anchor=east, black] {$\textbf{\scriptsize{1}}$};
	\def\a{1.25}
	\def\b{0}
	\draw[thick] (-1+\a, 0+\b) -- (1+\a, 0+\b);
	\draw[thick] (0+\a, -1+\b) -- (0+\a, 1+\b);
	\filldraw[red] (-0.5+\a,0+\b) circle (2pt) node[anchor=north, black] {$\textbf{\scriptsize{2}}$};
	\filldraw[red] (0+\a, -0.5+\b) circle (2pt) node[anchor=west, black] {$\textbf{\scriptsize{1}}$};
	\filldraw[red] (0.5+\a,0+\b) circle (2pt) node[anchor=south, black] {$\textbf{\scriptsize{1}}$};
	\filldraw[red] (0+\a, 0.5+\b) circle (2pt) node[anchor=east, black] {$\textbf{\scriptsize{2}}$};
}
&A_{v}^{(6)}&:
\tikz[baseline=-0.5ex, scale=0.8]{
	\def\a{-1.25}
	\def\b{0}
	\draw[thick] (-1+\a, 0+\b) -- (1+\a, 0+\b);
	\draw[thick] (0+\a, -1+\b) -- (0+\a, 1+\b);
	\filldraw[red] (-0.5+\a,0+\b) circle (2pt) node[anchor=north, black] {$\textbf{\scriptsize{1}}$};
	\filldraw[red] (0+\a, -0.5+\b) circle (2pt) node[anchor=west, black] {$\textbf{\scriptsize{2}}$};
	\filldraw[red] (0.5+\a,0+\b) circle (2pt) node[anchor=south, black] {$\textbf{\scriptsize{1}}$};
	\filldraw[red] (0+\a, 0.5+\b) circle (2pt) node[anchor=east, black] {$\textbf{\scriptsize{1}}$};
	\def\a{1.25}
	\def\b{0}
	\draw[thick] (-1+\a, 0+\b) -- (1+\a, 0+\b);
	\draw[thick] (0+\a, -1+\b) -- (0+\a, 1+\b);
	\filldraw[red] (-0.5+\a,0+\b) circle (2pt) node[anchor=north, black] {$\textbf{\scriptsize{2}}$};
	\filldraw[red] (0+\a, -0.5+\b) circle (2pt) node[anchor=west, black] {$\textbf{\scriptsize{1}}$};
	\filldraw[red] (0.5+\a,0+\b) circle (2pt) node[anchor=south, black] {$\textbf{\scriptsize{2}}$};
	\filldraw[red] (0+\a, 0.5+\b) circle (2pt) node[anchor=east, black] {$\textbf{\scriptsize{2}}$};
}\\
	A_{v}^{(3)}&:
\tikz[baseline=-0.5ex, scale=0.8]{
	\def\a{-1.25}
	\def\b{0}
	\draw[thick] (-1+\a, 0+\b) -- (1+\a, 0+\b);
	\draw[thick] (0+\a, -1+\b) -- (0+\a, 1+\b);
	\filldraw[red] (-0.5+\a,0+\b) circle (2pt) node[anchor=north, black] {$\textbf{\scriptsize{1}}$};
	\filldraw[red] (0+\a, -0.5+\b) circle (2pt) node[anchor=west, black] {$\textbf{\scriptsize{1}}$};
	\filldraw[red] (0.5+\a,0+\b) circle (2pt) node[anchor=south, black] {$\textbf{\scriptsize{2}}$};
	\filldraw[red] (0+\a, 0.5+\b) circle (2pt) node[anchor=east, black] {$\textbf{\scriptsize{1}}$};
	\def\a{1.25}
	\def\b{0}
	\draw[thick] (-1+\a, 0+\b) -- (1+\a, 0+\b);
	\draw[thick] (0+\a, -1+\b) -- (0+\a, 1+\b);
	\filldraw[red] (-0.5+\a,0+\b) circle (2pt) node[anchor=north, black] {$\textbf{\scriptsize{2}}$};
	\filldraw[red] (0+\a, -0.5+\b) circle (2pt) node[anchor=west, black] {$\textbf{\scriptsize{2}}$};
	\filldraw[red] (0.5+\a,0+\b) circle (2pt) node[anchor=south, black] {$\textbf{\scriptsize{1}}$};
	\filldraw[red] (0+\a, 0.5+\b) circle (2pt) node[anchor=east, black] {$\textbf{\scriptsize{2}}$};
}
&A_{v}^{(7)}&:
\tikz[baseline=-0.5ex, scale=0.8]{
	\def\a{-1.25}
	\def\b{0}
	\draw[thick] (-1+\a, 0+\b) -- (1+\a, 0+\b);
	\draw[thick] (0+\a, -1+\b) -- (0+\a, 1+\b);
	\filldraw[red] (-0.5+\a,0+\b) circle (2pt) node[anchor=north, black] {$\textbf{\scriptsize{2}}$};
	\filldraw[red] (0+\a, -0.5+\b) circle (2pt) node[anchor=west, black] {$\textbf{\scriptsize{1}}$};
	\filldraw[red] (0.5+\a,0+\b) circle (2pt) node[anchor=south, black] {$\textbf{\scriptsize{2}}$};
	\filldraw[red] (0+\a, 0.5+\b) circle (2pt) node[anchor=east, black] {$\textbf{\scriptsize{1}}$};
	\def\a{1.25}
	\def\b{0}
	\draw[thick] (-1+\a, 0+\b) -- (1+\a, 0+\b);
	\draw[thick] (0+\a, -1+\b) -- (0+\a, 1+\b);
	\filldraw[red] (-0.5+\a,0+\b) circle (2pt) node[anchor=north, black] {$\textbf{\scriptsize{1}}$};
	\filldraw[red] (0+\a, -0.5+\b) circle (2pt) node[anchor=west, black] {$\textbf{\scriptsize{2}}$};
	\filldraw[red] (0.5+\a,0+\b) circle (2pt) node[anchor=south, black] {$\textbf{\scriptsize{1}}$};
	\filldraw[red] (0+\a, 0.5+\b) circle (2pt) node[anchor=east, black] {$\textbf{\scriptsize{2}}$};
}\\
\end{align*}
    \caption{The configurations on which the vertex operators act non-trivially.}
    \label{fig:vertexconfigswithvertexaction}
\end{figure}

Just like the vertex operators on vertices with valency two, the four valence vertex operators in Eq. \ref{eq:vertexopsqubitversionvalence4} act as the $\mathbb{Z}_2$ toric code vertex operators but after checking if the four corners of the vertices are either in a matched or a mismatched configuration. We summarize the action of these operators in Fig. \ref{fig:vertexconfigswithvertexaction}.

Having exhausted the mutually orthogonal set of vertex projectors we turn to the face operators on the square lattice. To this end we pick a convention for measuring the holonomy of a face as
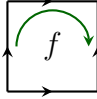
\tikz[scale = 1.2, baseline = 2.5ex, decoration={
	markings,
	mark=at position 0.5 with {\arrow{>}}}]{
	\draw[postaction={decorate}, thick] (0,0) -- (1, 0);
	\draw[postaction={decorate}, thick] (1,0) -- (1, 1);
	\draw[postaction={decorate}, thick] (0,0) -- (0, 1);
	\draw[postaction={decorate}, thick] (0,1) -- (1, 1);
	\node at (0.5, 0.5) {$f$};
	\draw [->, thick, black!60!green] (0.1,0.5)  arc (180:0:0.4);
}. The holonomy is computed clockwise starting from the vertex in the bottom-left corner. For example a configuration whose holonomy is $1_1$ is given by 
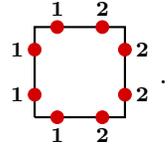
\tikz[scale=1.2, baseline=2.5ex]{
		\draw[thick] (0, 0) -- (1, 0) -- (1, 1) -- (0, 1) -- (0, 0);
		\filldraw[red] (0.25,0) circle (2pt) node[anchor=north, black] {$\textbf{\scriptsize{1}}$};
		\filldraw[red] (0.75,0) circle (2pt) node[anchor=north, black] {$\textbf{\scriptsize{2}}$};
		\filldraw[red] (1,0.25) circle (2pt) node[anchor=west, black] {$\textbf{\scriptsize{2}}$};
		\filldraw[red] (1,0.75) circle (2pt) node[anchor=west, black] {$\textbf{\scriptsize{2}}$};
		\filldraw[red] (0.25,1) circle (2pt) node[anchor=south, black] {$\textbf{\scriptsize{1}}$};
		\filldraw[red] (0.75,1) circle (2pt) node[anchor=south, black] {$\textbf{\scriptsize{2}}$};
		\filldraw[red] (0,0.25) circle (2pt) node[anchor=east, black] {$\textbf{\scriptsize{1}}$};
		\filldraw[red] (0,0.75) circle (2pt) node[anchor=east, black] {$\textbf{\scriptsize{1}}$};
	}.
We will first construct the face operators whose holonomies are the four morphisms, $1_1$, $1_2$, $a$ and $a^{-1}$, in the qubit space. To do this we need to check for the parity (match or mismatch) of the configurations on  three corners of each face; 
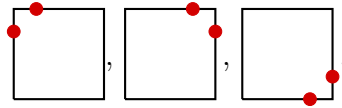
\tikz[scale=1.2, baseline=2.5ex]{
	\draw[thick] (0, 0) -- (1, 0) -- (1, 1) -- (0, 1) -- (0, 0);
 \filldraw[red] (0, 0.75) circle (2pt);
 \filldraw[red] (0.25, 1) circle (2pt);
}, \tikz[scale=1.2, baseline=2.5ex]{
	\draw[thick] (0, 0) -- (1, 0) -- (1, 1) -- (0, 1) -- (0, 0);
 \filldraw[red] (1, 0.75) circle (2pt);
 \filldraw[red] (0.75, 1) circle (2pt);
}, \tikz[scale=1.2, baseline=2.5ex]{
	\draw[thick] (0, 0) -- (1, 0) -- (1, 1) -- (0, 1) -- (0, 0);
 \filldraw[red] (0.75, 0) circle (2pt);
 \filldraw[red] (1, 0.25) circle (2pt);
}.

The operators that measure the parity on the corners are 
\begin{align}\label{eq:cornerfaceprojectors}
	F^{NW, \pm} & = \frac{1}{2}\left[1 \pm \tikz[baseline=2.5ex]{
		\draw[thick] (0, 0) -- (1, 0) -- (1, 1) -- (0, 1) -- (0, 0);
		\filldraw[red] (0.75,0) circle (2pt) node[anchor=north, black] {$\textbf{\scriptsize{}}$};
		\filldraw[red] (1,0.25) circle (2pt) node[anchor=west, black] {$\textbf{\scriptsize{}}$};
		\filldraw[red] (1,0.75) circle (2pt) node[anchor=west, black] {$\textbf{\scriptsize{}}$};
		\filldraw[red] (0.75,1) circle (2pt) node[anchor=south, black] {$\textbf{\scriptsize{}}$};
		\filldraw[red] (0.25,1) circle (2pt) node[anchor=south, black] {$\textbf{\scriptsize{Z}}$};
		\filldraw[red] (0,0.75) circle (2pt) node[anchor=east, black] {$\textbf{\scriptsize{Z}}$};
        \node at (0.5, 0.5) {$f$};
	}\right]\nonumber\\
	F^{NE, \pm} & = \frac{1}{2}\left[1 \pm \tikz[baseline=2.5ex]{
	\draw[thick] (0, 0) -- (1, 0) -- (1, 1) -- (0, 1) -- (0, 0);
	\filldraw[red] (0.75,0) circle (2pt) node[anchor=north, black] {$\textbf{\scriptsize{}}$};
	\filldraw[red] (1,0.25) circle (2pt) node[anchor=west, black] {$\textbf{\scriptsize{}}$};
	\filldraw[red] (1,0.75) circle (2pt) node[anchor=west, black] {$\textbf{\scriptsize{Z}}$};
	\filldraw[red] (0.75,1) circle (2pt) node[anchor=south, black] {$\textbf{\scriptsize{Z}}$};
	\filldraw[red] (0.25,1) circle (2pt) node[anchor=south, black] {$\textbf{\scriptsize{}}$};
	\filldraw[red] (0,0.75) circle (2pt) node[anchor=east, black] {$\textbf{\scriptsize{}}$};
    \node at (0.5, 0.5) {$f$};
	}\right]\nonumber\\
	F^{SE, \pm} & = \frac{1}{2}\left[1 \pm \tikz[baseline=2.5ex]{
	\draw[thick] (0, 0) -- (1, 0) -- (1, 1) -- (0, 1) -- (0, 0);
	\filldraw[red] (0.75,0) circle (2pt) node[anchor=north, black] {$\textbf{\scriptsize{Z}}$};
	\filldraw[red] (1,0.25) circle (2pt) node[anchor=west, black] {$\textbf{\scriptsize{Z}}$};
	\filldraw[red] (1,0.75) circle (2pt) node[anchor=west, black] {$\textbf{\scriptsize{}}$};
	\filldraw[red] (0.75,1) circle (2pt) node[anchor=south, black] {$\textbf{\scriptsize{}}$};
	\filldraw[red] (0.25,1) circle (2pt) node[anchor=south, black] {$\textbf{\scriptsize{}}$};
	\filldraw[red] (0,0.75) circle (2pt) node[anchor=east, black] {$\textbf{\scriptsize{}}$};
    \node at (0.5, 0.5) {$f$};
	}\right]
\end{align}

Using these we can write down the face operators projecting to the four holonomies as
\begin{align}
	B_{f}^{1_{1}} &= F_{f}^{NW, +}F_{f}^{NE, +}F_{f}^{SE, +}\frac{1}{2}\Bigg[1 + \tikz[baseline=2.5ex]{
		\draw[thick] (0, 0) -- (1, 0) -- (1, 1) -- (0, 1) -- (0, 0);
		\filldraw[red] (0,0.25) circle (2pt) node[anchor=east, black] {$\textbf{\scriptsize{Z}}$};
		\filldraw[red] (0.25, 0) circle (2pt);
		\node at (0.5, 0.5) {$f$};
	}\Bigg]\frac{1}{2}\Bigg[1 + \tikz[baseline=2.5ex]{
	\draw[thick] (0, 0) -- (1, 0) -- (1, 1) -- (0, 1) -- (0, 0);
	\filldraw[red] (0.25,0) circle (2pt) node[anchor=north, black] {$\textbf{\scriptsize{Z}}$};
	\filldraw[red] (0, 0.25) circle (2pt);
	\node at (0.5, 0.5) {$f$};
	}\Bigg]\nonumber\\
	B_{f}^{1_{2}} &= F_{f}^{NW, +}F_{f}^{NE, +}F_{f}^{SE, +}\frac{1}{2}\Bigg[1 - \tikz[baseline=2.5ex]{
	\draw[thick] (0, 0) -- (1, 0) -- (1, 1) -- (0, 1) -- (0, 0);
	\filldraw[red] (0,0.25) circle (2pt) node[anchor=east, black] {$\textbf{\scriptsize{Z}}$};
	\filldraw[red] (0.25, 0) circle (2pt);
	\node at (0.5, 0.5) {$f$};
	}\Bigg]\frac{1}{2}\Bigg[1 - \tikz[baseline=2.5ex]{
		\draw[thick] (0, 0) -- (1, 0) -- (1, 1) -- (0, 1) -- (0, 0);
		\filldraw[red] (0.25,0) circle (2pt) node[anchor=north, black] {$\textbf{\scriptsize{Z}}$};
		\filldraw[red] (0, 0.25) circle (2pt);
		\node at (0.5, 0.5) {$f$};
	}\Bigg]\nonumber\\
	B_{f}^{a} &= F_{f}^{NW, +}F_{f}^{NE, +}F_{f}^{SE, +}\frac{1}{2}\Bigg[1 + \tikz[baseline=2.5ex]{
	\draw[thick] (0, 0) -- (1, 0) -- (1, 1) -- (0, 1) -- (0, 0);
	\filldraw[red] (0,0.25) circle (2pt) node[anchor=east, black] {$\textbf{\scriptsize{Z}}$};
	\filldraw[red] (0.25, 0) circle (2pt);
	\node at (0.5, 0.5) {$f$};
	}\Bigg]\frac{1}{2}\Bigg[1 - \tikz[baseline=2.5ex]{
		\draw[thick] (0, 0) -- (1, 0) -- (1, 1) -- (0, 1) -- (0, 0);
		\filldraw[red] (0.25,0) circle (2pt) node[anchor=north, black] {$\textbf{\scriptsize{Z}}$};
		\filldraw[red] (0, 0.25) circle (2pt);
		\node at (0.5, 0.5) {$f$};
	}\Bigg]\nonumber\\
	B_{f}^{a^{-1}} &= F_{f}^{NW, +}F_{f}^{NE, +}F_{f}^{SE, +}\frac{1}{2}\Bigg[1 - \tikz[baseline=2.5ex]{
	\draw[thick] (0, 0) -- (1, 0) -- (1, 1) -- (0, 1) -- (0, 0);
	\filldraw[red] (0,0.25) circle (2pt) node[anchor=east, black] {$\textbf{\scriptsize{Z}}$};
	\filldraw[red] (0.25, 0) circle (2pt);
	\node at (0.5, 0.5) {$f$};
	}\Bigg]\frac{1}{2}\Bigg[1 + \tikz[baseline=2.5ex]{
		\draw[thick] (0, 0) -- (1, 0) -- (1, 1) -- (0, 1) -- (0, 0);
		\filldraw[red] (0.25,0) circle (2pt) node[anchor=north, black] {$\textbf{\scriptsize{Z}}$};
		\filldraw[red] (0, 0.25) circle (2pt);
		\node at (0.5, 0.5) {$f$};
	}\Bigg]
\end{align}

Note that these operators do not commute with the vertex operators in Eq. \ref{eq:vertexopsqubitversionvalence4}. However the combinations,
\begin{eqnarray}\label{eq:bf1112aam1}
    B_f^{1_1} + B_f^{1_2} & = & F_f^{NW, +}F_f^{NE, +}F_f^{SE, +}F_f^{SW, +}, \nonumber \\
    B_f^{a} + B_f^{a^{-1}} & = & F_f^{NW, +}F_f^{NE, +}F_f^{SE, +}F_f^{SW, -},
\end{eqnarray}
commute with the vertex operators in Eq. \ref{eq:vertexopsqubitversionvalence4}. Next we get to the final face operator, the projector to the zero state. This is obtained when there is at least one mismatch in one of the three corners (NW, NE, SE). With the corner projectors of Eq. \ref{eq:cornerfaceprojectors} this operator is easily obtained as 
\begin{eqnarray}\label{eq:bf0}
    B_f^0 & = & F_f^{NW,-}F_f^{NE,+}F_f^{SE,+} + F_f^{NW,+}F_f^{NE,-}F_f^{SE,+} + F_f^{NW,+}F_f^{NE,+}F_f^{SE,-} + F_f^{NW,-}F_f^{NE,-}F_f^{SE,-} \nonumber \\ & + & F_f^{NW,-}F_f^{NE,-}F_f^{SE,+} + F_f^{NW,-}F_f^{NE,+}F_f^{SE,-} + F_f^{NW,+}F_f^{NE,-}F_f^{SE,-}.
\end{eqnarray}
This exhausts all the orthogonal face projectors for a given face and they satisfy,
\begin{equation}
    B_f^{1_1}+B_f^{1_2}+B_f^{a}+B_f^{a^{-1}}+B_f^0=1.
\end{equation}

Thus to summarize the local operators on a square lattice we have a total of eight plus eight, sixteen orthogonal vertex operators for each vertex and a total of five mutually orthogonal face operators for each face. We will use these operators to construct Hamiltonians with interesting properties in the next section.

\section{Hamiltonian}
\label{sec:hamiltonian1}
The set of operators constructed in Sec. \ref{sec:equivalentqubitsystem} can be used to construct several exactly solvable Hamiltonians of the $\mathbb{Z}_2$ toric code type. Among the many possibilities we will be interested in those that have an extensive ground state degeneracy (GSD). We will write down and analyze two such models in this section.

We choose the Hamiltonian with the vertex operator given by 
\begin{equation}
    A_v = A_v^{(1)} + A_v^{(2)} + A_v^{(3)} + A_v^{(7)}.
\end{equation}
Some simple algebra reduces this expression to
\begin{align}\label{eq:avmodel1}
A_{v}=\frac{1}{2}\left[1+\tikz[baseline=-0.5ex, scale=0.8]{
	\def\a{0}
	\def\b{0}
	\draw[thick] (-1+\a, 0+\b) -- (1+\a, 0+\b);
	\draw[thick] (0+\a, -1+\b) -- (0+\a, 1+\b);
	\node[anchor=north west] at (-0.1, 0.1) {$v$};
	\filldraw[red] (-0.5+\a,0+\b) circle (2pt) node[anchor=north, black] {$\textbf{\scriptsize{X}}$};
	\filldraw[red] (0.5+\a,0+\b) circle (2pt) node[anchor=south, black] {$\textbf{\scriptsize{X}}$};
	\filldraw[red] (0+\a, -0.5+\b) circle (2pt) node[anchor=north west, black] {$\textbf{\scriptsize{X}}$};
	\filldraw[red] (0+\a, 0.5+\b) circle (2pt) node[anchor=south east, black] {$\textbf{\scriptsize{X}}$};
}\right]
\frac{1}{2}\left[1-\tikz[baseline=-0.5ex, scale=0.8]{
	\def\a{0}
	\def\b{0}
	\draw[thick] (-1+\a, 0+\b) -- (1+\a, 0+\b);
	\draw[thick] (0+\a, -1+\b) -- (0+\a, 1+\b);
	\node[anchor=north west] at (-0.1, 0.1) {$v$};
	\filldraw[red] (-0.5+\a,0+\b) circle (2pt) node[anchor=north, black] {$\textbf{\scriptsize{}}$};
	\filldraw[red] (0+\a, -0.5+\b) circle (2pt) node[anchor=north west, black] {$\textbf{\scriptsize{}}$};
	\filldraw[red] (0.5+\a,0+\b) circle (2pt) node[anchor=south, black] {$\textbf{\scriptsize{Z}}$};
	\filldraw[red] (0+\a, 0.5+\b) circle (2pt) node[anchor=south east, black] {$\textbf{\scriptsize{Z}}$};
}\right]
\end{align}
which is an operator acting only when the NE corner of a vertex is mismatched. Such configs are shown in Fig. \ref{fig:vertexconfigswithvertexaction}. Note that every vertex operator is indexed by the vertex of the square lattice and acts only on the sites, carrying the qubit space, surrounding this vertex. As a consequence neighboring vertex operators do not have common support unlike the group toric code case where they share the common edge.  This will play a role in the GSD of this model as we shall soon see.

The face operator is chosen as the one which projects to the zero state. However we do not choose the entire $B_f^0$ operator in Eq. \ref{eq:bf0} but instead we pick the configurations that have an odd number of mismatches in the holonomy, i. e.,
\begin{equation}
  B_f  =  F_f^{NW,-}F_f^{NE,+}F_f^{SE,+} + F_f^{NW,+}F_f^{NE,-}F_f^{SE,+} + F_f^{NW,+}F_f^{NE,+}F_f^{SE,-} + F_f^{NW,-}F_f^{NE,-}F_f^{SE,-}.  
\end{equation}
This operator simplifies to
\begin{align}\label{eq:bfmodel1}
    B_{f}=\frac{1}{2}\left[1 - \tikz[baseline=2ex]{
		\draw[thick] (0, 0) -- (1, 0) -- (1, 1) -- (0, 1) -- (0, 0);
		\filldraw[red] (0.75,0) circle (2pt) node[anchor=north, black] {$\textbf{\scriptsize{Z}}$};
		\filldraw[red] (1,0.25) circle (2pt) node[anchor=west, black] {$\textbf{\scriptsize{Z}}$};
		\filldraw[red] (1,0.75) circle (2pt) node[anchor=west, black] {$\textbf{\scriptsize{Z}}$};
		\filldraw[red] (0.75,1) circle (2pt) node[anchor=south, black] {$\textbf{\scriptsize{Z}}$};
		\filldraw[red] (0.25,1) circle (2pt) node[anchor=south, black] {$\textbf{\scriptsize{Z}}$};
		\filldraw[red] (0,0.75) circle (2pt) node[anchor=east, black] {$\textbf{\scriptsize{Z}}$};
        \node at (0.5, 0.5) {$f$};
	}\right].
\end{align}
Thus the Hamiltonian can now be written as 
\begin{equation}\label{eq:hmodel1}
    H = -\sum\limits_{v=1}^{N_v}~A_v -\sum\limits_{f=1}^{N_f}~B_f,
\end{equation}
with $A_v$ and $B_f$ given by Eqs. \ref{eq:avmodel1} and \ref{eq:bfmodel1} respectively. It is easily seen that both these operators are projectors and they commute with each other for all vertices and faces of the square lattice. The definition of the model on other lattices has to be appropriately modified which we will address separately. Thus this model is exactly solvable like the $\mathbb{Z}_2$-toric code Hamiltonian. We will follow those techniques to analyze the ground states and excitations of the Hamiltonian in Eq. \ref{eq:hmodel1}. 

\subsection{Ground state degeneracy}
\label{subsec:gsd}
The coefficients in the Hamiltonian in Eq. \ref{eq:hmodel1} have been chosen such that the ground state energy on a closed square lattice is $-2N_v$ and the ground states satisfy the relations,
\begin{equation}
    \ket{G} = A_v~\ket{G} = B_f~\ket{G},~\forall~v, f.
\end{equation}
As a consequence we find that the projector to the ground state manifold is given by the operator,
\begin{equation}
    U = \prod\limits_{v=1}^{N_v}~A_v\prod\limits_{f=1}^{N_f}~B_f.
\end{equation}
It then follows that the GSD can be computed as
\begin{equation}
    GSD = \Tr~U = \Tr~\prod\limits_{v=1}^{N_v}~A_v\prod\limits_{f=1}^{N_f}~B_f.
\end{equation}
For the vertex and face operators in Eqs. \ref{eq:avmodel1} and \ref{eq:bfmodel1}, this is easily computed by using the facts that the Pauli matrices are traceless and that the trace of the tensor product equals the product of the traces. Thus we need to count the number of identity operators appearing in the products of the vertex and face operators. All the other terms in this product will have at least one of the Pauli matrices on the qubit spaces and such terms will not contribute to the trace. We see that there are just two identity operators appearing in the product of the vertex and face operators. The first one, is the trivial one, obtained by multiplying the identity operators in all the vertex and plaquette operators. The second identity is a result of multiplying all the \tikz[baseline=2ex]{
		\draw[thick] (0, 0) -- (1, 0) -- (1, 1) -- (0, 1) -- (0, 0);
		\filldraw[red] (0.75,0) circle (2pt) node[anchor=north, black] {$\textbf{\scriptsize{Z}}$};
		\filldraw[red] (1,0.25) circle (2pt) node[anchor=west, black] {$\textbf{\scriptsize{Z}}$};
		\filldraw[red] (1,0.75) circle (2pt) node[anchor=west, black] {$\textbf{\scriptsize{Z}}$};
		\filldraw[red] (0.75,1) circle (2pt) node[anchor=south, black] {$\textbf{\scriptsize{Z}}$};
		\filldraw[red] (0.25,1) circle (2pt) node[anchor=south, black] {$\textbf{\scriptsize{Z}}$};
		\filldraw[red] (0,0.75) circle (2pt) node[anchor=east, black] {$\textbf{\scriptsize{Z}}$};
        \node at (0.5, 0.5) {$f$};
	} terms from the face operators along with the \tikz[baseline=-0.5ex, scale=0.8]{
	\def\a{0}
	\def\b{0}
	\draw[thick] (-1+\a, 0+\b) -- (1+\a, 0+\b);
	\draw[thick] (0+\a, -1+\b) -- (0+\a, 1+\b);
	\node[anchor=north west] at (-0.1, 0.1) {$v$};
	\filldraw[red] (-0.5+\a,0+\b) circle (2pt) node[anchor=north, black] {$\textbf{\scriptsize{}}$};
	\filldraw[red] (0+\a, -0.5+\b) circle (2pt) node[anchor=north west, black] {$\textbf{\scriptsize{}}$};
	\filldraw[red] (0.5+\a,0+\b) circle (2pt) node[anchor=south, black] {$\textbf{\scriptsize{Z}}$};
	\filldraw[red] (0+\a, 0.5+\b) circle (2pt) node[anchor=south east, black] {$\textbf{\scriptsize{Z}}$};
} from the vertex operators. Thus the GSD can now be evaluated as 
\begin{eqnarray}\label{eq:gsdmodel1}
    GSD & = & \Tr~\prod\limits_{v=1}^{N_v}~A_v\prod\limits_{f=1}^{N_f}~B_f \nonumber \\
    & = & \frac{1}{2^{2N_v}}\frac{1}{2^{N_f}}\times \Tr~\left[2\times \prod_{j=1}^{4N_v}~\left(1_{2\times 2}\right)_j\right] + \textrm{traceless terms} \nonumber \\
    & = & 2\times 2^{N_v}.
\end{eqnarray}
The factors $\frac{1}{2^{2N_v}}$ and $\frac{1}{2^{N_f}}$ are those appearing from the product of vertex and face operators respectively. We have used the fact the $N_v=N_f$ for a square lattice. The operator $1_{2\times 2}$ is size 2 identity operator acting on the qubit space. Note that this computation depends on the chosen lattice, in particular the valency of the vertices. Thus we see that the GSD of this model is extensive in the lattice size. We will now account for this degeneracy by identifying the appropriate symmetries. 

\subsection{Symmetries accounting for the GSD}
\label{subsec:symmetries}
It is easily verified that the global operator,
\begin{align}\label{eq:globalz2}
    \prod_{v=1}^{N_{v}}\tikz[baseline=-0.5ex, scale=0.8]{
	\def\a{0}
	\def\b{0}
	\draw[thick] (-1+\a, 0+\b) -- (1+\a, 0+\b);
	\draw[thick] (0+\a, -1+\b) -- (0+\a, 1+\b);
	\node[anchor=north west] at (-0.1, 0.1) {$v$};
	\filldraw[red] (-0.5+\a,0+\b) circle (2pt) node[anchor=north, black] {$\textbf{\scriptsize{}}$};
	\filldraw[red] (0+\a, -0.5+\b) circle (2pt) node[anchor=north west, black] {$\textbf{\scriptsize{}}$};
	\filldraw[red] (0.5+\a,0+\b) circle (2pt) node[anchor=south, black] {$\textbf{\scriptsize{X}}$};
	\filldraw[red] (0+\a, 0.5+\b) circle (2pt) node[anchor=south east, black] {$\textbf{\scriptsize{X}}$};
}
\end{align}
is a symmetry of the Hamiltonian in Eq. \ref{eq:hmodel1}. This accounts for the factor of 2 in the GSD, Eq. \ref{eq:gsdmodel1}. This symmetry corresponds to globally changing the object indices in the NE corner of every vertex in the square lattice. The ground state pair connected by this global symmetry are separated by an infinite energy barrier in the continuum limit. This is due to the fact that flipping just a single NE corner creates a pair of flux excitations that the corner connects. This implies that while trying to convert one ground state to another we need to excite an infinite number of fluxes (face excitations). Hence they represent two different superselection sectors and the system will choose one of these two states breaking this global $\mathbb{Z}_2$ symmetry. Thus the first factor of $2$ in the GSD corresponds to a symmetry broken phase. 

The factor $2^{N_v}$ in the GSD, Eq. \ref{eq:gsdmodel1}, originates from operators supported on contractible and non-contractible loops. Each vertex of the square lattice is either enclosed or not by a contractible or non-contractible loop and thus we precisely obtain $2^{N_v}$ such operators. We now identify these loop operators by considering the constituents of such loops, that is the locations where the operators can be applied so that no vertex is excited in the process. These constituents are,
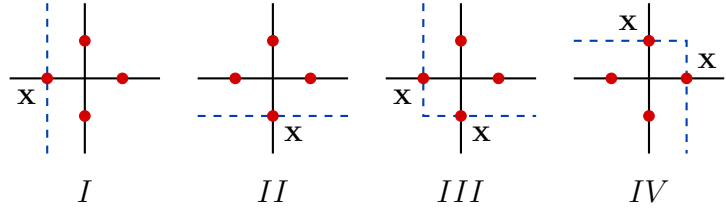
\begin{align*}
    \tikz[baseline=-0.5ex, scale=1]{
	\def\a{0}
	\def\b{0}
	\draw[thick] (-1+\a, 0+\b) -- (1+\a, 0+\b);
	\draw[thick] (0+\a, -1+\b) -- (0+\a, 1+\b);
    \draw[thick, dashed, blue] (-0.5+\a, -1+\b) -- (-0.5+\a, 1+\b);
	\filldraw[red] (-0.5+\a,0+\b) circle (2pt) node[anchor=north east, black] {$\textbf{\scriptsize{X}}$};
	\filldraw[red] (0+\a, -0.5+\b) circle (2pt) node[anchor=north west, black] {$\textbf{\scriptsize{}}$};
	\filldraw[red] (0.5+\a,0+\b) circle (2pt) node[anchor=south, black] {$\textbf{\scriptsize{}}$};
	\filldraw[red] (0+\a, 0.5+\b) circle (2pt) node[anchor=south east, black] {$\textbf{\scriptsize{}}$};
    \node at (0+\a, -1.5+\b) {$I$};
	\def\a{2.5}
	\def\b{0}
	\draw[thick] (-1+\a, 0+\b) -- (1+\a, 0+\b);
	\draw[thick] (0+\a, -1+\b) -- (0+\a, 1+\b);
    \draw[thick, dashed, blue] (-1+\a, -0.5+\b) -- (1+\a, -0.5+\b);
	\filldraw[red] (-0.5+\a,0+\b) circle (2pt) node[anchor=north east, black] {$\textbf{\scriptsize{}}$};
	\filldraw[red] (0+\a, -0.5+\b) circle (2pt) node[anchor=north west, black] {$\textbf{\scriptsize{X}}$};
	\filldraw[red] (0.5+\a,0+\b) circle (2pt) node[anchor=south, black] {$\textbf{\scriptsize{}}$};
	\filldraw[red] (0+\a, 0.5+\b) circle (2pt) node[anchor=south east, black] {$\textbf{\scriptsize{}}$};
    \node at (0+\a, -1.5+\b) {$II$};
    \def\a{5}
	\def\b{0}
	\draw[thick] (-1+\a, 0+\b) -- (1+\a, 0+\b);
	\draw[thick] (0+\a, -1+\b) -- (0+\a, 1+\b);
    \draw[thick, dashed, blue] (-0.5+\a, 1+\b) -- (-0.5+\a, -0.5+\b) -- (1+\a, -0.5+\b);
	\filldraw[red] (-0.5+\a,0+\b) circle (2pt) node[anchor=north east, black] {$\textbf{\scriptsize{X}}$};
	\filldraw[red] (0+\a, -0.5+\b) circle (2pt) node[anchor=north west, black] {$\textbf{\scriptsize{X}}$};
	\filldraw[red] (0.5+\a,0+\b) circle (2pt) node[anchor=south, black] {$\textbf{\scriptsize{}}$};
	\filldraw[red] (0+\a, 0.5+\b) circle (2pt) node[anchor=south east, black] {$\textbf{\scriptsize{}}$};
    \node at (0+\a, -1.5+\b) {$III$};
    \def\a{7.5}
	\def\b{0}
	\draw[thick] (-1+\a, 0+\b) -- (1+\a, 0+\b);
	\draw[thick] (0+\a, -1+\b) -- (0+\a, 1+\b);
    \draw[thick, dashed, blue] (-1+\a, 0.5+\b) -- (0.5+\a, 0.5+\b) -- (0.5+\a, -1+\b);
	\filldraw[red] (-0.5+\a,0+\b) circle (2pt) node[anchor=north east, black] {$\textbf{\scriptsize{}}$};
	\filldraw[red] (0+\a, -0.5+\b) circle (2pt) node[anchor=north west, black] {$\textbf{\scriptsize{}}$};
	\filldraw[red] (0.5+\a,0+\b) circle (2pt) node[anchor=south west, black] {$\textbf{\scriptsize{X}}$};
	\filldraw[red] (0+\a, 0.5+\b) circle (2pt) node[anchor=south east, black] {$\textbf{\scriptsize{X}}$};
    \node at (0+\a, -1.5+\b) {$IV$};
}
\end{align*}
We can form contractible and non-contractible loop symmetries using these configurations. Combining configuration $III$ and $IV$ around a single vertex is forbidden as that will result in the corresponding vertex operator and hence cannot give rise to a new ground state. Similarly configurations $I$, $II$ and $IV$ cannot be combined around a single vertex for the same reason. Moreover the configurations $III$ and $IV$ are equivalent to each other by their vertex operator and hence a state containing the configuration $III$ will also contain the configuration $IV$. Barring these exceptions all other combinations result in several loop operators that account for the factor $2^{N_v}$ in the GSD. An example of a contractible loop operator is,

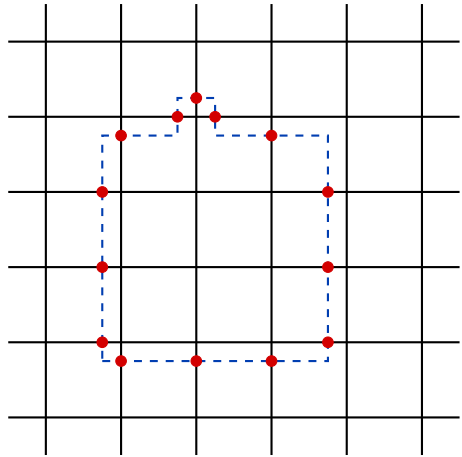
\begin{align*}
	\tikz{
		\foreach \x in {0, 1, 2, 3, 4, 5}{
			\draw[thick] (-0.5, \x) -- (5.5, \x);
		}
		\foreach \x in {0, 1, 2, 3, 4, 5}{
			\draw[thick] (\x, -0.5) -- (\x, 5.5);
		}
		\draw[thick, dashed, blue] (0.75, 0.75) -- (3.75, 0.75) -- (3.75, 3.75) -- (2.25, 3.75) -- (2.25, 4.25) -- (1.75, 4.25) -- (1.75, 3.75) -- (0.75, 3.75) -- (0.75, 0.75);
		\filldraw[red] (1,0.75) circle (2pt);
		\filldraw[red] (2,0.75) circle (2pt);
		\filldraw[red] (3,0.75) circle (2pt);
		\filldraw[red] (3.75,1) circle (2pt);
		\filldraw[red] (3.75,2) circle (2pt);
		\filldraw[red] (3.75,3) circle (2pt);
		\filldraw[red] (3,3.75) circle (2pt);
		\filldraw[red] (2.25,4) circle (2pt);
		\filldraw[red] (2,4.25) circle (2pt);
		\filldraw[red] (1.75,4) circle (2pt);
		\filldraw[red] (1,3.75) circle (2pt);
		\filldraw[red] (0.75,3) circle (2pt);
		\filldraw[red] (0.75,2) circle (2pt);
		\filldraw[red] (0.75,1) circle (2pt);
	}
\end{align*}
with the understanding that there is a $X$ operator acting on every qubit space. A non-contractible loop operator can appear as,
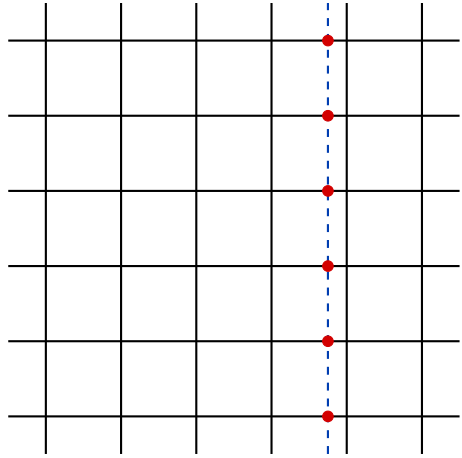
\begin{align*}
	\tikz{
		\foreach \x in {0, 1, 2, 3, 4, 5}{
			\draw[thick] (-0.5, \x) -- (5.5, \x);
		}
		\foreach \x in {0, 1, 2, 3, 4, 5}{
			\draw[thick] (\x, -0.5) -- (\x, 5.5);
		}
		\draw[thick, dashed, blue] (3.75, -0.5) -- (3.75, 5.5);
		\filldraw[red] (3.75,0) circle (2pt);
		\filldraw[red] (3.75,1) circle (2pt);
		\filldraw[red] (3.75,2) circle (2pt);
		\filldraw[red] (3.75,3) circle (2pt);
		\filldraw[red] (3.75,4) circle (2pt);
		\filldraw[red] (3.75,5) circle (2pt);
	}
\end{align*}
while a deformed version of the same can be obtained by taking a product with a contractible loop operator,
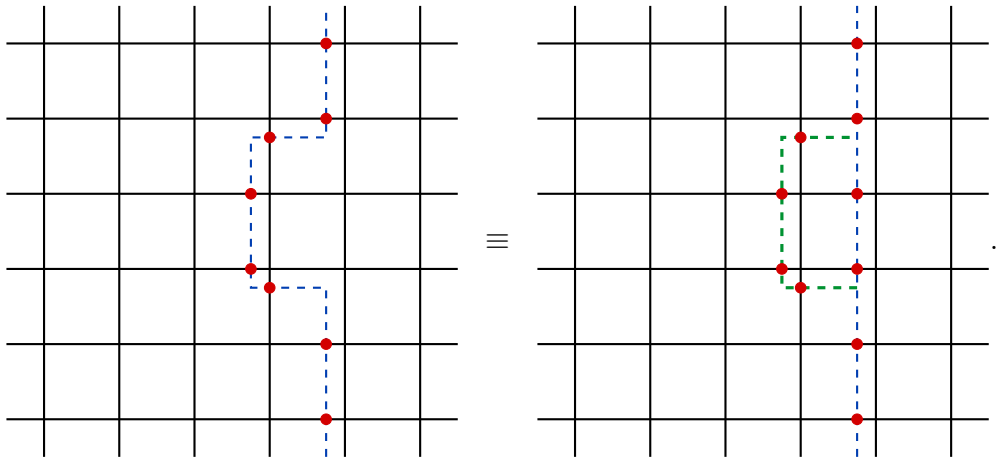
\begin{align*}
	\tikz[baseline=12.5ex]{
		\foreach \x in {0, 1, 2, 3, 4, 5}{
			\draw[thick] (-0.5, \x) -- (5.5, \x);
		}
		\foreach \x in {0, 1, 2, 3, 4, 5}{
			\draw[thick] (\x, -0.5) -- (\x, 5.5);
		}
		\draw[thick, dashed, blue] (3.75, -0.5) -- (3.75, 1.75) -- (2.75, 1.75) -- (2.75, 3.75) -- (3.75, 3.75) -- (3.75, 5.5);
		\filldraw[red] (3.75,0) circle (2pt);
		\filldraw[red] (3.75,1) circle (2pt);
		\filldraw[red] (2.75,2) circle (2pt);
		\filldraw[red] (2.75,3) circle (2pt);
		\filldraw[red] (3,1.75) circle (2pt);
		\filldraw[red] (3,3.75) circle (2pt);
		\filldraw[red] (3.75,4) circle (2pt);
		\filldraw[red] (3.75,5) circle (2pt);
	}\;\; \equiv \;\;
		\tikz[baseline=12.5ex]{
			\foreach \x in {0, 1, 2, 3, 4, 5}{
				\draw[thick] (-0.5, \x) -- (5.5, \x);
			}
			\foreach \x in {0, 1, 2, 3, 4, 5}{
				\draw[thick] (\x, -0.5) -- (\x, 5.5);
			}
			\draw[thick, dashed, blue] (3.75, -0.5) -- (3.75, 5.5);
			\draw[very thick, dashed, ForestGreen] (3.75, 1.75) -- (2.75, 1.75) -- (2.75, 3.75) -- (3.75, 3.75);
			\filldraw[red] (3.75,0) circle (2pt);
			\filldraw[red] (3.75,1) circle (2pt);
			\filldraw[red] (3.75,2) circle (2pt);
			\filldraw[red] (3.75,3) circle (2pt);
			\filldraw[red] (2.75,2) circle (2pt);
			\filldraw[red] (3,1.75) circle (2pt);
			\filldraw[red] (3,3.75) circle (2pt);
			\filldraw[red] (2.75,3) circle (2pt);
			\filldraw[red] (3.75,4) circle (2pt);
			\filldraw[red] (3.75,5) circle (2pt);
		}.
\end{align*}
While `large' deformations of non-contractible loops result in different ground states, we are allowed to make `local' deformations around a single vertex,

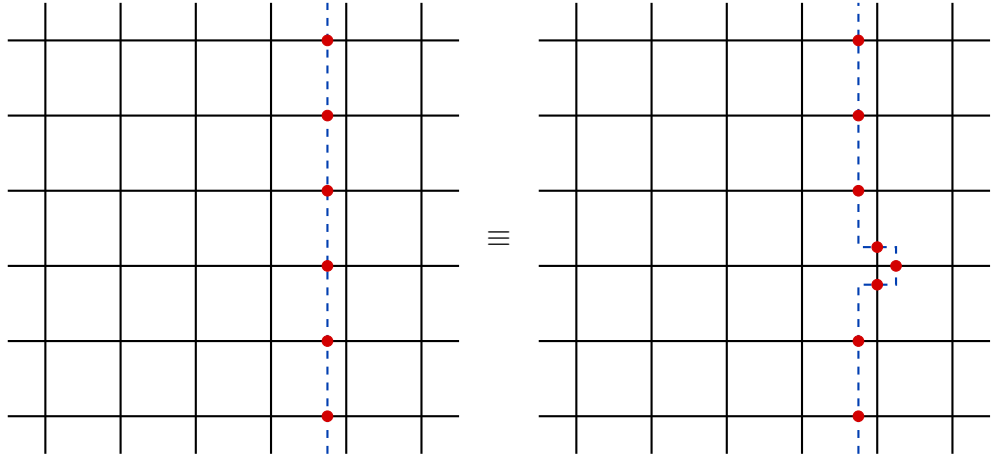
\begin{align*}
	\tikz[baseline=12.5ex]{
		\foreach \x in {0, 1, 2, 3, 4, 5}{
			\draw[thick] (-0.5, \x) -- (5.5, \x);
		}
		\foreach \x in {0, 1, 2, 3, 4, 5}{
			\draw[thick] (\x, -0.5) -- (\x, 5.5);
		}
		\draw[thick, dashed, blue] (3.75, -0.5) -- (3.75, 5.5);
		\filldraw[red] (3.75,0) circle (2pt);
		\filldraw[red] (3.75,1) circle (2pt);
		\filldraw[red] (3.75,2) circle (2pt);
		\filldraw[red] (3.75,3) circle (2pt);
		\filldraw[red] (3.75,4) circle (2pt);
		\filldraw[red] (3.75,5) circle (2pt);
	}\;\;\equiv \;\;
 \tikz[baseline=12.5ex]{
		\foreach \x in {0, 1, 2, 3, 4, 5}{
			\draw[thick] (-0.5, \x) -- (5.5, \x);
		}
		\foreach \x in {0, 1, 2, 3, 4, 5}{
			\draw[thick] (\x, -0.5) -- (\x, 5.5);
		}
		\draw[thick, dashed, blue] (3.75, -0.5) -- (3.75, 1.75) -- (4.25, 1.75) -- (4.25, 2.25) -- (3.75, 2.25) -- (3.75, 5.5);
		\filldraw[red] (3.75,0) circle (2pt);
		\filldraw[red] (3.75,1) circle (2pt);
        \filldraw[red] (4,1.75) circle (2pt);
        \filldraw[red] (4.25,2) circle (2pt);
        \filldraw[red] (4,2.25) circle (2pt);
		\filldraw[red] (3.75,3) circle (2pt);
		\filldraw[red] (3.75,4) circle (2pt);
		\filldraw[red] (3.75,5) circle (2pt);
	}
\end{align*}

Thus unlike the group toric code the different deformed non-contractible loops are not equivalent to each other and they give rise to new ground states. This is seen as a consequence of the fact that the neighboring vertex operators in the $\mathcal{S}^2_1$-groupoid toric code do not have common support and hence they cannot be used to deform curves as it happens in the group case. 

\subsection{Organizing the ground states}
\label{subsec:organizegs}
Now we identify the symmetries that help us to distinguish between the different ground states obtained from the loop operators. This is done by specifying a $N_v$-tuple whose entries are either $\pm 1$, which are the eigenvalues of the local symmetry,
\tikz[baseline=-0.5ex, scale=0.8]{
	\def\a{0}
	\def\b{0}
	\draw[thick] (-1+\a, 0+\b) -- (1+\a, 0+\b);
	\draw[thick] (0+\a, -1+\b) -- (0+\a, 1+\b);
	\node[anchor=north west] at (-0.1, 0.1) {$v$};
	\filldraw[red] (-0.5+\a,0+\b) circle (2pt) node[anchor=north, black] {$\textbf{\scriptsize{Z}}$};
	\filldraw[red] (0+\a, -0.5+\b) circle (2pt) node[anchor=north west, black] {$\textbf{\scriptsize{Z}}$};
	\filldraw[red] (0.5+\a,0+\b) circle (2pt) node[anchor=south, black] {$\textbf{\scriptsize{}}$};
	\filldraw[red] (0+\a, 0.5+\b) circle (2pt) node[anchor=south east, black] {$\textbf{\scriptsize{}}$};
}. Clearly there are $2^{N_v}$ such tuples as at each vertex the local symmetry operator takes values $1(-1)$ for a match(mismatch).
By measuring this local symmetry at every vertex we can trace out the loop (contractible or non-contractible) that determines the different loop ground states. This is clear as at every vertex this operator distinguishes between 
the configurations,
\tikz[baseline=-0.5ex, scale=0.7]{
	\def\a{0}
	\def\b{0}
	\draw[thick] (-1+\a, 0+\b) -- (1+\a, 0+\b);
	\draw[thick] (0+\a, -1+\b) -- (0+\a, 1+\b);
    \draw[thick, dashed, blue] (-0.5+\a, -1+\b) -- (-0.5+\a, 1+\b);
	\filldraw[red] (-0.5+\a,0+\b) circle (2pt) node[anchor=north east, black] {$\textbf{\scriptsize{X}}$};
	\filldraw[red] (0+\a, -0.5+\b) circle (2pt) node[anchor=north west, black] {$\textbf{\scriptsize{}}$};
	\filldraw[red] (0.5+\a,0+\b) circle (2pt) node[anchor=south, black] {$\textbf{\scriptsize{}}$};
	\filldraw[red] (0+\a, 0.5+\b) circle (2pt) node[anchor=south east, black] {$\textbf{\scriptsize{}}$};
    \node at (0+\a, -1.5+\b) {$I$};
	\def\a{2.5}
	\def\b{0}
	\draw[thick] (-1+\a, 0+\b) -- (1+\a, 0+\b);
	\draw[thick] (0+\a, -1+\b) -- (0+\a, 1+\b);
    \draw[thick, dashed, blue] (-1+\a, -0.5+\b) -- (1+\a, -0.5+\b);
	\filldraw[red] (-0.5+\a,0+\b) circle (2pt) node[anchor=north east, black] {$\textbf{\scriptsize{}}$};
	\filldraw[red] (0+\a, -0.5+\b) circle (2pt) node[anchor=north west, black] {$\textbf{\scriptsize{X}}$};
	\filldraw[red] (0.5+\a,0+\b) circle (2pt) node[anchor=south, black] {$\textbf{\scriptsize{}}$};
	\filldraw[red] (0+\a, 0.5+\b) circle (2pt) node[anchor=south east, black] {$\textbf{\scriptsize{}}$};
    \node at (0+\a, -1.5+\b) {$II$};
} for a mismatch and the configuration
\tikz[baseline=-0.5ex, scale=0.7]{
\def\a{0}
	\def\b{0}
	\draw[thick] (-1+\a, 0+\b) -- (1+\a, 0+\b);
	\draw[thick] (0+\a, -1+\b) -- (0+\a, 1+\b);
    \draw[thick, dashed, blue] (-0.5+\a, 1+\b) -- (-0.5+\a, -0.5+\b) -- (1+\a, -0.5+\b);
	\filldraw[red] (-0.5+\a,0+\b) circle (2pt) node[anchor=north east, black] {$\textbf{\scriptsize{X}}$};
	\filldraw[red] (0+\a, -0.5+\b) circle (2pt) node[anchor=north west, black] {$\textbf{\scriptsize{X}}$};
	\filldraw[red] (0.5+\a,0+\b) circle (2pt) node[anchor=south, black] {$\textbf{\scriptsize{}}$};
	\filldraw[red] (0+\a, 0.5+\b) circle (2pt) node[anchor=south east, black] {$\textbf{\scriptsize{}}$};
    \node at (0+\a, -1.5+\b) {$III$};
} for a match. However this operator cannot detect the presence of the loop configuration
\tikz[baseline=-0.5ex, scale=0.7]{
    \def\a{0}
	\def\b{0}
	\draw[thick] (-1+\a, 0+\b) -- (1+\a, 0+\b);
	\draw[thick] (0+\a, -1+\b) -- (0+\a, 1+\b);
    \draw[thick, dashed, blue] (-1+\a, 0.5+\b) -- (0.5+\a, 0.5+\b) -- (0.5+\a, -1+\b);
	\filldraw[red] (-0.5+\a,0+\b) circle (2pt) node[anchor=north east, black] {$\textbf{\scriptsize{}}$};
	\filldraw[red] (0+\a, -0.5+\b) circle (2pt) node[anchor=north west, black] {$\textbf{\scriptsize{}}$};
	\filldraw[red] (0.5+\a,0+\b) circle (2pt) node[anchor=south west, black] {$\textbf{\scriptsize{X}}$};
	\filldraw[red] (0+\a, 0.5+\b) circle (2pt) node[anchor=south east, black] {$\textbf{\scriptsize{X}}$};
    \node at (0+\a, -1.5+\b) {$IV$};
}. Anyhow there is no need to distinguish this loop configuration as it is equivalent to the configuration $III$ by the vertex operator as explained in Sec. \ref{subsec:symmetries}. As a final remark we observe that though the symmetry is local, we need to measure the corresponding value at every vertex of the square lattice to completely identify the loop state. Thus these operators distinguish the $2^{N_v}$ ground states corresponding to the loop operators modulo the global $\mathbb{Z}_2$ symmetry of Eq. \ref{eq:globalz2}. 

We are now left with finding an operator that will distinguish states having the same values in the $N_v$-tuple but are connected by the global $\mathbb{Z}_2$ symmetry operator. To find this we first look at the structure of the ground states by developing a pictorial notation to construct the ground states of the Hamiltonian in Eq. \ref{eq:hmodel1}. At every vertex we denote a match by
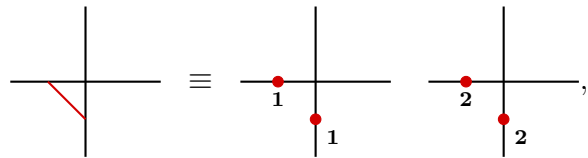
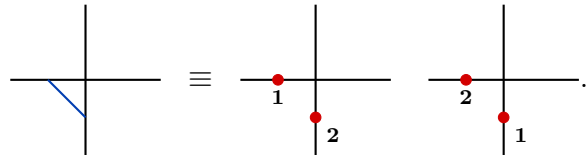
\begin{align*}
    \tikz[baseline=-0.5ex, scale=1]{
	\def\a{0}
	\def\b{0}
	\draw[thick] (-1+\a, 0+\b) -- (1+\a, 0+\b);
	\draw[thick] (0+\a, -1+\b) -- (0+\a, 1+\b);
	\draw[thick, red] (-0.5+\a, \b) -- (\a, -0.5+\b);
}\;\;\equiv\;\;
    \tikz[baseline=-0.5ex, scale=1]{
	\def\a{0}
	\def\b{0}
	\draw[thick] (-1+\a, 0+\b) -- (1+\a, 0+\b);
	\draw[thick] (0+\a, -1+\b) -- (0+\a, 1+\b);
	\filldraw[red] (-0.5+\a,0+\b) circle (2pt) node[anchor=north, black] {$\textbf{\scriptsize{1}}$};
	\filldraw[red] (0+\a, -0.5+\b) circle (2pt) node[anchor=north west, black] {$\textbf{\scriptsize{1}}$};
    \def\a{2.5}
	\def\b{0}
	\draw[thick] (-1+\a, 0+\b) -- (1+\a, 0+\b);
	\draw[thick] (0+\a, -1+\b) -- (0+\a, 1+\b);
	\filldraw[red] (-0.5+\a,0+\b) circle (2pt) node[anchor=north, black] {$\textbf{\scriptsize{2}}$};
	\filldraw[red] (0+\a, -0.5+\b) circle (2pt) node[anchor=north west, black] {$\textbf{\scriptsize{2}}$};
},
\end{align*}
and a mismatch by
\begin{align*}
    \tikz[baseline=-0.5ex, scale=1]{
	\def\a{0}
	\def\b{0}
	\draw[thick] (-1+\a, 0+\b) -- (1+\a, 0+\b);
	\draw[thick] (0+\a, -1+\b) -- (0+\a, 1+\b);
	\draw[thick, blue] (-0.5+\a, \b) -- (\a, -0.5+\b);
}\;\;\equiv\;\;
    \tikz[baseline=-0.5ex, scale=1]{
	\def\a{0}
	\def\b{0}
	\draw[thick] (-1+\a, 0+\b) -- (1+\a, 0+\b);
	\draw[thick] (0+\a, -1+\b) -- (0+\a, 1+\b);
	\filldraw[red] (-0.5+\a,0+\b) circle (2pt) node[anchor=north, black] {$\textbf{\scriptsize{1}}$};
	\filldraw[red] (0+\a, -0.5+\b) circle (2pt) node[anchor=north west, black] {$\textbf{\scriptsize{2}}$};
    \def\a{2.5}
	\def\b{0}
	\draw[thick] (-1+\a, 0+\b) -- (1+\a, 0+\b);
	\draw[thick] (0+\a, -1+\b) -- (0+\a, 1+\b);
	\filldraw[red] (-0.5+\a,0+\b) circle (2pt) node[anchor=north, black] {$\textbf{\scriptsize{2}}$};
	\filldraw[red] (0+\a, -0.5+\b) circle (2pt) node[anchor=north west, black] {$\textbf{\scriptsize{1}}$};
}.
\end{align*}

With these notations the allowed local vertex and face configurations in the ground state sector of the Hamiltonian in Eq. \ref{eq:hmodel1} are
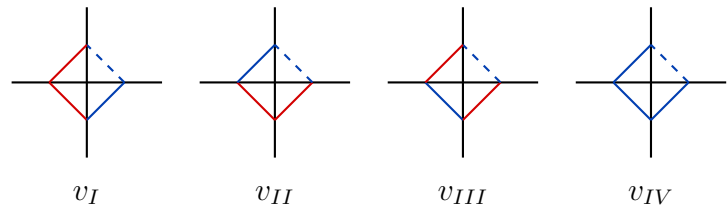
\begin{align*}
    \tikz[baseline=-0.5ex, scale=1]{
	\def\a{0}
	\def\b{0}
	\draw[thick] (-1+\a, 0+\b) -- (1+\a, 0+\b);
	\draw[thick] (0+\a, -1+\b) -- (0+\a, 1+\b);
	\draw[thick, red] (-0.5+\a, \b) -- (\a, -0.5+\b);
    \draw[thick, blue] (\a, -0.5+\b) -- (0.5+\a, \b);
    \draw[thick, blue, dashed] (0.5+\a, \b) -- (\a, 0.5+\b);
    \draw[thick, red] (\a, 0.5+\b) -- (-0.5+\a, \b);
    \node at (\a, -1.5) {$v_{I}$};
    \def\a{2.5}
	\def\b{0}
	\draw[thick] (-1+\a, 0+\b) -- (1+\a, 0+\b);
	\draw[thick] (0+\a, -1+\b) -- (0+\a, 1+\b);
	\draw[thick, red] (-0.5+\a, \b) -- (\a, -0.5+\b);
    \draw[thick, red] (\a, -0.5+\b) -- (0.5+\a, \b);
    \draw[thick, blue, dashed] (0.5+\a, \b) -- (\a, 0.5+\b);
    \draw[thick, blue] (\a, 0.5+\b) -- (-0.5+\a, \b);
    \node at (\a, -1.5) {$v_{II}$};
    \def\a{5}
	\def\b{0}
	\draw[thick] (-1+\a, 0+\b) -- (1+\a, 0+\b);
	\draw[thick] (0+\a, -1+\b) -- (0+\a, 1+\b);
	\draw[thick, blue] (-0.5+\a, \b) -- (\a, -0.5+\b);
    \draw[thick, red] (\a, -0.5+\b) -- (0.5+\a, \b);
    \draw[thick, blue, dashed] (0.5+\a, \b) -- (\a, 0.5+\b);
    \draw[thick, red] (\a, 0.5+\b) -- (-0.5+\a, \b);
    \node at (\a, -1.5) {$v_{III}$};
    \def\a{7.5}
	\def\b{0}
	\draw[thick] (-1+\a, 0+\b) -- (1+\a, 0+\b);
	\draw[thick] (0+\a, -1+\b) -- (0+\a, 1+\b);
	\draw[thick, blue] (-0.5+\a, \b) -- (\a, -0.5+\b);
    \draw[thick, blue] (\a, -0.5+\b) -- (0.5+\a, \b);
    \draw[thick, blue, dashed] (0.5+\a, \b) -- (\a, 0.5+\b);
    \draw[thick, blue] (\a, 0.5+\b) -- (-0.5+\a, \b);
    \node at (\a, -1.5) {$v_{IV}$};
}
\end{align*}
and
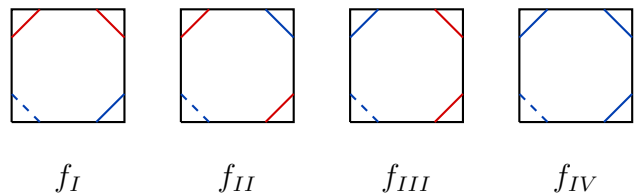
\begin{align*}
    \tikz[baseline=-0.5ex, scale=1.5]{
	\def\a{0}
	\def\b{0}
    \draw[thick] (\a, \b) -- (\a+1, \b) -- (\a+1, \b+1) -- (\a, \b+1) -- (\a, \b);
    \draw[thick, blue, dashed] (0.25+\a, \b) -- (\a, 0.25+\b);
    \draw[thick, blue] (0.75+\a, \b) -- (1+\a, 0.25+\b);
    \draw[thick, red] (1+\a, 0.75+\b) -- (0.75+\a, 1+\b);
    \draw[thick, red] (0.25+\a, 1+\b) -- (\a, 0.75+\b);
    \node at (\a+0.5, -0.5) {$f_{I}$};
    \def\a{1.5}
	\def\b{0}
    \draw[thick] (\a, \b) -- (\a+1, \b) -- (\a+1, \b+1) -- (\a, \b+1) -- (\a, \b);
    \draw[thick, blue, dashed] (0.25+\a, \b) -- (\a, 0.25+\b);
    \draw[thick, red] (0.75+\a, \b) -- (1+\a, 0.25+\b);
    \draw[thick, blue] (1+\a, 0.75+\b) -- (0.75+\a, 1+\b);
    \draw[thick, red] (0.25+\a, 1+\b) -- (\a, 0.75+\b);
    \node at (\a+0.5, -0.5) {$f_{II}$};
    \def\a{3}
	\def\b{0}
    \draw[thick] (\a, \b) -- (\a+1, \b) -- (\a+1, \b+1) -- (\a, \b+1) -- (\a, \b);
    \draw[thick, blue, dashed] (0.25+\a, \b) -- (\a, 0.25+\b);
    \draw[thick, red] (0.75+\a, \b) -- (1+\a, 0.25+\b);
    \draw[thick, red] (1+\a, 0.75+\b) -- (0.75+\a, 1+\b);
    \draw[thick, blue] (0.25+\a, 1+\b) -- (\a, 0.75+\b);
    \node at (\a+0.5, -0.5) {$f_{III}$};
    \def\a{4.5}
	\def\b{0}
    \draw[thick] (\a, \b) -- (\a+1, \b) -- (\a+1, \b+1) -- (\a, \b+1) -- (\a, \b);
    \draw[thick, blue, dashed] (0.25+\a, \b) -- (\a, 0.25+\b);
    \draw[thick, blue] (0.75+\a, \b) -- (1+\a, 0.25+\b);
    \draw[thick, blue] (1+\a, 0.75+\b) -- (0.75+\a, 1+\b);
    \draw[thick, blue] (0.25+\a, 1+\b) -- (\a, 0.75+\b);
    \node at (\a+0.5, -0.5) {$f_{IV}$};
    }
\end{align*}
respectively.
The vertex operator in Eq. \ref{eq:avmodel1} preserves the vertex configurations in the ground state sector. Thus any ground state will be a sum of states where each state is built by tiling the square lattice with the face configurations, $f_I$, $f_{II}$, $f_{III}$ and $f_{IV}$, such that when four faces are joined at a vertex we obtain one of the vertex configurations, $v_I$, $v_{II}$, $v_{III}$ or $v_{IV}$. Using this language a state with the $N_v$-tuple, $\{-1, -1, \cdots, -1\}$, is given by
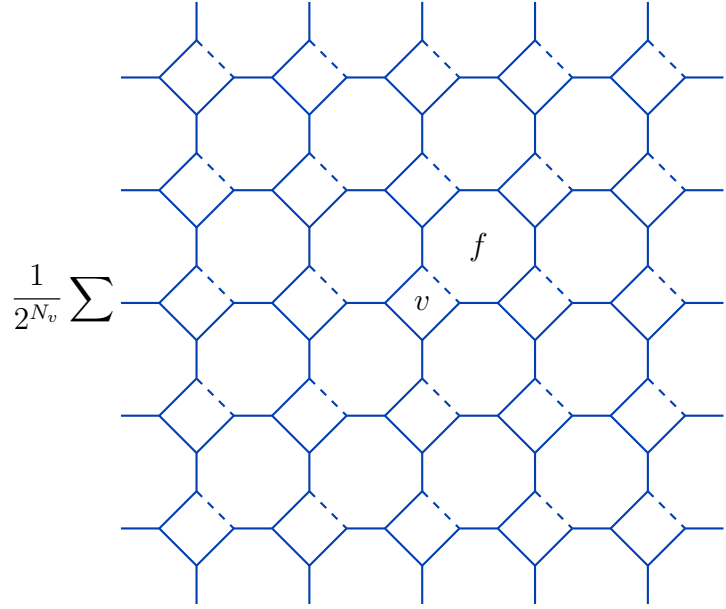
\begin{align*}
    \frac{1}{2^{N_{v}}}\sum \tikz[scale=1.5, baseline = 24.5ex]{
    \foreach \a in {0, 1, 2, 3, 4}{
    \foreach \b in {0, 1, 2, 3, 4} {
    \draw[thick, blue] (\a+1, \b+0.33) -- (\a+1, 0.67+\b) -- (\a+0.67, \b+1) -- (\a+0.33, 1+\b);
    \draw[thick, blue] (\a+1, 0.67+\b) -- (\a+1.33, \b+1);
    \draw[thick, blue] (\a+0.67, 1+\b) -- (\a+1, \b+1.33);
    \draw[thick, blue, dashed] (\a+1, \b+1.33) -- (\a+1.33, \b+1);
    }
    \draw[thick, blue] (\a+1, 5+0.33) -- (\a+1, 5+0.67);
    }
    \foreach \b in  {0, 1, 2, 3, 4}{
    \draw[thick, blue] (5.33, \b+1) -- (5.67, \b+1);
    }
    \node at (3, 3) {$v$};
    \node at (3.5, 3.5) {$f$};
    }.
\end{align*}
Here the summation is over all possible mismatches for the three corners (NW, SW, SE) at each vertex. The vertex, $v$ and face, $f$ of the square lattice are indicated as the centers of the squares and octagons in this state. The vertical blue line merely connect neighboring squares and have no meaning in terms of matched or mismatched configuration. Upon acting the global $\mathbb{Z}_2$ symmetry operator of Eq. \ref{eq:globalz2} on this state we obtain the other $N_v$-tuple labelled by $\{-1, -1, \cdots, -1\}$,
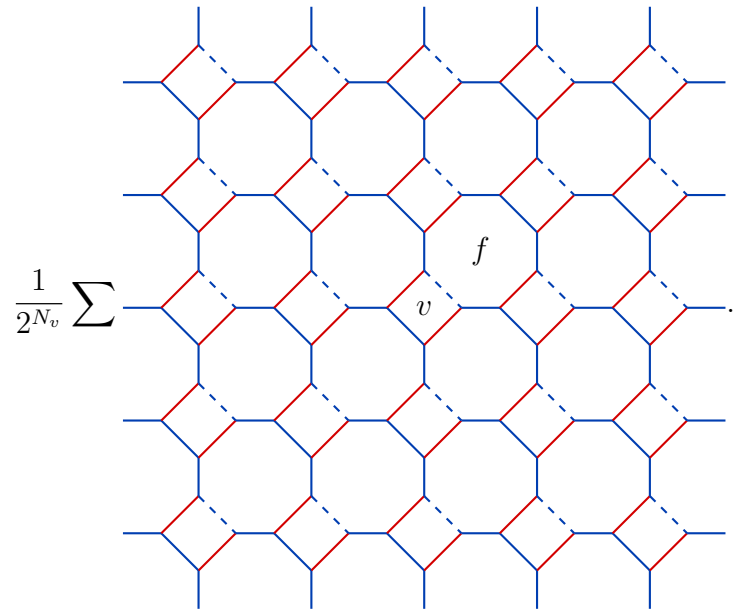
\begin{align*}
    \frac{1}{2^{N_{v}}}\sum \tikz[scale=1.5, baseline = 24.5ex]{
    \foreach \a in {0, 1, 2, 3, 4}{
    \foreach \b in {0, 1, 2, 3, 4} {
    \draw[thick, blue] (\a+1, \b+0.33) -- (\a+1, 0.67+\b) -- (\a+0.67, \b+1) -- (\a+0.33, 1+\b);
    \draw[thick, red] (\a+1, 0.67+\b) -- (\a+1.33, \b+1);
    \draw[thick, red] (\a+0.67, 1+\b) -- (\a+1, \b+1.33);
    \draw[thick, blue, dashed] (\a+1, \b+1.33) -- (\a+1.33, \b+1);
    }
    \draw[thick, blue] (\a+1, 5+0.33) -- (\a+1, 5+0.67);
    }
    \foreach \b in  {0, 1, 2, 3, 4}{
    \draw[thick, blue] (5.33, \b+1) -- (5.67, \b+1);
    }
    \node at (3, 3) {$v$};
    \node at (3.5, 3.5) {$f$};
    }.
\end{align*}
An example of a state obtained by applying a non-contractible loop operator on the state with $N_v$-tuple labelled by $\{-1, -1, \cdots, -1\}$ is
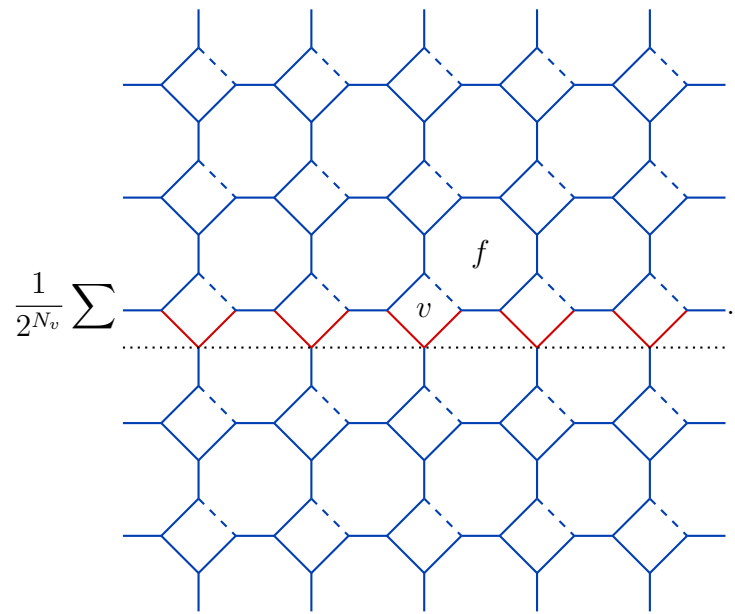
\begin{align*}
    \frac{1}{2^{N_{v}}}\sum \tikz[scale=1.5, baseline = 24.5ex]{
    \foreach \a in {0, 1, 2, 3, 4}{
    \foreach \b in {0, 1, 3, 4} {
    \draw[thick, blue] (\a+1, \b+0.33) -- (\a+1, 0.67+\b) -- (\a+0.67, \b+1) -- (\a+0.33, 1+\b);
    \draw[thick, blue] (\a+1, 0.67+\b) -- (\a+1.33, \b+1);
    \draw[thick, blue] (\a+0.67, 1+\b) -- (\a+1, \b+1.33);
    \draw[thick, blue, dashed] (\a+1, \b+1.33) -- (\a+1.33, \b+1);
    }
    \def\b{2};
    \draw[thick, blue] (\a+1, \b+0.33) -- (\a+1, 0.67+\b);
    \draw[thick, red] (\a+1, 0.67+\b) -- (\a+0.67, \b+1);
    \draw[thick, blue] (\a+0.67, \b+1) -- (\a+0.33, 1+\b);
    \draw[thick, red] (\a+1, 0.67+\b) -- (\a+1.33, \b+1);
    \draw[thick, blue] (\a+0.67, 1+\b) -- (\a+1, \b+1.33);
    \draw[thick, blue, dashed] (\a+1, \b+1.33) -- (\a+1.33, \b+1);
    \draw[thick, blue] (\a+1, 5+0.33) -- (\a+1, 5+0.67);
    }
    \foreach \b in  {0, 1, 2, 3, 4}{
    \draw[thick, blue] (5.33, \b+1) -- (5.67, \b+1);
    }
    \node at (3, 3) {$v$};
    \node at (3.5, 3.5) {$f$};
    \draw[thick, black, dotted] (0.33, 2+0.67) -- (0.67+5, 2+0.67);
    }.
\end{align*}
This state is paired with
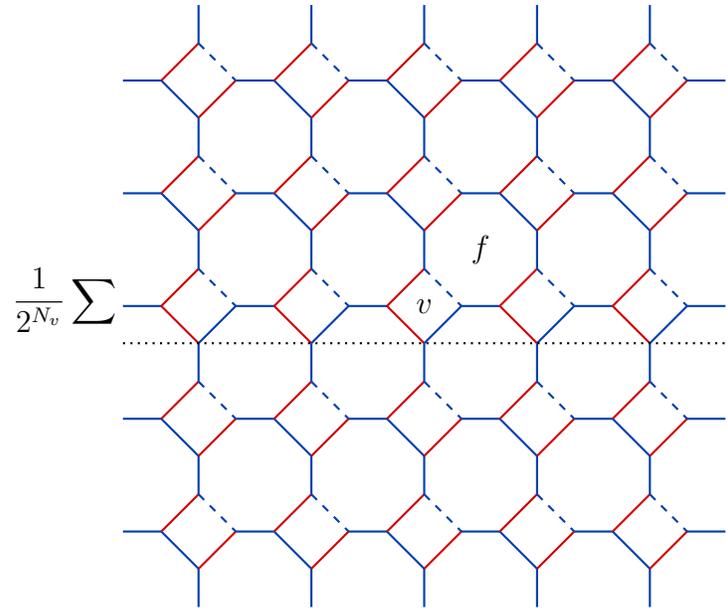
\begin{align*}
	\frac{1}{2^{N_{v}}}\sum \tikz[scale=1.5, baseline = 24.5ex]{
		\foreach \a in {0, 1, 2, 3, 4}{
			\foreach \b in {0, 1, 3, 4} {
				\draw[thick, blue] (\a+1, \b+0.33) -- (\a+1, 0.67+\b);
				\draw[thick, blue] (\a+1, 0.67+\b) -- (\a+0.67, \b+1);
				\draw[thick, blue] (\a+0.67, \b+1) -- (\a+0.33, 1+\b);
				\draw[thick, red] (\a+1, 0.67+\b) -- (\a+1.33, \b+1);
				\draw[thick, red] (\a+0.67, 1+\b) -- (\a+1, \b+1.33);
				\draw[thick, blue, dashed] (\a+1, \b+1.33) -- (\a+1.33, \b+1);
			}
			\draw[thick, blue] (\a+1, 5+0.33) -- (\a+1, 5+0.67);
			\def\b{2}
			\draw[thick, blue] (\a+1, \b+0.33) -- (\a+1, 0.67+\b);
			\draw[thick, red] (\a+1, 0.67+\b) -- (\a+0.67, \b+1);
			\draw[thick, blue] (\a+0.67, \b+1) -- (\a+0.33, 1+\b);
			\draw[thick, blue] (\a+1, 0.67+\b) -- (\a+1.33, \b+1);
			\draw[thick, red] (\a+0.67, 1+\b) -- (\a+1, \b+1.33);
			\draw[thick, blue, dashed] (\a+1, \b+1.33) -- (\a+1.33, \b+1);
		}
		\foreach \b in  {0, 1, 2, 3, 4}{
			\draw[thick, blue] (5.33, \b+1) -- (5.67, \b+1);
		}
		\node at (3, 3) {$v$};
		\node at (3.5, 3.5) {$f$};
		\draw[thick, black, dotted] (0.33, 2+0.67) -- (0.67+5, 2+0.67);
	},
\end{align*}
by the global $\mathbb{Z}_2$ symmetry of Eq. \ref{eq:globalz2}. An example of a state obtained by applying a contractible loop operator on the state with $N_v$-tuple labelled by $\{-1, -1, \cdots, -1\}$ is
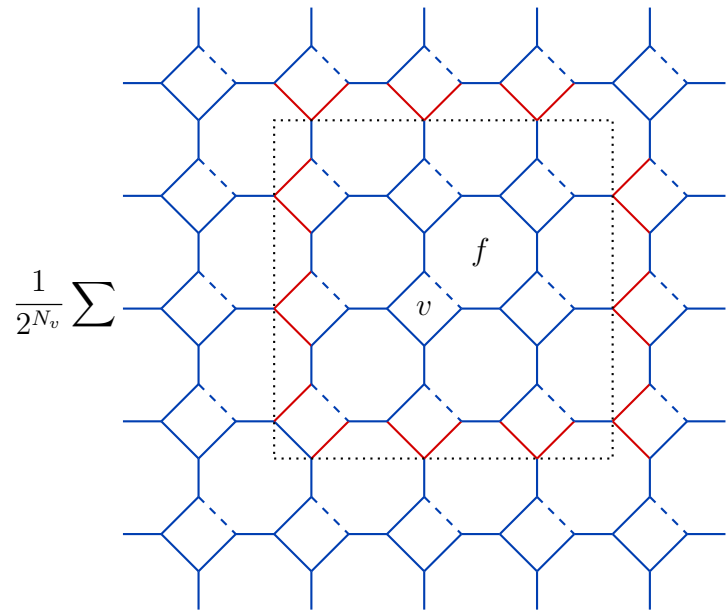
\begin{align*}
	\frac{1}{2^{N_{v}}}\sum \tikz[scale=1.5, baseline = 24.5ex]{
		\foreach \a in {0, 4}{
		\foreach \b in {0, 4}{
			\draw[thick, blue] (\a+1, \b+0.33) -- (\a+1, 0.67+\b);
			\draw[thick, blue] (\a+1, 0.67+\b) -- (\a+0.67, \b+1);
			\draw[thick, blue] (\a+0.67, \b+1) -- (\a+0.33, 1+\b);
			\draw[thick, blue] (\a+1, 0.67+\b) -- (\a+1.33, \b+1);
			\draw[thick, blue] (\a+0.67, 1+\b) -- (\a+1, \b+1.33);
			\draw[thick, blue, dashed] (\a+1, \b+1.33) -- (\a+1.33, \b+1);
		}
		}
		\draw[thick, blue] (1, 5.33) -- (1, 5.67);
		\draw[thick, blue] (5, 5.33) -- (5, 5.67);
		\foreach \a in {1, 2, 3}{
			\def\b{0}
				\draw[thick, blue] (\a+1, \b+0.33) -- (\a+1, 0.67+\b);
				\draw[thick, blue] (\a+1, 0.67+\b) -- (\a+0.67, \b+1);
				\draw[thick, blue] (\a+0.67, \b+1) -- (\a+0.33, 1+\b);
				\draw[thick, blue] (\a+1, 0.67+\b) -- (\a+1.33, \b+1);
				\draw[thick, blue] (\a+0.67, 1+\b) -- (\a+1, \b+1.33);
				\draw[thick, blue, dashed] (\a+1, \b+1.33) -- (\a+1.33, \b+1);
			\def\b{4}
				\draw[thick, blue] (\a+1, \b+0.33) -- (\a+1, 0.67+\b);
				\draw[thick, red] (\a+1, 0.67+\b) -- (\a+0.67, \b+1);
				\draw[thick, blue] (\a+0.67, \b+1) -- (\a+0.33, 1+\b);
				\draw[thick, red] (\a+1, 0.67+\b) -- (\a+1.33, \b+1);
				\draw[thick, blue] (\a+0.67, 1+\b) -- (\a+1, \b+1.33);
				\draw[thick, blue, dashed] (\a+1, \b+1.33) -- (\a+1.33, \b+1);
			\draw[thick, blue] (\a+1, 5+0.33) -- (\a+1, 5+0.67);
		}
		\foreach \b in {1, 2, 3}{
			\def\a{0}
				\draw[thick, blue] (\a+1, \b+0.33) -- (\a+1, 0.67+\b);
				\draw[thick, blue] (\a+1, 0.67+\b) -- (\a+0.67, \b+1);
				\draw[thick, blue] (\a+0.67, \b+1) -- (\a+0.33, 1+\b);
				\draw[thick, blue] (\a+1, 0.67+\b) -- (\a+1.33, \b+1);
				\draw[thick, blue] (\a+0.67, 1+\b) -- (\a+1, \b+1.33);
				\draw[thick, blue, dashed] (\a+1, \b+1.33) -- (\a+1.33, \b+1);
			\def\a{4}
				\draw[thick, blue] (\a+1, \b+0.33) -- (\a+1, 0.67+\b);
				\draw[thick, red] (\a+1, 0.67+\b) -- (\a+0.67, \b+1);
				\draw[thick, blue] (\a+0.67, \b+1) -- (\a+0.33, 1+\b);
				\draw[thick, blue] (\a+1, 0.67+\b) -- (\a+1.33, \b+1);
				\draw[thick, red] (\a+0.67, 1+\b) -- (\a+1, \b+1.33);
				\draw[thick, blue, dashed] (\a+1, \b+1.33) -- (\a+1.33, \b+1);
			}
		\foreach \a in {2, 3}{
			\def \b{1}
			\draw[thick, blue] (\a+1, \b+0.33) -- (\a+1, 0.67+\b);
			\draw[thick, red] (\a+1, 0.67+\b) -- (\a+0.67, \b+1);
			\draw[thick, blue] (\a+0.67, \b+1) -- (\a+0.33, 1+\b);
			\draw[thick, red] (\a+1, 0.67+\b) -- (\a+1.33, \b+1);
			\draw[thick, blue] (\a+0.67, 1+\b) -- (\a+1, \b+1.33);
			\draw[thick, blue, dashed] (\a+1, \b+1.33) -- (\a+1.33, \b+1);
		\foreach \b in {2, 3}{
			\draw[thick, blue] (\a+1, \b+0.33) -- (\a+1, 0.67+\b);
			\draw[thick, blue] (\a+1, 0.67+\b) -- (\a+0.67, \b+1);
			\draw[thick, blue] (\a+0.67, \b+1) -- (\a+0.33, 1+\b);
			\draw[thick, blue] (\a+1, 0.67+\b) -- (\a+1.33, \b+1);
			\draw[thick, blue] (\a+0.67, 1+\b) -- (\a+1, \b+1.33);
			\draw[thick, blue, dashed] (\a+1, \b+1.33) -- (\a+1.33, \b+1);
	}}
		\foreach \b in {2, 3}{
			\def\a{1}
			\draw[thick, blue] (\a+1, \b+0.33) -- (\a+1, 0.67+\b);
			\draw[thick, red] (\a+1, 0.67+\b) -- (\a+0.67, \b+1);
			\draw[thick, blue] (\a+0.67, \b+1) -- (\a+0.33, 1+\b);
			\draw[thick, blue] (\a+1, 0.67+\b) -- (\a+1.33, \b+1);
			\draw[thick, red] (\a+0.67, 1+\b) -- (\a+1, \b+1.33);
			\draw[thick, blue, dashed] (\a+1, \b+1.33) -- (\a+1.33, \b+1);
		}
	\def\a{1}
	\def\b{1}
			\draw[thick, blue] (\a+1, \b+0.33) -- (\a+1, 0.67+\b);
			\draw[thick, blue] (\a+1, 0.67+\b) -- (\a+0.67, \b+1);
			\draw[thick, blue] (\a+0.67, \b+1) -- (\a+0.33, 1+\b);
			\draw[thick, red] (\a+1, 0.67+\b) -- (\a+1.33, \b+1);
			\draw[thick, red] (\a+0.67, 1+\b) -- (\a+1, \b+1.33);
			\draw[thick, blue, dashed] (\a+1, \b+1.33) -- (\a+1.33, \b+1);
		\foreach \b in  {0, 1, 2, 3, 4}{
			\draw[thick, blue] (5.33, \b+1) -- (5.67, \b+1);
		}
		\draw[thick, dotted] (1.67, 1.67) -- (4.67, 1.67) -- (4.67, 4.67) -- (1.67, 4.67) -- (1.67, 1.67);
		\node at (3, 3) {$v$};
		\node at (3.5, 3.5) {$f$};
	},
\end{align*}
and its global $\mathbb{Z}_2$ symmetric pair is given by,
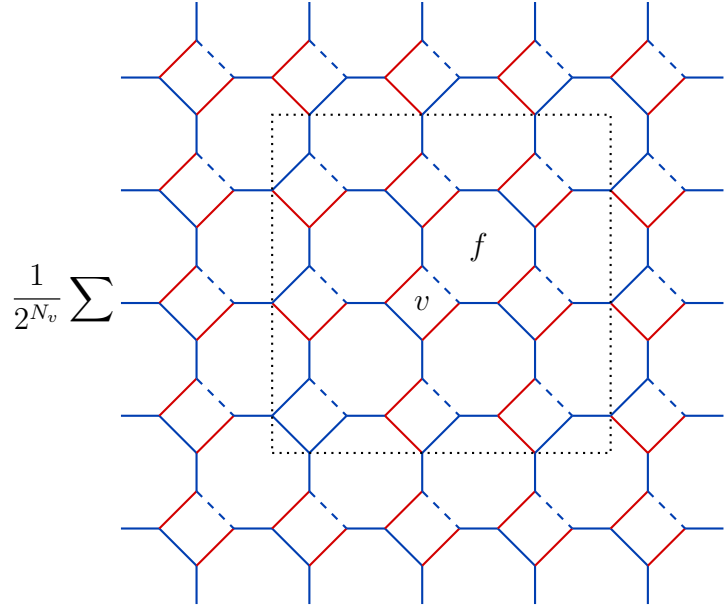
\begin{align*}
	\frac{1}{2^{N_{v}}}\sum \tikz[scale=1.5, baseline = 24.5ex]{
		\foreach \a in {0, 4}{
		\foreach \b in {0, 4}{
			\draw[thick, blue] (\a+1, \b+0.33) -- (\a+1, 0.67+\b);
			\draw[thick, blue] (\a+1, 0.67+\b) -- (\a+0.67, \b+1);
			\draw[thick, blue] (\a+0.67, \b+1) -- (\a+0.33, 1+\b);
			\draw[thick, red] (\a+1, 0.67+\b) -- (\a+1.33, \b+1);
			\draw[thick, red] (\a+0.67, 1+\b) -- (\a+1, \b+1.33);
			\draw[thick, blue, dashed] (\a+1, \b+1.33) -- (\a+1.33, \b+1);
		}
		}
		\draw[thick, blue] (1, 5.33) -- (1, 5.67);
		\draw[thick, blue] (5, 5.33) -- (5, 5.67);
		\foreach \a in {1, 2, 3}{
			\def\b{0}
				\draw[thick, blue] (\a+1, \b+0.33) -- (\a+1, 0.67+\b);
				\draw[thick, blue] (\a+1, 0.67+\b) -- (\a+0.67, \b+1);
				\draw[thick, blue] (\a+0.67, \b+1) -- (\a+0.33, 1+\b);
				\draw[thick, red] (\a+1, 0.67+\b) -- (\a+1.33, \b+1);
				\draw[thick, red] (\a+0.67, 1+\b) -- (\a+1, \b+1.33);
				\draw[thick, blue, dashed] (\a+1, \b+1.33) -- (\a+1.33, \b+1);
			\def\b{4}
				\draw[thick, blue] (\a+1, \b+0.33) -- (\a+1, 0.67+\b);
				\draw[thick, red] (\a+1, 0.67+\b) -- (\a+0.67, \b+1);
				\draw[thick, blue] (\a+0.67, \b+1) -- (\a+0.33, 1+\b);
				\draw[thick, blue] (\a+1, 0.67+\b) -- (\a+1.33, \b+1);
				\draw[thick, red] (\a+0.67, 1+\b) -- (\a+1, \b+1.33);
				\draw[thick, blue, dashed] (\a+1, \b+1.33) -- (\a+1.33, \b+1);
			\draw[thick, blue] (\a+1, 5+0.33) -- (\a+1, 5+0.67);
		}
		\foreach \b in {1, 2, 3}{
			\def\a{0}
				\draw[thick, blue] (\a+1, \b+0.33) -- (\a+1, 0.67+\b);
				\draw[thick, blue] (\a+1, 0.67+\b) -- (\a+0.67, \b+1);
				\draw[thick, blue] (\a+0.67, \b+1) -- (\a+0.33, 1+\b);
				\draw[thick, red] (\a+1, 0.67+\b) -- (\a+1.33, \b+1);
				\draw[thick, red] (\a+0.67, 1+\b) -- (\a+1, \b+1.33);
				\draw[thick, blue, dashed] (\a+1, \b+1.33) -- (\a+1.33, \b+1);
			\def\a{4}
				\draw[thick, blue] (\a+1, \b+0.33) -- (\a+1, 0.67+\b);
				\draw[thick, red] (\a+1, 0.67+\b) -- (\a+0.67, \b+1);
				\draw[thick, blue] (\a+0.67, \b+1) -- (\a+0.33, 1+\b);
				\draw[thick, red] (\a+1, 0.67+\b) -- (\a+1.33, \b+1);
				\draw[thick, blue] (\a+0.67, 1+\b) -- (\a+1, \b+1.33);
				\draw[thick, blue, dashed] (\a+1, \b+1.33) -- (\a+1.33, \b+1);
			}
		\foreach \a in {2, 3}{
			\def \b{1}
			\draw[thick, blue] (\a+1, \b+0.33) -- (\a+1, 0.67+\b);
			\draw[thick, red] (\a+1, 0.67+\b) -- (\a+0.67, \b+1);
			\draw[thick, blue] (\a+0.67, \b+1) -- (\a+0.33, 1+\b);
			\draw[thick, blue] (\a+1, 0.67+\b) -- (\a+1.33, \b+1);
			\draw[thick, red] (\a+0.67, 1+\b) -- (\a+1, \b+1.33);
			\draw[thick, blue, dashed] (\a+1, \b+1.33) -- (\a+1.33, \b+1);
		\foreach \b in {2, 3}{
			\draw[thick, blue] (\a+1, \b+0.33) -- (\a+1, 0.67+\b);
			\draw[thick, blue] (\a+1, 0.67+\b) -- (\a+0.67, \b+1);
			\draw[thick, blue] (\a+0.67, \b+1) -- (\a+0.33, 1+\b);
			\draw[thick, red] (\a+1, 0.67+\b) -- (\a+1.33, \b+1);
			\draw[thick, red] (\a+0.67, 1+\b) -- (\a+1, \b+1.33);
			\draw[thick, blue, dashed] (\a+1, \b+1.33) -- (\a+1.33, \b+1);
	}}
		\foreach \b in {2, 3}{
			\def\a{1}
			\draw[thick, blue] (\a+1, \b+0.33) -- (\a+1, 0.67+\b);
			\draw[thick, red] (\a+1, 0.67+\b) -- (\a+0.67, \b+1);
			\draw[thick, blue] (\a+0.67, \b+1) -- (\a+0.33, 1+\b);
			\draw[thick, red] (\a+1, 0.67+\b) -- (\a+1.33, \b+1);
			\draw[thick, blue] (\a+0.67, 1+\b) -- (\a+1, \b+1.33);
			\draw[thick, blue, dashed] (\a+1, \b+1.33) -- (\a+1.33, \b+1);
		}
	\def\a{1}
	\def\b{1}
			\draw[thick, blue] (\a+1, \b+0.33) -- (\a+1, 0.67+\b);
			\draw[thick, blue] (\a+1, 0.67+\b) -- (\a+0.67, \b+1);
			\draw[thick, blue] (\a+0.67, \b+1) -- (\a+0.33, 1+\b);
			\draw[thick, blue] (\a+1, 0.67+\b) -- (\a+1.33, \b+1);
			\draw[thick, blue] (\a+0.67, 1+\b) -- (\a+1, \b+1.33);
			\draw[thick, blue, dashed] (\a+1, \b+1.33) -- (\a+1.33, \b+1);
		\foreach \b in  {0, 1, 2, 3, 4}{
			\draw[thick, blue] (5.33, \b+1) -- (5.67, \b+1);
		}
		\draw[thick, dotted] (1.67, 1.67) -- (4.67, 1.67) -- (4.67, 4.67) -- (1.67, 4.67) -- (1.67, 1.67);
		\node at (3, 3) {$v$};
		\node at (3.5, 3.5) {$f$};
	}.
\end{align*}
To distinguish the two states having the $N_v$-tuple labelled by $\{-1, -1, \cdots, -1\}$ we use the local operators,
\tikz[baseline=-0.5ex, scale=0.7]{
	\def\a{0}
	\def\b{0}
	\draw[thick] (-1+\a, 0+\b) -- (1+\a, 0+\b);
	\draw[thick] (0+\a, -1+\b) -- (0+\a, 1+\b);
	\filldraw[red] (-0.5+\a,0+\b) circle (2pt) node[anchor=north, black] {$\textbf{\scriptsize{Z}}$};
	\filldraw[red] (0+\a, 0.5+\b) circle (2pt) node[anchor=south east, black] {$\textbf{\scriptsize{Z}}$};
    \def\a{2.5}
	\def\b{0}
	\draw[thick] (-1+\a, 0+\b) -- (1+\a, 0+\b);
	\draw[thick] (0+\a, -1+\b) -- (0+\a, 1+\b);
	\filldraw[red] (0+\a, -0.5+\b) circle (2pt) node[anchor=north west, black] {$\textbf{\scriptsize{Z}}$};
	\filldraw[red] (0.5+\a,0+\b) circle (2pt) node[anchor=south, black] {$\textbf{\scriptsize{Z}}$};
}, 
that measure if the NW and the SE corners of some vertex are in a mismatch-mismatch state or in a match-match state. Thus measuring this local order parameter distinguishes the two states with the same $N_v$ tuple. When this local operator is applied on the states obtained by applying a non-contractible loop on the state with $N_v$-tuple labelled by $\{-1, -1, \cdots, -1\}$ we can obtain either $(-1,-1)$ (mismatch-mismatch) or $(+1,+1)$ (match-match) when the vertex on which the local operator is applied is not next to the non-contractible loop and we obtain $(-1,+1)$ (mismatch-match) or $(+1,-1)$ (match-mismatch) when the vertex on which the local operator is applied is next to the non-contractible loop. The same logic applies to the states obtained by applying a contractible loop operator on the states with $N_v$-tuple labelled by $\{-1, -1, \cdots, -1\}$. Thus we need to measure the operator
\tikz[baseline=-0.5ex, scale=0.7]{
	\def\a{0}
	\def\b{0}
	\draw[thick] (-1+\a, 0+\b) -- (1+\a, 0+\b);
	\draw[thick] (0+\a, -1+\b) -- (0+\a, 1+\b);
	\node[anchor=north west] at (-0.1, 0.1) {$v$};
	\filldraw[red] (-0.5+\a,0+\b) circle (2pt) node[anchor=north, black] {$\textbf{\scriptsize{Z}}$};
	\filldraw[red] (0+\a, -0.5+\b) circle (2pt) node[anchor=north west, black] {$\textbf{\scriptsize{Z}}$};
}
at every vertex to discern among the $2^{N_v}$ loop states using the $N_v$-tuple and then the local operator
\tikz[baseline=-0.5ex, scale=0.7]{
	\def\a{0}
	\def\b{0}
	\draw[thick] (-1+\a, 0+\b) -- (1+\a, 0+\b);
	\draw[thick] (0+\a, -1+\b) -- (0+\a, 1+\b);
	\node[anchor=north west] at (-0.1+\a, 0.1+\b) {$v$};
	\filldraw[red] (-0.5+\a,0+\b) circle (2pt) node[anchor=north, black] {$\textbf{\scriptsize{Z}}$};
	\filldraw[red] (0+\a, 0.5+\b) circle (2pt) node[anchor=south east, black] {$\textbf{\scriptsize{Z}}$};
    \def\a{2.5}
	\def\b{0}
	\draw[thick] (-1+\a, 0+\b) -- (1+\a, 0+\b);
	\draw[thick] (0+\a, -1+\b) -- (0+\a, 1+\b);
	\node[anchor=north west] at (-0.1+\a, 0.1+\b) {$v$};
	\filldraw[red] (0+\a, -0.5+\b) circle (2pt) node[anchor=north west, black] {$\textbf{\scriptsize{Z}}$};
	\filldraw[red] (0.5+\a,0+\b) circle (2pt) node[anchor=south, black] {$\textbf{\scriptsize{Z}}$};
}
to distinguish between the two states connected by the global symmetry operator.

For completion we also write down the states labelled by the $N_v$-tuple $\{+1, +1, \cdots, +1\}$, 
\begin{align*}
    \frac{1}{2^{N_{v}}}\sum \tikz[scale=1.5, baseline = 24.5ex]{
    \foreach \a in {0, 1, 2, 3, 4}{
    \foreach \b in {0, 1, 2, 3, 4} {
    \draw[thick, blue] (\a+1, \b+0.33) -- (\a+1, 0.67+\b);
    \draw[thick, red] (\a+1, 0.67+\b) -- (\a+0.67, \b+1);
    \draw[thick, blue] (\a+0.67, \b+1) -- (\a+0.33, 1+\b);
    \draw[thick, blue] (\a+1, 0.67+\b) -- (\a+1.33, \b+1);
    \draw[thick, red] (\a+0.67, 1+\b) -- (\a+1, \b+1.33);
    \draw[thick, blue, dashed] (\a+1, \b+1.33) -- (\a+1.33, \b+1);
    }
    \draw[thick, blue] (\a+1, 5+0.33) -- (\a+1, 5+0.67);
    }
    \foreach \b in  {0, 1, 2, 3, 4}{
    \draw[thick, blue] (5.33, \b+1) -- (5.67, \b+1);
    }
    \node at (3, 3) {$v$};
    \node at (3.5, 3.5) {$f$};
    },
\end{align*}
and
\begin{align*}
    \frac{1}{2^{N_{v}}}\sum \tikz[scale=1.5, baseline = 24.5ex]{
    \foreach \a in {0, 1, 2, 3, 4}{
    \foreach \b in {0, 1, 2, 3, 4} {
    \draw[thick, blue] (\a+1, \b+0.33) -- (\a+1, 0.67+\b);
    \draw[thick, red] (\a+1, 0.67+\b) -- (\a+0.67, \b+1);
    \draw[thick, blue] (\a+0.67, \b+1) -- (\a+0.33, 1+\b);
    \draw[thick, red] (\a+1, 0.67+\b) -- (\a+1.33, \b+1);
    \draw[thick, blue] (\a+0.67, 1+\b) -- (\a+1, \b+1.33);
    \draw[thick, blue, dashed] (\a+1, \b+1.33) -- (\a+1.33, \b+1);
    }
    \draw[thick, blue] (\a+1, 5+0.33) -- (\a+1, 5+0.67);
    }
    \foreach \b in  {0, 1, 2, 3, 4}{
    \draw[thick, blue] (5.33, \b+1) -- (5.67, \b+1);
    }
    \node at (3, 3) {$v$};
    \node at (3.5, 3.5) {$f$};
    }.
\end{align*}

\subsection{Excitations}
\label{subsec:excitations}
Now we turn to the excitations of the Hamiltonian in Sec. \ref{eq:hmodel1} made of a sum of commuting projectors, the $A_v$ of Eq. \ref{eq:avmodel1} and the $B_f$ of Eq. \ref{eq:bfmodel1}. The analysis closely follows that of the $\mathbb{Z}_2$ toric code whose Hamiltonian has the same structure of a sum of commuting projectors. As we have already seen, the ground state sector is obtained when all the vertex and face operators take the eigenvalue +1, making the ground state energy $-2N_v$. Excitations are obtained when some of these projectors take the eigenvalue 0. Henceforth we will use the language that a vertex or face operator is excited when its eigenvalue is 0.

The ground states $\ket{G}$ take the form,
\begin{equation}\label{eq:canonicalgs}
    \ket{G} = \prod\limits_{v=1}^{N_v}~A_v~\ket{s},
\end{equation}
where $\ket{s}$ is an eigenstate of the $B_f$ operators. We create excitations by acting on this state. We will see that the excitations are either immobile in the case of the vertex excitations or have restricted mobility in the case of the flux excitations. These mimic the behavior of fracton models. 

\subsubsection*{Immobile vertex excitations - } All the face operators commute with $Z$ applied on any one of the qubit spaces and hence cannot be excited. The vertex operators, Eq. \ref{eq:avmodel1} contain a part checking parity and a part which transforms the qubits. While the former commutes with $Z$, the latter get excited by the $Z$ operator acting on a single qubit space. As neighboring vertex operators do not have common support, a single application of a $Z$ operator excites just the vertex operator associated with that qubit space. The resulting state is given by 
\begin{equation}
\ket{v} = A_{v_1}^{\perp}\prod\limits_{v\neq v_1}^{N_v}~A_v\ket{s},    
\end{equation}
where $A_v^{\perp}$ is obtained by replacing $\frac{1}{2}\left[1+\tikz[baseline=-0.5ex, scale=0.5]{
		\def\a{0}
		\def\b{0}
		\draw[thick] (-1+\a, 0+\b) -- (1+\a, 0+\b);
		\draw[thick] (0+\a, -1+\b) -- (0+\a, 1+\b);
		\node[anchor=north west] at (-0.1, 0.1) {$v$};
		\filldraw[red] (-0.5+\a,0+\b) circle (2pt) node[anchor=north, black] {$\textbf{\scriptsize{X}}$};
		\filldraw[red] (0+\a, -0.5+\b) circle (2pt) node[anchor=north west, black] {$\textbf{\scriptsize{X}}$};
		\filldraw[red] (0.5+\a,0+\b) circle (2pt) node[anchor=south, black] {$\textbf{\scriptsize{X}}$};
		\filldraw[red] (0+\a, 0.5+\b) circle (2pt) node[anchor=south east, black] {$\textbf{\scriptsize{X}}$};
	}\right]$
with
	$\frac{1}{2}\left[1-\tikz[baseline=-0.5ex, scale=0.5]{
		\def\a{0}
		\def\b{0}
		\draw[thick] (-1+\a, 0+\b) -- (1+\a, 0+\b);
		\draw[thick] (0+\a, -1+\b) -- (0+\a, 1+\b);
		\node[anchor=north west] at (-0.1, 0.1) {$v$};
		\filldraw[red] (-0.5+\a,0+\b) circle (2pt) node[anchor=north, black] {$\textbf{\scriptsize{X}}$};
		\filldraw[red] (0+\a, -0.5+\b) circle (2pt) node[anchor=north west, black] {$\textbf{\scriptsize{X}}$};
		\filldraw[red] (0.5+\a,0+\b) circle (2pt) node[anchor=south, black] {$\textbf{\scriptsize{X}}$};
		\filldraw[red] (0+\a, 0.5+\b) circle (2pt) node[anchor=south east, black] {$\textbf{\scriptsize{X}}$};
	}\right],$ in $A_v$ of Eq. \ref{eq:avmodel1}.
Applying more $Z$ operators on neighboring vertices just excites all of them and hence we conclude that these excitations are {\it isolated} and {\it immobile} unlike the vertex excitations of the $\mathbb{Z}_2$ toric code which occur in pairs on a closed surface, are deconfined, meaning they are free to move without an energy cost. Applying two $Z$ operators on qubit spaces associated to the same vertex does not change the ground state $\ket{G}$ as this is a local $\mathbb{Z}_2$ symmetry.

\subsubsection*{Deconfined fluxes with restricted motion - } The fluxes or face excitations of the $\mathbb{Z}_2$ toric code are deconfined and free to move along any direction of the lattice with no energy cost. In the groupoid version the face excitations are only deconfined along certain directions as we shall now illustrate. The paths that allow the fluxes to be deconfined are
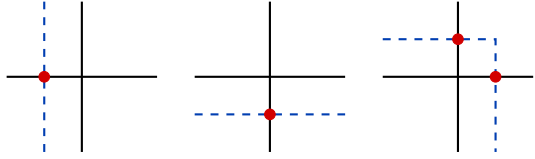
\begin{align*}
    \tikz[baseline=-0.5ex, scale=1]{
	\def\a{0}
	\def\b{0}
	\draw[thick] (-1+\a, 0+\b) -- (1+\a, 0+\b);
	\draw[thick] (0+\a, -1+\b) -- (0+\a, 1+\b);
    \draw[thick, dashed, blue] (-0.5+\a, -1+\b) -- (-0.5+\a, 1+\b);
	\filldraw[red] (-0.5+\a,0+\b) circle (2pt) node[anchor=north east, black] {$\textbf{\scriptsize{}}$};
	\def\a{2.5}
	\def\b{0}
	\draw[thick] (-1+\a, 0+\b) -- (1+\a, 0+\b);
	\draw[thick] (0+\a, -1+\b) -- (0+\a, 1+\b);
    \draw[thick, dashed, blue] (-1+\a, -0.5+\b) -- (1+\a, -0.5+\b);
	\filldraw[red] (0+\a, -0.5+\b) circle (2pt) node[anchor=north west, black] {$\textbf{\scriptsize{}}$};
    \def\a{5}
	\def\b{0}
	\draw[thick] (-1+\a, 0+\b) -- (1+\a, 0+\b);
	\draw[thick] (0+\a, -1+\b) -- (0+\a, 1+\b);
    \draw[thick, dashed, blue] (-1+\a, 0.5+\b) -- (0.5+\a, 0.5+\b) -- (0.5+\a, -1+\b);
	\filldraw[red] (0.5+\a,0+\b) circle (2pt) node[anchor=south west, black] {$\textbf{\scriptsize{}}$};
	\filldraw[red] (0+\a, 0.5+\b) circle (2pt) node[anchor=south east, black] {$\textbf{\scriptsize{}}$};
}
\end{align*}
Note that these are precisely the same configurations that help us construct the loop symmetries that are responsible for the GSD of the model. Alternately we can think of creating the deconfined fluxes by cutting open the loops of Sec. \ref{subsec:symmetries}. It is easy to see that they do not excite the associated vertex operators as well. An example of a pair of deconfined fluxes, $f_1$, $f_2$ is
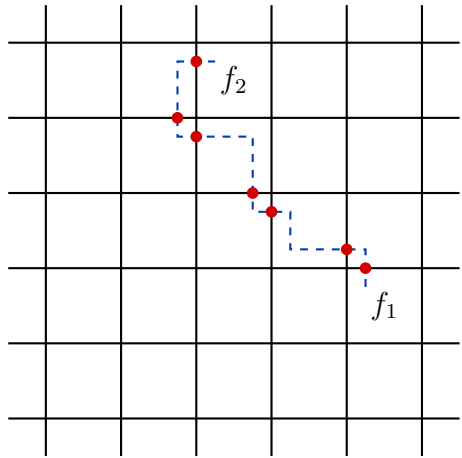
\begin{align*}
    	\tikz[baseline=12.5ex]{
		\foreach \x in {0, 1, 2, 3, 4, 5}{
			\draw[thick] (-0.5, \x) -- (5.5, \x);
		}
		\foreach \x in {0, 1, 2, 3, 4, 5}{
			\draw[thick] (\x, -0.5) -- (\x, 5.5);
		}
		\draw[thick, dashed, blue] (4.25, 1.75) -- (4.25, 2.25) -- (3.25, 2.25) -- (3.25, 2.75) -- (2.75, 2.75) -- (2.75, 3.75) -- (1.75, 3.75) -- (1.75, 4.75) -- (2.25, 4.75);
		\filldraw[red] (4.25,2) circle (2pt);
		\filldraw[red] (4,2.25) circle (2pt);
		\filldraw[red] (3,2.75) circle (2pt);
		\filldraw[red] (2.75,3) circle (2pt);
		\filldraw[red] (2,3.75) circle (2pt);
		\filldraw[red] (1.75,4) circle (2pt);
        \filldraw[red] (2,4.75) circle (2pt);
        \node at (4.5, 1.5) {$f_{1}$};
        \node at (2.5, 4.5) {$f_{2}$};
	}
\end{align*}
The restrictions to the motion of these fluxes occur when the paths cross either of the edges in the NE corner of a vertex,
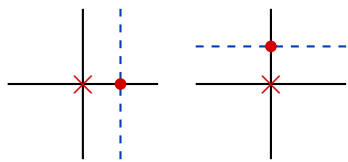
\begin{align*}
    \tikz[baseline=-0.5ex, scale=1]{
	\def\a{0}
	\def\b{0}
	\draw[thick] (-1+\a, 0+\b) -- (1+\a, 0+\b);
	\draw[thick] (0+\a, -1+\b) -- (0+\a, 1+\b);
    \draw[thick, dashed, blue] (0.5+\a, -1+\b) -- (0.5+\a, 1+\b);
	\filldraw[red] (0.5+\a,0+\b) circle (2pt) node[anchor=south, black] {$\textbf{\scriptsize{}}$};
    \node[thick, red] at (\a, \b) {\large{$\times$}};
	\def\a{2.5}
	\def\b{0}
	\draw[thick] (-1+\a, 0+\b) -- (1+\a, 0+\b);
	\draw[thick] (0+\a, -1+\b) -- (0+\a, 1+\b);
    \draw[thick, dashed, blue] (-1+\a, 0.5+\b) -- (1+\a, 0.5+\b);
	\filldraw[red] (0+\a, 0.5+\b) circle (2pt) node[anchor=south east, black] {$\textbf{\scriptsize{}}$};
    \node[thick, red] at (\a, \b) {\large{$\times$}};
 }
\end{align*}
as these paths excite the associated vertex operator as shown by the cross symbol. Note that the associated face operator is not excited leading to an interpretation where these paths convert a face excitation to a vertex excitation,
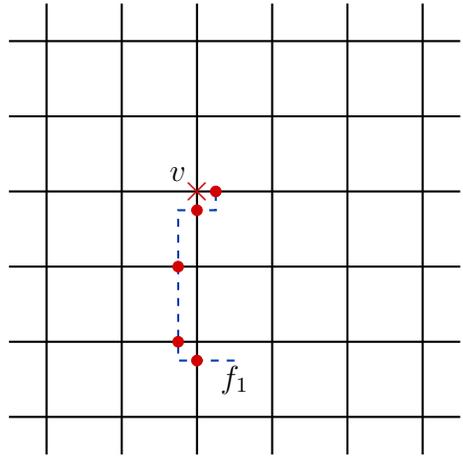
\begin{align*}
	\tikz[baseline=12.5ex]{
		\foreach \x in {0, 1, 2, 3, 4, 5}{
			\draw[thick] (-0.5, \x) -- (5.5, \x);
		}
		\foreach \x in {0, 1, 2, 3, 4, 5}{
			\draw[thick] (\x, -0.5) -- (\x, 5.5);
		}
		\draw[thick, dashed, blue] (2.5, 0.75) -- (1.75, 0.75) -- (1.75, 2.75) -- (2.25, 2.75) -- (2.25, 3);
		\filldraw[red] (2, 0.75) circle (2pt);
		\filldraw[red] (1.75, 1) circle (2pt);
		\filldraw[red] (1.75, 2) circle (2pt);
		\filldraw[red] (2, 2.75) circle (2pt);
		\filldraw[red] (2.25, 3) circle (2pt);
		\node at (2.5, 0.5) {$f_{1}$};
		\node[anchor=south east] at (2, 3) {$v$};
		\node[thick, red] at (2, 3) {\large{$\times$}};
	},
\end{align*}
resulting in a {\it face-vertex excitation pair}. Further extension of the string operator in the vertical direction creates new face-vertex pairs in addition to the existing one,
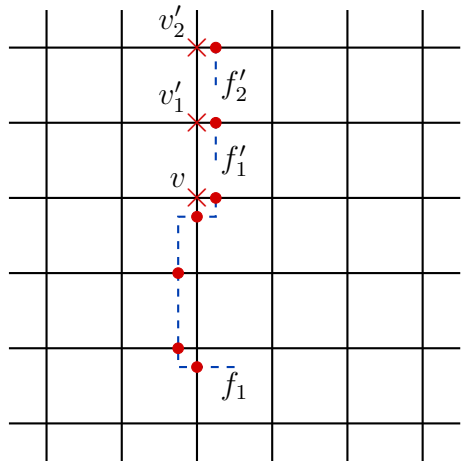
\begin{align*}
	\tikz[baseline=12.5ex]{
		\foreach \x in {0, 1, 2, 3, 4, 5}{
			\draw[thick] (-0.5, \x) -- (5.5, \x);
		}
		\foreach \x in {0, 1, 2, 3, 4, 5}{
			\draw[thick] (\x, -0.5) -- (\x, 5.5);
		}
		\draw[thick, dashed, blue] (2.5, 0.75) -- (1.75, 0.75) -- (1.75, 2.75) -- (2.25, 2.75) -- (2.25, 3);
		\filldraw[red] (2, 0.75) circle (2pt);
		\filldraw[red] (1.75, 1) circle (2pt);
		\filldraw[red] (1.75, 2) circle (2pt);
		\filldraw[red] (2, 2.75) circle (2pt);
		\filldraw[red] (2.25, 3) circle (2pt);
		\draw[thick, dashed, blue] (2.25, 3.5) -- (2.25, 4);
		\filldraw[red] (2.25, 4) circle (2pt);
		\draw[thick, dashed, blue] (2.25, 4.5) -- (2.25, 5);
		\filldraw[red] (2.25, 5) circle (2pt);
		\node at (2.5, 0.5) {$f_{1}$};
		\node at (2.5, 3.5) {$f_{1}'$};
		\node at (2.5, 4.5) {$f_{2}'$};
		\node[anchor=south east] at (2, 3) {$v$};
		\node[anchor=south east] at (2, 4) {$v_{1}'$};
		\node[anchor=south east] at (2, 5) {$v_{2}'$};
		\node[thick, red] at (2, 3) {\large{$\times$}};
		\node[thick, red] at (2, 4) {\large{$\times$}};
		\node[thick, red] at (2, 5) {\large{$\times$}};
	},
\end{align*}
thus confining the fluxes along this direction. The confinement also occurs along the horizontal direction,
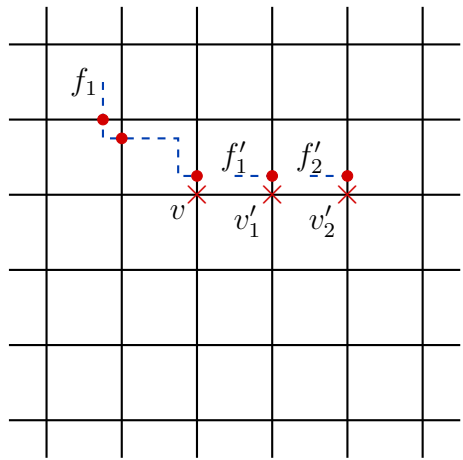
\begin{align*}
	\tikz[baseline=12.5ex]{
		\foreach \x in {0, 1, 2, 3, 4, 5}{
			\draw[thick] (-0.5, \x) -- (5.5, \x);
		}
		\foreach \x in {0, 1, 2, 3, 4, 5}{
			\draw[thick] (\x, -0.5) -- (\x, 5.5);
		}
		\draw[thick, dashed, blue] (0.75, 4.5) -- (0.75, 3.75) -- (1.75, 3.75) -- (1.75, 3.25) -- (2, 3.25);
		\filldraw[red] (0.75, 4) circle (2pt);
		\filldraw[red] (1, 3.75) circle (2pt);
		\filldraw[red] (2, 3.25) circle (2pt);
		\draw[thick, dashed, blue] (2.5, 3.25) -- (3, 3.25);
		\filldraw[red] (3, 3.25) circle (2pt);
		\draw[thick, dashed, blue] (3.5, 3.25) -- (4, 3.25);
		\filldraw[red] (4, 3.25) circle (2pt);
		\node at (0.5, 4.5) {$f_{1}$};
		\node at (2.5, 3.5) {$f_{1}'$};
		\node at (3.5, 3.5) {$f_{2}'$};
		\node[anchor=north east] at (2, 3) {$v$};
		\node[anchor=north east] at (3, 3) {$v_{1}'$};
		\node[anchor=north east] at (4, 3) {$v_{2}'$};
		\node[thick, red] at (2, 3) {\large{$\times$}};
		\node[thick, red] at (3, 3) {\large{$\times$}};
		\node[thick, red] at (4, 3) {\large{$\times$}};
	}.
\end{align*}
This feature is reminiscent of a mixing of low and high energy modes or UV/IR mixing. This seems to be another property of fracton models.

\subsubsection*{Immobile dyons - }
Unlike the $\mathbb{Z}_2$ toric code there are no deconfined dyonic excitations in the groupoid case as the vertex excitations are immobile. We can however think of the face-vertex pairs in horizontal $(f_1',v_1')$ and vertical directions $(f_2',v_2')$,
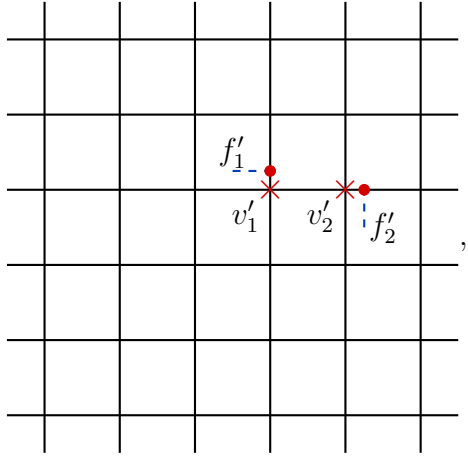
\begin{align*}
	\tikz[baseline=12.5ex]{
		\foreach \x in {0, 1, 2, 3, 4, 5}{
			\draw[thick] (-0.5, \x) -- (5.5, \x);
		}
		\foreach \x in {0, 1, 2, 3, 4, 5}{
			\draw[thick] (\x, -0.5) -- (\x, 5.5);
		}
		\draw[thick, dashed, blue] (2.5, 3.25) -- (3, 3.25);
		\filldraw[red] (3, 3.25) circle (2pt);
		\draw[thick, dashed, blue] (4.25, 2.5) -- (4.25, 3);
		\filldraw[red] (4.25, 3) circle (2pt);
		\node at (2.5, 3.5) {$f_{1}'$};
		\node at (4.5, 2.5) {$f_{2}'$};
		\node[anchor=north east] at (3, 3) {$v_{1}'$};
		\node[anchor=north east] at (4, 3) {$v_{2}'$};
		\node[thick, red] at (3, 3) {\large{$\times$}};
		\node[thick, red] at (4, 3) {\large{$\times$}};
	},
\end{align*}
as {\it immobile dyonic excitations}. Note that these are obtained by applying just a $X$ operator on the appropriate qubit spaces. 

\section{Other Hamiltonians}
\label{sec:hamiltonian2}
We can construct more exactly solvable Hamiltonians in the Hilbert space of the $\mathcal{S}^2_1$-groupoid using the vertex operators in Eq. \ref{eq:vertexopsqubitversionvalence4} and the $B_f^0$ operator in Eq. \ref{eq:bf0}. We list some possibilities here.

\subsubsection*{Model similar to Eq. \ref{eq:hmodel1} - } Consider the model with vertex operator obtained as 
\begin{equation}
    A_v = A_v^{(0)} + A_v^{(4)} + A_v^{(5)} + A_v^{(6)}. 
\end{equation}
This vertex operator simplifies to 
\begin{align}\label{eq:avmodel2}
A_{v}=\frac{1}{2}\left[1+\tikz[baseline=-0.5ex, scale=0.8]{
	\def\a{0}
	\def\b{0}
	\draw[thick] (-1+\a, 0+\b) -- (1+\a, 0+\b);
	\draw[thick] (0+\a, -1+\b) -- (0+\a, 1+\b);
	\node[anchor=north west] at (-0.1, 0.1) {$v$};
	\filldraw[red] (-0.5+\a,0+\b) circle (2pt) node[anchor=north, black] {$\textbf{\scriptsize{X}}$};
	\filldraw[red] (0.5+\a,0+\b) circle (2pt) node[anchor=south, black] {$\textbf{\scriptsize{X}}$};
	\filldraw[red] (0+\a, -0.5+\b) circle (2pt) node[anchor=north west, black] {$\textbf{\scriptsize{X}}$};
	\filldraw[red] (0+\a, 0.5+\b) circle (2pt) node[anchor=south east, black] {$\textbf{\scriptsize{X}}$};
}\right]
\frac{1}{2}\left[1+\tikz[baseline=-0.5ex, scale=0.8]{
	\def\a{0}
	\def\b{0}
	\draw[thick] (-1+\a, 0+\b) -- (1+\a, 0+\b);
	\draw[thick] (0+\a, -1+\b) -- (0+\a, 1+\b);
	\node[anchor=north west] at (-0.1, 0.1) {$v$};
	\filldraw[red] (-0.5+\a,0+\b) circle (2pt) node[anchor=north, black] {$\textbf{\scriptsize{}}$};
	\filldraw[red] (0+\a, -0.5+\b) circle (2pt) node[anchor=north west, black] {$\textbf{\scriptsize{}}$};
	\filldraw[red] (0.5+\a,0+\b) circle (2pt) node[anchor=south, black] {$\textbf{\scriptsize{Z}}$};
	\filldraw[red] (0+\a, 0.5+\b) circle (2pt) node[anchor=south east, black] {$\textbf{\scriptsize{Z}}$};
}\right].
\end{align}
The face operator is constructed out of the $B_f^0$ operator which have zero holonomy as a result of two mismatches in the corners,
\begin{equation}
  B_f  =  F_f^{NW,+}F_f^{NE,+}F_f^{SE,+} + F_f^{NW,-}F_f^{NE,-}F_f^{SE,+} + F_f^{NW,-}F_f^{NE,+}F_f^{SE,-} + F_f^{NW,+}F_f^{NE,-}F_f^{SE,-}.  
\end{equation}
This operator simplifies to
\begin{align}\label{eq:bfmodel2}
    B_{f}=\frac{1}{2}\left[1 + \tikz[baseline=2ex]{
		\draw[thick] (0, 0) -- (1, 0) -- (1, 1) -- (0, 1) -- (0, 0);
		\filldraw[red] (0.75,0) circle (2pt) node[anchor=north, black] {$\textbf{\scriptsize{Z}}$};
		\filldraw[red] (1,0.25) circle (2pt) node[anchor=west, black] {$\textbf{\scriptsize{Z}}$};
		\filldraw[red] (1,0.75) circle (2pt) node[anchor=west, black] {$\textbf{\scriptsize{Z}}$};
		\filldraw[red] (0.75,1) circle (2pt) node[anchor=south, black] {$\textbf{\scriptsize{Z}}$};
		\filldraw[red] (0.25,1) circle (2pt) node[anchor=south, black] {$\textbf{\scriptsize{Z}}$};
		\filldraw[red] (0,0.75) circle (2pt) node[anchor=east, black] {$\textbf{\scriptsize{Z}}$};
        \node at (0.5, 0.5) {$f$};
	}\right].
\end{align}
The resulting Hamiltonian takes the same form as in Eq. \ref{eq:hmodel1} but with the above vertex and face operators. The GSD of this model is similar to the model analyzed in Sec. \ref{sec:hamiltonian1} and the origin of the degeneracy is exactly the same as the before. As a result the nature of the excitations of this system is also similar to the model studied in Sec. \ref{sec:hamiltonian1}.

\subsubsection*{Model with an exponentially enlarged degeneracy - }
Now we turn to a model where the vertex operator acts as the gauge transformation when the configuration matches on one of either the $NW$, $SW$ or $SE$ corners of the vertex. This operator is obtained from 
\begin{equation}
    A_v = A_v^{(0)} + A_v^{(1)} + A_v^{(3)} + A_v^{(5)}, 
\end{equation}
which simplifies to 
\begin{align}\label{eq:avmodel3}
A_{v}=\frac{1}{2}\left[1+\tikz[baseline=-0.5ex, scale=0.8]{
	\def\a{0}
	\def\b{0}
	\draw[thick] (-1+\a, 0+\b) -- (1+\a, 0+\b);
	\draw[thick] (0+\a, -1+\b) -- (0+\a, 1+\b);
	\node[anchor=north west] at (-0.1, 0.1) {$v$};
	\filldraw[red] (-0.5+\a,0+\b) circle (2pt) node[anchor=north, black] {$\textbf{\scriptsize{X}}$};
	\filldraw[red] (0.5+\a,0+\b) circle (2pt) node[anchor=south, black] {$\textbf{\scriptsize{X}}$};
	\filldraw[red] (0+\a, -0.5+\b) circle (2pt) node[anchor=north west, black] {$\textbf{\scriptsize{X}}$};
	\filldraw[red] (0+\a, 0.5+\b) circle (2pt) node[anchor=south east, black] {$\textbf{\scriptsize{X}}$};
}\right]
\frac{1}{2}\left[1+\tikz[baseline=-0.5ex, scale=0.8]{
	\def\a{0}
	\def\b{0}
	\draw[thick] (-1+\a, 0+\b) -- (1+\a, 0+\b);
	\draw[thick] (0+\a, -1+\b) -- (0+\a, 1+\b);
	\node[anchor=north west] at (-0.1, 0.1) {$v$};
	\filldraw[red] (-0.5+\a,0+\b) circle (2pt) node[anchor=north, black] {$\textbf{\scriptsize{Z}}$};
	\filldraw[red] (0+\a, -0.5+\b) circle (2pt) node[anchor=north west, black] {$\textbf{\scriptsize{Z}}$};
	\filldraw[red] (0.5+\a,0+\b) circle (2pt) node[anchor=south, black] {$\textbf{\scriptsize{}}$};
	\filldraw[red] (0+\a, 0.5+\b) circle (2pt) node[anchor=south east, black] {$\textbf{\scriptsize{}}$};
}\right].
\end{align}
For the face operator we choose the configuration from the $B_f^0$ operator that projects the $NE$ corner to a matched configuration, that is $F^{NE,+}_f$,
\begin{align}\label{eq:bfmodel3}
    B_{f}=\frac{1}{2}\left[1 + \tikz[baseline=2ex]{
		\draw[thick] (0, 0) -- (1, 0) -- (1, 1) -- (0, 1) -- (0, 0);
		\filldraw[red] (1,0.75) circle (2pt) node[anchor=west, black] {$\textbf{\scriptsize{Z}}$};
		\filldraw[red] (0.75,1) circle (2pt) node[anchor=south, black] {$\textbf{\scriptsize{Z}}$};
        \node at (0.5, 0.5) {$f$};
	}\right].
\end{align}
As always the Hamiltonian for these choices takes the form of Eq. \ref{eq:hmodel1} and the GSD can be computed using the methods of Sec. \ref{subsec:gsd} as follows,
\begin{eqnarray}
    GSD  & = & \Tr~\prod\limits_{v=1}^{N_v}~A_v\prod\limits_{f=1}^{N_f}~B_f \nonumber \\
    & = & \frac{1}{2^{2N_v}}\frac{1}{2^{N_f}}~2^{N_v}\times 2^{4N_v} \nonumber \\ 
    & = & 2^{N_v}\times 2^{N_v}.
\end{eqnarray}
While the origin of most factors in this computation are exactly the same as the ones obtained in Sec. \ref{subsec:gsd}, there is an extra factor of $2^{N_v}$ which counts the number of identity operators in the product of the vertex and face operators. This results an exponential increase in the GSD of this model. At every vertex this model has three types of local symmetry operators each of which generates a $\mathbb{Z}_2$, 
\tikz[baseline=-0.5ex, scale=0.7]{
	\def\a{0}
	\def\b{0}
	\draw[thick] (-1+\a, 0+\b) -- (1+\a, 0+\b);
	\draw[thick] (0+\a, -1+\b) -- (0+\a, 1+\b);
	\filldraw[red] (-0.5+\a,0+\b) circle (2pt) node[anchor=north, black] {$\textbf{\scriptsize{X}}$};
	\filldraw[red] (0+\a, -0.5+\b) circle (2pt) node[anchor=west, black] {$\textbf{\scriptsize{X}}$};
}
\tikz[baseline=-0.5ex, scale=0.7]{
	\def\a{0}
	\def\b{0}
	\draw[thick] (-1+\a, 0+\b) -- (1+\a, 0+\b);
	\draw[thick] (0+\a, -1+\b) -- (0+\a, 1+\b);
	\filldraw[red] (0+\a, 0.5+\b) circle (2pt) node[anchor=east, black] {$\textbf{\scriptsize{X}}$};
}
\tikz[baseline=-0.5ex, scale=0.7]{
	\def\a{0}
	\def\b{0}
	\draw[thick] (-1+\a, 0+\b) -- (1+\a, 0+\b);
	\draw[thick] (0+\a, -1+\b) -- (0+\a, 1+\b);
	\filldraw[red] (0.5+\a, \b) circle (2pt) node[anchor=south, black] {$\textbf{\scriptsize{X}}$};
}.
However only two of these three operators are independent as the the third can be obtained by using the local vertex operator, and this accounts for the GSD of $2^{N_v}\times 2^{N_v}$. 

This model supports only immobile excitations. This is seen through the application of a $X$ operator on either the south or west of every vertex creating a pair of vertex and face excitations. Note that the faces cannot be separately excited in this system. An application of a $Z$ operator on any of the directions around a given vertex creates an isolated and immobile vertex excitation. 

\subsubsection*{Models where the fluxes are deconfined only along the horizontal or vertical directions - }
Consider a model where the vertex operator flips the qubits only when the configurations on the horizontal (W-E directions) match,
\begin{align}\label{eq:avmodel5}
A_{v}=\frac{1}{2}\left[1+\tikz[baseline=-0.5ex, scale=0.8]{
	\def\a{0}
	\def\b{0}
	\draw[thick] (-1+\a, 0+\b) -- (1+\a, 0+\b);
	\draw[thick] (0+\a, -1+\b) -- (0+\a, 1+\b);
	\node[anchor=north west] at (-0.1, 0.1) {$v$};
	\filldraw[red] (-0.5+\a,0+\b) circle (2pt) node[anchor=north, black] {$\textbf{\scriptsize{X}}$};
	\filldraw[red] (0.5+\a,0+\b) circle (2pt) node[anchor=south, black] {$\textbf{\scriptsize{X}}$};
	\filldraw[red] (0+\a, -0.5+\b) circle (2pt) node[anchor=north west, black] {$\textbf{\scriptsize{X}}$};
	\filldraw[red] (0+\a, 0.5+\b) circle (2pt) node[anchor=south east, black] {$\textbf{\scriptsize{X}}$};
}\right]
\frac{1}{2}\left[1+\tikz[baseline=-0.5ex, scale=0.8]{
	\def\a{0}
	\def\b{0}
	\draw[thick] (-1+\a, 0+\b) -- (1+\a, 0+\b);
	\draw[thick] (0+\a, -1+\b) -- (0+\a, 1+\b);
	\node[anchor=north west] at (-0.1, 0.1) {$v$};
	\filldraw[red] (-0.5+\a,0+\b) circle (2pt) node[anchor=north, black] {$\textbf{\scriptsize{Z}}$};
	\filldraw[red] (0+\a, -0.5+\b) circle (2pt) node[anchor=north west, black] {$\textbf{\scriptsize{}}$};
	\filldraw[red] (0.5+\a,0+\b) circle (2pt) node[anchor=south, black] {$\textbf{\scriptsize{Z}}$};
	\filldraw[red] (0+\a, 0.5+\b) circle (2pt) node[anchor=south east, black] {$\textbf{\scriptsize{}}$};
}\right].
\end{align}
This operator can also be obtained as
\begin{equation}
    A_v = A_v^{(0)} + A_v^{(1)} + A_v^{(6)} + A_v^{(7)},
\end{equation}
from the orthogonal set of vertex operators in Eq. \ref{eq:avmodel1}.
We choose the face operator as, 
\begin{align}\label{eq:bfmodel5}
    B_{f}=\frac{1}{2}\left[1 + \tikz[baseline=2ex]{
		\draw[thick] (0, 0) -- (1, 0) -- (1, 1) -- (0, 1) -- (0, 0);
		\filldraw[red] (1,0.75) circle (2pt) node[anchor=west, black] {$\textbf{\scriptsize{Z}}$};
		\filldraw[red] (0.75,1) circle (2pt) node[anchor=south, black] {$\textbf{\scriptsize{Z}}$};
		\filldraw[red] (0.25,1) circle (2pt) node[anchor=south, black] {$\textbf{\scriptsize{Z}}$};
		\filldraw[red] (0,0.75) circle (2pt) node[anchor=east, black] {$\textbf{\scriptsize{Z}}$};
        \node at (0.5, 0.5) {$f$};
	}\right],
\end{align}
which can be obtained as 
\begin{eqnarray}
    B_f & = &  F_f^{NW,+}F_f^{NE,+}F_f^{SE,+} + F_f^{NW,+}F_f^{NE,+}F_f^{SE,-} \nonumber \\  & + & F_f^{NW,-}F_f^{NE,-}F_f^{SE,+} + F_f^{NW,-}F_f^{NE,-}F_f^{SE,-},
\end{eqnarray}
using the face corner projectors in Eq. \ref{eq:cornerfaceprojectors}. 
With the techniques of Sec. \ref{subsec:gsd} the GSD of this model is found to be $2^k\times 2^{N_v}$ where $k$ is the number of horizontal non-contractible loops. Alternatively $k$ can be determined as follows. Consider a lattice with $m$ columns and $n$ rows (See Appendix \ref{app:finitegdtc}), the number of vertices equals $mn$ which also equals the number of faces. Now $k=m$ and hence the GSD becomes $2^m\times 2^{mn}$. The origin of $2^{mn}$ in this degeneracy is accounted for by \tikz[baseline=-0.5ex, scale=0.7]{
	\def\a{0}
	\def\b{0}
	\draw[thick] (-1+\a, 0+\b) -- (1+\a, 0+\b);
	\draw[thick] (0+\a, -1+\b) -- (0+\a, 1+\b);
	\filldraw[red] (0+\a, 0.5+\b) circle (2pt) node[anchor=east, black] {$\textbf{\scriptsize{X}}$};
} which generates a local $\mathbb{Z}_2$ symmetry at every vertex of the square lattice. The factor $2^m$ in the GSD originates from horizontal non-contractible loops,
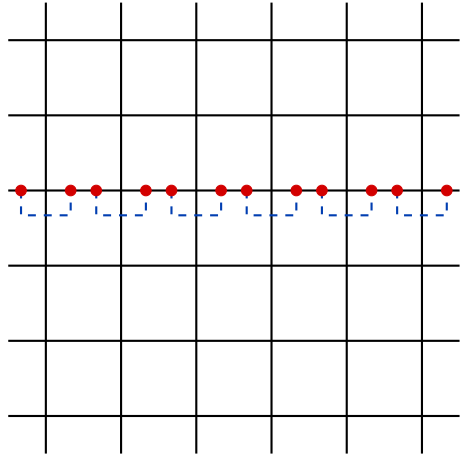
\begin{align*}
	\tikz[scale=1]{
		\foreach \a in {0, 1, 2, 3, 4, 5}{
			\draw[thick] (\a, -0.5) -- (\a, 5.5);
		}
		\foreach \b in {0, 1, 2, 3, 4, 5}{
			\draw[thick] (-0.5, \b) -- (5.5, \b);
		}
		\foreach \a in {0, 1, 2, 3, 4, 5}{
			\draw[thick, dashed, blue] (\a-0.33, 3) -- (\a - 0.33, 2.67) -- (\a + 0.33, 2.67) -- (\a + 0.33, 3);
			\filldraw[red] (\a-0.33, 3) circle (2pt);
			\filldraw[red] (\a+0.33, 3) circle (2pt);
		}
	}
\end{align*}
which can also be obtained from
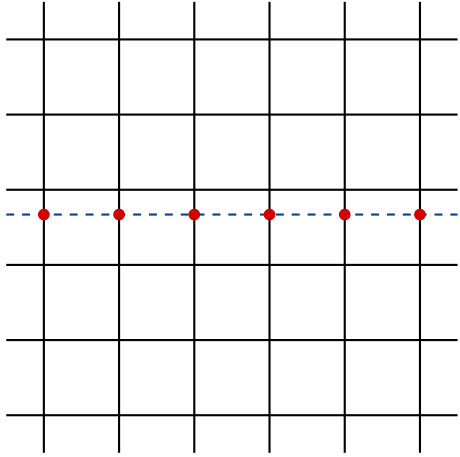
\begin{align*}
	\tikz[scale=1]{
		\foreach \a in {0, 1, 2, 3, 4, 5}{
			\draw[thick] (\a, -0.5) -- (\a, 5.5);
		}
		\foreach \b in {0, 1, 2, 3, 4, 5}{
			\draw[thick] (-0.5, \b) -- (5.5, \b);
		}
		\draw[thick, blue, dashed] (-0.5, 2.67) -- (5.5, 2.67);
		\foreach \a in {0, 1, 2, 3, 4, 5}{
			\filldraw[red] (\a, 2.67) circle (2pt);
		}
	}
\end{align*}
by the product of adjacent vertex transformations and the local $\mathbb{Z}_2$'s \tikz[baseline=-0.5ex, scale=0.7]{
	\def\a{0}
	\def\b{0}
	\draw[thick] (-1+\a, 0+\b) -- (1+\a, 0+\b);
	\draw[thick] (0+\a, -1+\b) -- (0+\a, 1+\b);
	\filldraw[red] (0+\a, 0.5+\b) circle (2pt) node[anchor=east, black] {$\textbf{\scriptsize{X}}$};
} on the same vertices. Thus we obtain two types of deconfined fluxes by cutting open either of these horizontal loops. It is easily seen that the fluxes get confined if they move out of this horizontal direction. This model also supports isolated and immobile vertex excitations obtained as a result of applying $Z$ on any of the qubit spaces around a vertex.

In a similar manner we write down the model that contains deconfined fluxes only along the vertical direction with the vertex operator,
\begin{align}\label{eq:avmodel6}
A_{v}=\frac{1}{2}\left[1+\tikz[baseline=-0.5ex, scale=0.8]{
	\def\a{0}
	\def\b{0}
	\draw[thick] (-1+\a, 0+\b) -- (1+\a, 0+\b);
	\draw[thick] (0+\a, -1+\b) -- (0+\a, 1+\b);
	\node[anchor=north west] at (-0.1, 0.1) {$v$};
	\filldraw[red] (-0.5+\a,0+\b) circle (2pt) node[anchor=north, black] {$\textbf{\scriptsize{X}}$};
	\filldraw[red] (0.5+\a,0+\b) circle (2pt) node[anchor=south, black] {$\textbf{\scriptsize{X}}$};
	\filldraw[red] (0+\a, -0.5+\b) circle (2pt) node[anchor=north west, black] {$\textbf{\scriptsize{X}}$};
	\filldraw[red] (0+\a, 0.5+\b) circle (2pt) node[anchor=south east, black] {$\textbf{\scriptsize{X}}$};
}\right]
\frac{1}{2}\left[1+\tikz[baseline=-0.5ex, scale=0.8]{
	\def\a{0}
	\def\b{0}
	\draw[thick] (-1+\a, 0+\b) -- (1+\a, 0+\b);
	\draw[thick] (0+\a, -1+\b) -- (0+\a, 1+\b);
	\node[anchor=north west] at (-0.1, 0.1) {$v$};
	\filldraw[red] (-0.5+\a,0+\b) circle (2pt) node[anchor=north, black] {$\textbf{\scriptsize{}}$};
	\filldraw[red] (0+\a, -0.5+\b) circle (2pt) node[anchor=north west, black] {$\textbf{\scriptsize{Z}}$};
	\filldraw[red] (0.5+\a,0+\b) circle (2pt) node[anchor=south, black] {$\textbf{\scriptsize{}}$};
	\filldraw[red] (0+\a, 0.5+\b) circle (2pt) node[anchor=south east, black] {$\textbf{\scriptsize{Z}}$};
}\right],
\end{align}
and the face operator,
\begin{align}\label{eq:bfmodel6}
    B_{f}=\frac{1}{2}\left[1 + \tikz[baseline=2ex]{
		\draw[thick] (0, 0) -- (1, 0) -- (1, 1) -- (0, 1) -- (0, 0);
		\filldraw[red] (0.75,0) circle (2pt) node[anchor=north, black] {$\textbf{\scriptsize{Z}}$};
		\filldraw[red] (1,0.25) circle (2pt) node[anchor=west, black] {$\textbf{\scriptsize{Z}}$};
		\filldraw[red] (1,0.75) circle (2pt) node[anchor=west, black] {$\textbf{\scriptsize{Z}}$};
		\filldraw[red] (0.75,1) circle (2pt) node[anchor=south, black] {$\textbf{\scriptsize{Z}}$};
        \node at (0.5, 0.5) {$f$};
	}\right].
\end{align}
The GSD of this model becomes $2^n\times 2^{mn}$ with local $\mathbb{Z}_2$ symmetries, \tikz[baseline=-0.5ex, scale=0.7]{
	\def\a{0}
	\def\b{0}
	\draw[thick] (-1+\a, 0+\b) -- (1+\a, 0+\b);
	\draw[thick] (0+\a, -1+\b) -- (0+\a, 1+\b);
	\filldraw[red] (-0.5+\a,0+\b) circle (2pt) node[anchor=north, black] {$\textbf{\scriptsize{X}}$};
} at every vertex and non-contractible loops along the vertical directions,
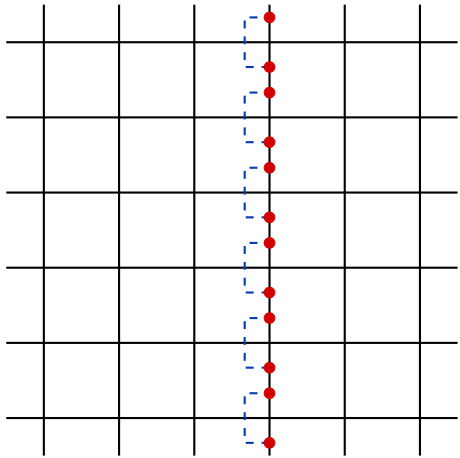
\begin{align*}
	\tikz[scale=1]{
		\foreach \a in {0, 1, 2, 3, 4, 5}{
			\draw[thick] (\a, -0.5) -- (\a, 5.5);
		}
		\foreach \b in {0, 1, 2, 3, 4, 5}{
			\draw[thick] (-0.5, \b) -- (5.5, \b);
		}
		\foreach \b in {0, 1, 2, 3, 4, 5}{
			\draw[thick, blue, dashed] (3, \b - 0.33) -- (2.67, \b - 0.33) -- (2.67, \b+0.33) -- (3, \b+0.33);
			\filldraw[red] (3, \b - 0.33) circle (2pt);
			\filldraw[red] (3, \b + 0.33) circle (2pt);
		}
	}
\end{align*}
The fluxes are now deconfined only along the vertical directions in this case. 

\subsubsection*{A non-degenerate model - }
We obtain a model with a unique ground state by picking the Hamiltonian,
\begin{equation}\label{eq:hmodel4}
    H = - \sum\limits_{v=1}^{N_v}~A_v^{(0)} - \sum\limits_{f=1}^{N_f}~\left(B_f^{1_1}+B_f^{1_2}\right),
\end{equation}
where $B_f^{1_1}+B_f^{1_2}$ is given by Eq. \ref{eq:bf1112aam1}. The number of ground states for this model can be counted as follows. Consider the ways to obtain a holonomy of either $1_1$ or $1_2$ for each face of the square lattice. This is achieved by having a matched configuration for all the corners of every face. In order for the vertex operator, $A_v^{(0)}$ to have a non-trivial action on this state we require four faces to join at a vertex in such a way that all the corners have the same matched configuration. This implies that if we populate the qubit spaces around every vertex of the square lattice with either 1 or 2 we will obtain a state where the holonomy across each face is either $1_1$ or $1_2$. The total number of such states is precisely $2^{N_v}$. Since the vertex operators act only on the qubit spaces surrounding the chosen vertex, we conclude that the number of gauge equivalent states to a configuration, whose holonomy across every face is either $1_1$ or $1_2$, is precisely $2^{N_v}$ as well. Thus there is just a single orbit under the action of the vertex operators on the +1 eigenstate of $B_f^{1_1}+B_f^{1_2}$ on all faces, leading to a non-degenerate ground state for the Hamiltonian in Eq. \ref{eq:hmodel4}. 

There are no deconfined excitations in this model. It is easy to see that the application of a $X$ operator on any qubit space creates a pair of flux excitations along with the vertex excitation adjacent to the qubit space. This triplet of excitations are immobile in all directions. Moving them with local operators will create more excitations on the lattice. Applying $Z$ operators on a single qubit space creates isolated and immobile vertex excitations as in the other examples. 

\section{Groupoid analogs of the $\mathbb{Z}_N$ toric code}
\label{sec:zngdtc}
We will now generalize the $\mathcal{S}^2_1$-groupoid toric code to a $\mathcal{S}^n_1$-groupoid case. The latter is a groupoid made of $n$ objects with precisely one morphism (and its inverse) between any two objects. If we denote the objects by the indices, $i$, $j$ taking values from $\{1,2,\cdots, n\}$, then the morphisms are written using the notation, $x_{ij}$ whose source and target are the objects $i$ and $j$ respectively. The partial identities are obtained when $i=j$ and there are $n$ of them. The morphisms obey the relations of the SIS in Eq. \ref{eq:SISmultiplication}, and hence we denote this groupoid toric code as the $\mathcal{S}_1^n$-toric code. The Hilbert space on each edge of the oriented lattice is spanned by the $n^2$ morphisms of the $\mathcal{S}^n_1$-groupoid and this is isomorphic to $\mathbb{C}^{n^2}$. As in the $n=2$ case this local space can be mapped to a $\mathbb{C}^n\otimes\mathbb{C}^n$ space located on two $n$-bit spaces adjacent to the vertices at the endpoints of this edge,
\begin{align*}
	\tikz[scale = 1.2, baseline = -0.5ex, decoration={
	markings,
	mark=at position 0.5 with {\arrow{>}}}]{
	\def\a{0}
	\def\b{0}
	\draw[thick, postaction={decorate}] (-3+\a, \b) -- (-1+\a, \b);
	\node[anchor=south] at (-2+\a, \b) {$x_{ij}$};
	\node at (\a, \b) {$\equiv$};
	\draw[thick] (1+\a, \b) -- (3+\a, \b);
	\filldraw[red] (1.5+\a, \b) circle (2pt) node[anchor=south, black] {$i$};
	\filldraw[red] (2.5+\a, \b) circle (2pt) node[anchor=south, black] {$j$};
},
\end{align*}
for $i,j\in\{1,2,\cdots, n\}$. Each $n$-bit space, $\mathbb{C}^n$ is spanned by the basis elements, $\{\ket{i}, i\in\{1,2,\cdots, n\}$, which are the object indices of the $\mathcal{S}^n_1$-groupoid. We use $\textrm{mod}~n$ arithmetic while doing the computations and identify $\ket{0}\equiv\ket{n}$ to be consistent with the indices used for the objects of the $\mathcal{S}^n_1$-groupoid.
Next we write down the orthogonal set of vertex and face operators for the square lattice acting on these $n$-bit spaces using the clock and shift operators,
\begin{equation}
    Z~\ket{i} = \omega_n^i~\ket{i},~~X~\ket{i} = \ket{i+1~\textrm{mod}~n},
\end{equation}
which satisfy the relations,
\begin{equation}\label{eq:nclockshiftrelations}
    X^{j_1}Z^{j_2} = \omega^{-j_1j_2}_n~Z^{j_2}X^{j_1},~~X^n=Z^n=1.
\end{equation}
In these expressions $\omega_n=e^{\frac{2\pi\mathrm{i}}{n}}$ is the $n$th root of unity. 

\subsubsection*{Face Operators - }
The face operators measure the holonomies around a face in the language of morphisms and this translates to checking if the corners around a face are either in a matched or mismatched configuration when interpreted in terms of the object indices of the $\mathcal{S}^n_1$-groupoid. Unlike the $n=2$ or the $\mathcal{S}^2_1$-groupoid case there are more mismatches possible for a higher $n$. In particular we can measure the mismatch by $k<n$ steps, that is when the configuration across a corner are the indices $j$, $j+k$. The projectors to such mismatches are,
\begin{align}
	F_{f}^{NW, k}&=\frac{1}{n}\sum_{j=1}^{n}\omega_{n}^{kj}\tikz[baseline=2ex, scale=1.2]{
		\draw[thick] (0, 0) -- (1, 0) -- (1, 1) -- (0, 1) -- (0, 0);
		\filldraw[red] (0.25,1) circle (2pt) node[anchor=south, black] {$\scriptstyle Z^{-j}$};
		\filldraw[red] (0,0.75) circle (2pt) node[anchor=east, black] {$\scriptstyle Z^{j}$};
		\node at (0.5, 0.5) {$f$};
	},&
	F_{f}^{NE, k}&=\frac{1}{n}\sum_{j=1}^{n}\omega_{n}^{kj}\tikz[baseline=2.5ex, scale=1.2]{
		\draw[thick] (0, 0) -- (1, 0) -- (1, 1) -- (0, 1) -- (0, 0);
		\filldraw[red] (0.75,1) circle (2pt) node[anchor=south, black] {$\scriptstyle Z^{j}$};
		\filldraw[red] (1,0.75) circle (2pt) node[anchor=west, black] {$\scriptstyle Z^{-j}$};
		\node at (0.5, 0.5) {$f$};
	},\nonumber\\
	F_{f}^{SE, k}&=\frac{1}{n}\sum_{j=1}^{n}\omega_{n}^{kj}\tikz[baseline=2.5ex, scale=1.2]{
	\draw[thick] (0, 0) -- (1, 0) -- (1, 1) -- (0, 1) -- (0, 0);
	\filldraw[red] (1,0.25) circle (2pt) node[anchor=west, black] {$\scriptstyle Z^{j}$};
	\filldraw[red] (0.75, 0) circle (2pt) node[anchor=north, black] {$\scriptstyle Z^{-j}$};
	\node at (0.5, 0.5) {$f$};
	},&
	F_{f}^{SW, k}&=\frac{1}{n}\sum_{j=1}^{n}\omega_{n}^{kj}\tikz[baseline=2.5ex, scale=1.2]{
	\draw[thick] (0, 0) -- (1, 0) -- (1, 1) -- (0, 1) -- (0, 0);
	\filldraw[red] (0.25,0) circle (2pt) node[anchor=north, black] {$\scriptstyle Z^{j}$};
	\filldraw[red] (0,0.25) circle (2pt) node[anchor=east, black] {$\scriptstyle Z^{-j}$};
	\node at (0.5, 0.5) {$f$};
	},
\end{align}
with $k\in\{1,2,\cdots, n\}$. The projector to the matched configurations across corners occur for $k=n$. Using these operators we can write down the projectors, $B_f^{x_{ii}}$, $B_f^{x_{ik}}$ and $B_f^0$, to the different morphism-valued holonomies as
\begin{align}
	B_{f}^{x_{ii}}&=F_{f}^{NW, n}F_{f}^{NE, n}F_{f}^{SE, n}\left[\frac{1}{n}\sum_{j=1}^{n}\omega_{n}^{-ij}\tikz[baseline=2.5ex, scale=1.2]{
		\draw[thick] (0, 0) -- (1, 0) -- (1, 1) -- (0, 1) -- (0, 0);
		\filldraw[red] (0.25,0) circle (2pt) node[anchor=north, black] {$\scriptstyle Z^{j}$};
		\node at (0.5, 0.5) {$f$};
	}\right]
	\left[\frac{1}{n}\sum_{j=1}^{n}\omega_{n}^{-ij}\tikz[baseline=2.5ex, scale=1.2]{
		\draw[thick] (0, 0) -- (1, 0) -- (1, 1) -- (0, 1) -- (0, 0);
		\filldraw[red] (0, 0.25) circle (2pt) node[anchor=east, black] {$\scriptstyle Z^{j}$};
		\node at (0.5, 0.5) {$f$};
		\node at (1, 1.1) {};
	}\right]\nonumber\\
	B_{f}^{x_{ik}}&=F_{f}^{NW, n}F_{f}^{NE, n}F_{f}^{SE, n}\left[\frac{1}{n}\sum_{j=1}^{n}\omega_{n}^{-ij}\tikz[baseline=2.5ex, scale=1.2]{
		\draw[thick] (0, 0) -- (1, 0) -- (1, 1) -- (0, 1) -- (0, 0);
		\filldraw[red] (0.25,0) circle (2pt) node[anchor=north, black] {$\scriptstyle Z^{j}$};
		\node at (0.5, 0.5) {$f$};
	}\right]
	\left[\frac{1}{n}\sum_{j=1}^{n}\omega_{n}^{-kj}\tikz[baseline=2.5ex, scale=1.2]{
		\draw[thick] (0, 0) -- (1, 0) -- (1, 1) -- (0, 1) -- (0, 0);
		\filldraw[red] (0, 0.25) circle (2pt) node[anchor=east, black] {$\scriptstyle Z^{j}$};
		\node at (0.5, 0.5) {$f$};
		\node at (1, 1.1) {};
	}\right],
\end{align}
for $i$, $k$ in $\{1,2,\cdots, n\}$, and 
\begin{eqnarray}
    B_f^0 & =& \sum\limits_{k_1,k_2,k_3=1}^{n-1}~F_f^{NW, k_1}F_f^{NE, k_2}F_f^{SE, k_3} \nonumber \\ & + & \sum\limits_{k_1,k_2=1}^{n-1}~F_f^{NW, k_1}F_f^{NE, k_2}F_f^{SE, n} + \sum\limits_{k_2,k_3=1}^{n-1}~F_f^{NW, n}F_f^{NE, k_2}F_f^{SE, k_3} + \sum\limits_{k_3,k_1=1}^{n-1}~F_f^{NW, k_1}F_f^{NE, n}F_f^{SE, k_3} \nonumber \\
    & + & \sum\limits_{k_1=1}^{n-1}~F_f^{NW, k_1}F_f^{NE, n}F_f^{SE, n} + \sum\limits_{k_2=1}^{n-1}~F_f^{NW, n}F_f^{NE, k_2}F_f^{SE, n} + \sum\limits_{k_3=1}^{n-1}~F_f^{NW, n}F_f^{NE, n}F_f^{SE, k_3}. \nonumber \\
\end{eqnarray}
The $B_f^0$ operator measures either one, two or three mismatches on the $NW$, $NE$ and $SE$ corners of the face $f$. These exhaust the set of orthogonal face operators satisfying,
\begin{equation}
    \sum\limits_{i=1}^n~B_f^{x_{ii}} + \sum\limits_{i, j=1}^n~B_f^{x_{ij}} + B_f^0 = 1.
\end{equation}

\subsubsection*{Vertex Operators - }
Following of the structure of the $n=2$ case we expect the vertex operators of the $\mathcal{S}^n_1$-toric code to shift the object indices after checking for matched and mismatched configurations on the $NW$, $SW$ and $SE$ corners of any vertex of the square lattice. These are done using the projectors,
\begin{align}
	V_{v}^{NW, k}&=\frac{1}{n}\sum_{j=1}^{n}\omega_{n}^{kj}\tikz[baseline=-0.5ex, scale=1]{
		\def\a{0}
		\def\b{0}
		\draw[thick] (-1+\a, 0+\b) -- (1+\a, 0+\b);
		\draw[thick] (0+\a, -1+\b) -- (0+\a, 1+\b);
		\filldraw[red] (-0.5+\a,0+\b) circle (2pt) node[anchor=north, black] {$\scriptstyle Z^{-j}$};
		\filldraw[red] (0+\a, 0.5+\b) circle (2pt) node[anchor=east, black] {$\scriptstyle Z^{j}$};
	}\nonumber\\
	V_{v}^{SW, k}&=\frac{1}{n}\sum_{j=1}^{n}\omega_{n}^{kj}\tikz[baseline=-0.5ex, scale=1]{
	\def\a{0}
	\def\b{0}
	\draw[thick] (-1+\a, 0+\b) -- (1+\a, 0+\b);
	\draw[thick] (0+\a, -1+\b) -- (0+\a, 1+\b);
	\filldraw[red] (-0.5+\a,0+\b) circle (2pt) node[anchor=north, black] {$\scriptstyle Z^{j}$};
	\filldraw[red] (0+\a, -0.5+\b) circle (2pt) node[anchor=west, black] {$\scriptstyle Z^{-j}$};
	}\nonumber\\
	V_{v}^{SE, k}&=\frac{1}{n}\sum_{j=1}^{n}\omega_{n}^{kj}\tikz[baseline=-0.5ex, scale=1]{
	\def\a{0}
	\def\b{0}
	\draw[thick] (-1+\a, 0+\b) -- (1+\a, 0+\b);
	\draw[thick] (0+\a, -1+\b) -- (0+\a, 1+\b);
	\filldraw[red] (\a, -0.5+\b) circle (2pt) node[anchor=west, black] {$\scriptstyle Z^{j}$};
	\filldraw[red] (0.5+\a, \b) circle (2pt) node[anchor=south, black] {$\scriptstyle Z^{-j}$};
	}.
\end{align}
The orthogonal set of shift operators at a vertex can be written as 
\begin{align}\label{eq:vertexshiftopn}
	X_{v}^{k}&=\frac{1}{n}\sum_{j=1}^{n}\omega_{n}^{kj}\tikz[baseline=-0.5ex, scale=1]{
		\def\a{0}
		\def\b{0}
		\draw[thick] (-1+\a, 0+\b) -- (1+\a, 0+\b);
		\draw[thick] (0+\a, -1+\b) -- (0+\a, 1+\b);
		\filldraw[red] (\a, -0.5+\b) circle (2pt) node[anchor=west, black] {$\scriptstyle X^{j}$};
		\filldraw[red] (0.5+\a, \b) circle (2pt) node[anchor=south, black] {$\scriptstyle X^{j}$};
		\filldraw[red] (\a, 0.5+\b) circle (2pt) node[anchor=east, black] {$\scriptstyle X^{j}$};
		\filldraw[red] (-0.5+\a, \b) circle (2pt) node[anchor=north, black] {$\scriptstyle X^{j}$};
	}
\end{align}
Using these operators we can write down the orthogonal set of vertex operators of the $\mathcal{S}^n_1$-toric code as 
\begin{equation}\label{eq:vertexopsetsn1}
    A_v^{k ; k_1,k_2,k_3} = X_v^k~V_v^{NW, k_1}V_v^{SW, k_2}V_v^{SE, k_3},
\end{equation}
for $\left(k, k_1, k_2, k_3\right)\in\{1,2,\cdots, n\}$. Thus we have a total of $n^4$ orthogonal operators satisfying
\begin{equation}
    \sum\limits_{k,k_1,k_2,k_3=1}^n~A_v^{k;k_1,k_2,k_3} =1.
\end{equation}

\subsubsection*{The $\mathcal{S}^n_1$-Groupoid Toric Code - }
The vertex operator for this model is a generalization of the vertex operator of Eq. \ref{eq:avmodel2},
\begin{align}
	A_{v}=X_{v}^{n}\left[\frac{1}{n}\sum_{j=1}^{n}\tikz[baseline=-0.5ex, scale=1]{
		\def\a{0}
		\def\b{0}
		\draw[thick] (-1+\a, 0+\b) -- (1+\a, 0+\b);
		\draw[thick] (0+\a, -1+\b) -- (0+\a, 1+\b);
		\filldraw[red] (\a, 0.5+\b) circle (2pt) node[anchor=east, black] {$\scriptstyle Z^{-j}$};
		\filldraw[red] (0.5+\a, \b) circle (2pt) node[anchor=south, black] {$\scriptstyle Z^{j}$};
	}\right]
\end{align}
where the $X_v^n$ is given by Eq. \ref{eq:vertexshiftopn}. The face operator generalizing Eq. \ref{eq:bfmodel2} is 
\begin{align}
	B_{f}=\frac{1}{n}\sum_{j=1}^{n}\tikz[baseline=3ex, scale=1.2]{
		\draw[thick] (0, 0) -- (1, 0) -- (1, 1) -- (0, 1) -- (0, 0);
		\filldraw[red] (0.75,0) circle (2pt) node[anchor=north, black] {$\scriptstyle Z^{-j}$};
		\filldraw[red] (1,0.25) circle (2pt) node[anchor=west, black] {$\scriptstyle Z^{j}$};
		\filldraw[red] (1,0.75) circle (2pt) node[anchor=west, black] {$\scriptstyle Z^{-j}$};
		\filldraw[red] (0.75,1) circle (2pt) node[anchor=south, black] {$\scriptstyle Z^{j}$};
		\filldraw[red] (0.25,1) circle (2pt) node[anchor=south, black] {$\scriptstyle Z^{-j}$};
		\filldraw[red] (0,0.75) circle (2pt) node[anchor=east, black] {$\scriptstyle Z^{j}$};
		\node at (0.5, 0.5) {$f$};
	}
\end{align}
The Hamiltonian takes the same form as in Eq. \ref{eq:hmodel1} and the GSD for this model is found to be 
\begin{equation}
    GSD = n\times n^{N_v},
\end{equation}
following the methods used in Sec. \ref{subsec:gsd}. This model has a $\mathbb{Z}_n$ global symmetry generated by $X$ of the form in Eq. \ref{eq:globalz2}. The factor of $n^{N_V}$ originates from loop operators as in the $n=2$ case. However the $X$ operator on the loops generate a $\mathbb{Z}_n$ in contrast to the $\mathbb{Z}_2$ in the earlier model. Breaking open these loop symmetries give us pairs of deconfined $\mathbb{Z}_n$ toric code fluxes. The restrictions on the position of these loops is the same as in the case of the $\mathcal{S}^2_1$-toric code. 

\section{Discussion}
\label{sec:discussion}
In this work we have replaced the finite group algebra with a finite groupoid algebra to construct exactly solvable Hamiltonians similar to the toric code and in general the quantum double models of Kitaev. This change modifies two of the crucial features of the group toric code models :
\begin{enumerate}
    \item The topological degeneracy of the group toric code gets replaced by an extensive degeneracy that is sensitive to lattice details.
    \item The excitations are either immobile or have restricted motion unlike the group toric code where they are all deconfined and are free to move.
    \item As a consequence of the restricted motion the braid statistics of these particles will also differ from the case of the group toric code. In particular there are a limited number of choices for the paths they can take while performing the exchange. 
\end{enumerate}
Some of the other considerations which we would like to address in future publications include :
\begin{enumerate}
    \item The groupoid systems developed in this paper are lattice dependent starting with the definition of the vertex and face operators that govern the dynamics of the model. This carries over to the model analyzed in Sec. \ref{sec:hamiltonian1}, especially the GSD computation of Sec. \ref{subsec:gsd} which is sensitive to the valency of the vertices in the lattice as is evident from the value of $\Tr \prod\limits_{\textrm{qubit spaces}}~1_{2\times 2}$ or the total dimension of the qubit Hilbert space. To check how the model changes with the underlying lattice we make a few observations about the properties of the model defined on a {\it hexagonal} lattice. To begin with we find that the orthogonal set of face operators for a given face is independent of its's shape and just depends on the number of morphisms included in the structure of the groupoid. For example in the $\mathcal{S}^2_1$ case, we have precisely five face operators in this orthogonal set, namely $B_f^{x_{11}}$, $B_f^{x_{22}}$, $B_f^{x_{12}}$, $B_f^{x_{21}}$ and $B_f^0$. This number becomes $n^2+1$ when we generalize to the $\mathcal{S}^n_1$ groupoid. To define each of these face operators we require the projectors to the matched and mismatched configurations at each corner of the face. For the hexagon there are five such projectors. This follows from the fact that there are six corners in a hexagonal face, denote them as the $W$, $NW$, $NE$, $E$, $SE$ and the $SW$ corners. If we measure holonomies across this face starting from the $SW$ corner and move clockwise, then we are left with exactly five corners.    
    As an example the projectors on the $W$ corner are,
    \begin{align*}
	F_{f}^{W, \pm}=\frac{1}{2}\left[1\pm \tikz[baseline=5ex, scale=2]{
		\draw[thick] (0.29, 0) -- (0.87, 0) -- (1.15, 0.5) -- (0.87, 1) -- (0.29, 1) -- (0, 0.5) -- (0.29, 0);
		\node at (0.57, 0.5) {$f$};
		\filldraw[red] (0.1, 0.33) circle (1.5pt) node[anchor = north east, black] {$\scriptstyle Z$};
		\filldraw[red] (0.1, 0.67) circle (1.5pt) node[anchor = south east, black] {$\scriptstyle Z$};
	}\right]
    \end{align*}
    with similar expressions for the other corners. Using these the face projectors to the holonomies can be written down. For example we have 
    \begin{align*}
	B_{f}^{a}=F_{f}^{W, +}F_{f}^{NW, +}F_{f}^{NE, +}F_{f}^{E, +}F_{f}^{SE, +}\frac{1}{2}\left[1+ \tikz[baseline=3.5ex, scale=1.5]{
		\draw[thick] (0.29, 0) -- (0.87, 0) -- (1.15, 0.5) -- (0.87, 1) -- (0.29, 1) -- (0, 0.5) -- (0.29, 0);
		\node at (0.57, 0.5) {$f$};
		\filldraw[red] (0.2, 0.15) circle (1.5pt) node[anchor = north east, black] {$\scriptstyle Z$};
	}\right]
	\frac{1}{2}\left[1- \tikz[baseline=3.5ex, scale=1.5]{
		\draw[thick] (0.29, 0) -- (0.87, 0) -- (1.15, 0.5) -- (0.87, 1) -- (0.29, 1) -- (0, 0.5) -- (0.29, 0);
		\node at (0.57, 0.5) {$f$};
		\filldraw[red] (0.45, 0) circle (1.5pt) node[anchor = north, black] {$\scriptstyle Z$};
	}\right]
    \end{align*}
    and similar expressions for the face projectors to the other holonomies. The projector to the zero holonomy is obtained when there is a mismatch in at least one of the five corners of the hexagon. Thus $B_f^0$ is a sum of terms, similar to Eq. \ref{eq:bf0}, counting all these mismatches, which can be either one, two, three, four or five. Having exhausted the set of orthogonal operators for a given face we can write down a face operator that mimics the one in Eq. \ref{eq:bfmodel1} for the square lattice,
    \begin{align*}
	B_{f}=\frac{1}{2}\left[1+\tikz[baseline=-0.5ex, scale=0.9]{
		\draw[thick] ({cos(0)}, {sin(0)}) -- ({cos(60)}, {sin(60)}) -- ({cos(120)}, {sin(120)}) -- ({cos(180)}, {sin(180)}) -- ({cos(240)}, {sin(240)}) -- ({cos(300)}, {sin(300)}) -- ({cos(0)}, {sin(0)});
		\foreach \x in {-75, -45, -15, 15, 45, 75, 105, 135, 165, 195}{
			\filldraw[red] ({0.89*cos(\x)}, {0.89*sin(\x)}) circle (2pt);
			\node at ({1.1*cos(\x)}, {1.1*sin(\x)}) {$\scriptstyle Z$};
		}
	}\right]
    \end{align*}
    These considerations show that such an operator can be written down for any polygon used to discretize the two dimensional surface. Now we move on to the orthogonal set of vertex operators for an arbitrary lattice. Consider a vertex of valency $k$. This vertex has $k$ corners and a configuration on this vertex can be fixed by specifying if there is a match or a mismatch on $k-1$ of these corners. Thus for the $\mathcal{S}^2_1$ toric code we will obtain $2^{k-1}$ orthogonal vertex operators analogous to the ones in Eq. \ref{eq:vertexopsqubitversionvalence4}. In total we will have $2^k$ \footnote{This number generalizes to $n^k$ for the $\mathcal{S}^n_1$ toric code.} orthogonal vertex operators for a vertex of valency $k$. For example in the hexagonal lattice we have four orthogonal vertex operators analogous to the ones in Eq. \ref{eq:vertexopsqubitversionvalence4}. For example,
    \begin{align*}
	A_{v}^{(0)} = \frac{1}{2}\left[1+\tikz[baseline=-0.5ex, scale=0.7]{
		\foreach \x in {60, 180, 300}{
			\draw[thick] (0, 0) -- ({cos(\x)}, {sin(\x)});
			\filldraw[red] ({0.5*cos(\x)}, {0.5*sin(\x)}) circle (2.5pt);
			\node at ({0.5*cos(\x + 40)}, {0.5*sin(\x + 40)}) {$\scriptstyle X$};
		}
	}\right]
	\frac{1}{2}\left[1+\tikz[baseline=-0.5ex, scale=0.7]{
		\foreach \x in {60, 180, 300}{
			\draw[thick] (0, 0) -- ({cos(\x)}, {sin(\x)});
			\filldraw[red] ({0.5*cos(\x)}, {0.5*sin(\x)}) circle (2.5pt);
		}
	\node at ({0.5*cos(60 + 40)}, {0.5*sin(60 + 40)}) {$\scriptstyle Z$};
	\node at ({0.5*cos(180 + 40)}, {0.5*sin(180 + 40)}) {$\scriptstyle Z$};
	\node at ({0.5*cos(300 + 40)}, {0.5*sin(300 + 40)}) {$\scriptstyle $};
	}\right]
	\frac{1}{2}\left[1+\tikz[baseline=-0.5ex, scale=0.7]{
	\foreach \x in {60, 180, 300}{
		\draw[thick] (0, 0) -- ({cos(\x)}, {sin(\x)});
		\filldraw[red] ({0.5*cos(\x)}, {0.5*sin(\x)}) circle (2.5pt);
	}
	\node at ({0.5*cos(300 + 40)}, {0.5*sin(300 + 40)}) {$\scriptstyle Z$};
	\node at ({0.5*cos(180 + 40)}, {0.5*sin(180 + 40)}) {$\scriptstyle Z$};
	}\right]
    \end{align*}
    and similarly for the other orthogonal vertex operators. This helps us write down the vertex operator analogous to Eq. \ref{eq:avmodel1} as,
    \begin{align*}
	A_{v} = \frac{1}{2}\left[1+\tikz[baseline=-0.5ex, scale=0.7]{
		\foreach \x in {60, 180, 300}{
			\draw[thick] (0, 0) -- ({cos(\x)}, {sin(\x)});
			\filldraw[red] ({0.5*cos(\x)}, {0.5*sin(\x)}) circle (2.5pt);
			\node at ({0.5*cos(\x + 40)}, {0.5*sin(\x + 40)}) {$\scriptstyle X$};
		}
	}\right]
	\frac{1}{2}\left[1+\tikz[baseline=-0.5ex, scale=0.7]{
		\foreach \x in {60, 180, 300}{
			\draw[thick] (0, 0) -- ({cos(\x)}, {sin(\x)});
			\filldraw[red] ({0.5*cos(\x)}, {0.5*sin(\x)}) circle (2.5pt);
		}
	\node at ({0.5*cos(60 + 40)}, {0.5*sin(60 + 40)}) {$\scriptstyle Z$};
	\node at ({0.5*cos(300 + 40)}, {0.5*sin(300 + 40)}) {$\scriptstyle Z$};
	}\right]
    \end{align*}
    We define these vertex operators for the vertices in the $SW$ and $NE$ corners of every face. Employing the techniques of Sec. \ref{subsec:gsd} we find that the Hamiltonian built out of these operators has a GSD of $2\times 2^{N_f}$. This degeneracy no longer has the interpretation of loop operators and thus requires a separate analysis. This example shows us that the $\mathcal{S}^n_1$ groupoid models considered in this paper is certainly sensitive to lattice details requiring a more careful study. In particular it would also be interesting to find examples of groupoid codes that are entirely topological.
    
    \item For a given groupoid $G_{\mathcal{C}}$ the set of morphisms whose source and target coincide, say $X\in Ob(\mathcal{C})$,\footnote{See Appendix \ref{app:groupoidreview} for a review of groupoids as categories.} form a group known as the {\it isotropy group} at $X$. In the examples considered in this paper, namely the $\mathcal{S}^n_1$ groupoids, the isotropy group of each of the $n$ objects is the trivial group made of just the identity element. This can be made non-trivial, for example with a groupoid that has two objects and two morphisms between them.
    
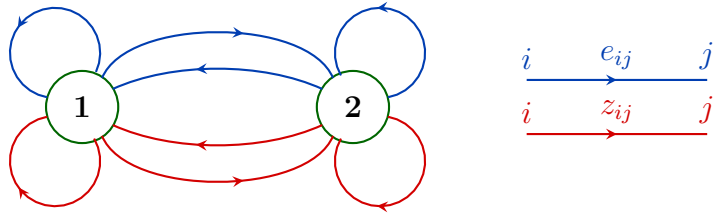
\begin{figure}[h!]
	\centering
	\tikz[scale = 1.2, baseline = 2.5ex, decoration={
		markings,
		mark=at position 0.6 with {\arrow{>}}}]{
		\draw [thick, black!60!green] (0.4,0)  arc (360:0:0.4);
		\node at (0, 0) {$\mathbf{1}$};
		\draw [thick, black!60!green] (3.4,0)  arc (360:0:0.4);
		\node at (3, 0) {$\mathbf{2}$};
		\draw [thick,blue,domain=-25:260, postaction={decorate}] plot ({-0.3+0.5*cos(\x)}, {0.6+0.5*sin(\x)});
		\draw [thick,red,domain=390:100, postaction={decorate}] plot ({-0.3+0.5*cos(\x)}, {-0.6+0.5*sin(\x)});
		\draw [thick,blue,domain=-80:210, postaction={decorate}] plot ({3.3+0.5*cos(\x)}, {0.6+0.5*sin(\x)});
		\draw [thick,red,domain=440:150, postaction={decorate}] plot ({3.3+0.5*cos(\x)}, {-0.6+0.5*sin(\x)});
		\draw [thick, blue, postaction={decorate}] (0.22,0.32) .. controls (0.5,1) and (2.5,1) .. (2.78,0.32);
		\draw [thick, blue, postaction={decorate}] (2.66,0.2) .. controls (2,0.5) and (1,0.5) .. (0.34, 0.2);
		\draw [thick, red, postaction={decorate}] (0.22, -0.32) .. controls (0.5, -1) and (2.5, -1) .. (2.78, -0.32);
		\draw [thick, red, postaction={decorate}] (2.66, -0.2) .. controls (2, -0.5) and (1, -0.5) .. (0.34, -0.2);
	}\tikz[scale = 1.2, baseline = 2.5ex, decoration={
	markings,
	mark=at position 0.5 with {\arrow{>}}}]{
		\node at (-1, 0) {};
		\def\a{0}
		\def\b{0.3}
		\draw[thick, blue, postaction={decorate}] (\a, \b) -- (\a+2, \b);
		\node[anchor=south, blue] at (\a, \b) {$i$};
		\node[anchor=south, blue] at (\a+1, \b) {$e_{ij}$};
		\node[anchor=south, blue] at (\a+2, \b) {$j$};
		\def\a{0}
		\def\b{-0.3}
		\draw[thick, red, postaction={decorate}] (\a, \b) -- (\a+2, \b);
		\node[anchor=south, red] at (\a, \b) {$i$};
		\node[anchor=south, red] at (\a+1, \b) {$z_{ij}$};
		\node[anchor=south, red] at (\a+2, \b) {$j$};
	}
\end{figure}

    The morphisms satisfy the relations,
    \begin{equation}
        e_{i_1,j_1}e_{i_2,j_2} = \delta_{j_1,i_2}e_{i_1,j_2},~e_{i_1,j_1}z_{i_2,j_2} = z_{i_1,j_1}e_{i_2,j_2}= \delta_{j_1,i_2}z_{i_1,j_2},~z_{i_1,j_1}z_{i_2,j_2} = \delta_{j_1,i_2}e_{i_1,j_2}.
    \end{equation}
    This makes the isotropy group at objects 1 and 2, $\mathbb{Z}_2$ \cite{https://doi.org/10.48550/arxiv.math/9602220, Brown1987FromGT, Higgins1971NotesOC}. It would be interesting to construct and study the features of toric code-like Hamiltonians for groupoids where the isotropy group of each object is non-trivial. In this case we can continue to use the mapping to the qubit space (space of object indices) used to simplify the groupoid model studied in this paper, with the addition of a {\it color} index to the object indices. These models continue to exhibit fracton-like features and we shall elaborate on them in an accompanying paper.
    
    Groups can be thought of as a special case of groupoids where there is just a single object and all the morphisms are precisely the group elements. This makes the Kitaev model built using the above groupoid with the isotropy group, $\mathbb{Z}_2$ reduce to the original $\mathbb{Z}_2$ toric code.

    \item The ground states of the group toric code models are known to obtain a topological entanglement entropy correction to the area law \cite{Kitaev2006TopologicalEE}. A similar computation for the ground states of the groupoid toric code can shed more light on the nature of the quantum phase they describe. It would be interesting to compare them to entanglement entropies of fracton models \cite{Ma2017TopologicalEE} and higher gauge models \cite{IbietaJimenez2020TopologicalEE}. 
   
    \item The Kitaev quantum double models based on finite groups are obtained from continuum gauge group using the Higgs field \cite{Propitius1995DiscreteGT}. It would be interesting to consider a similar mechanism to obtain the finite groupoid algebras considered here from a continuum gauge theory based on Lie groupoids. Along these lines it is also of interest to consider the effective low energy theories of the groupoid models studied here.  
    
    \item The rewriting of the $\mathcal{S}^2_1$-toric code in terms of qubits arranged around the vertex opens the possibility of using this model as a quantum error correcting code on the square-octagon lattice. However this model does not entirely sit in the framework of the stabilizer formalism as the vertex operators do not generate the Pauli group but instead the Pauli group algebra. The same is not true of the face operators which can be thought of as stabilizers generating a subgroup of the Pauli group. Regardless of this technical detail it would be interesting to check the performance of this code and weigh it against the likes of the color code and the surface code on the square-octagon lattice \cite{Bombin2006TopologicalQD, Fowler2012SurfaceCT}. An important reason for this is that the groupoid codes can hold more logical qubits than the color codes on the same Hilbert space and the logical Pauli operators can have an arbitrarily large distance depending on the size of the lattice making the chance of logical errors very low in such codes. Constructing the non-CSS versions of this code along the lines of \cite{Padmanabhan2021NonCSSCC} could further reduce the logical error rates in this code.
    
\end{enumerate}

\section*{Acknowledgemements}
We thank Paulo Teotonio Sobrinho, Ayan Mukhopadhyay and Fumihiko Sugino for many useful comments. IJ is partially supported by DST's INSPIRE Faculty Fellowship through the grant DST/INSPIRE/04/2019/000015. 

\appendix
\section{Review on groupoids}
\label{app:groupoidreview}
A {\it category}, $\mathcal{C}$ is an algebraic structure consisting of {\it objects} and {\it morphisms}. The objects, denoted by $A$, $B$, $\cdots$, belong to a set $Ob(\mathcal{C})$ and the morphisms are arrows between objects belonging to the set $\rm{Hom}(A, B)$. While the objects are fixed the morphisms obey certain axioms,
\begin{enumerate}
    \item For any objects $A$, $B$, $C$ there is a composition rule\footnote{Note that the convention used in the definition of the composition of morphisms in this appendix is opposite to the convention used for the examples in the main text.},
    \begin{eqnarray}
        \circ : \rm{Hom}(A,B) \times \rm{Hom}(B,C) & \rightarrow & \rm{Hom}(A,C) \nonumber \\
        (f,g) &\rightarrow & g\circ f \nonumber
    \end{eqnarray}  
    This implies the existence of a morphism $g\circ f\in\rm{Hom}(A,C)$ when there are morphisms $f\in\rm{Hom}(A,B)$ and $g\in\rm{Hom}(B,C)$.
    \begin{figure}[h!]
	\centering
	\tikz[scale = 1.2, baseline = 2.5ex, decoration={
		markings,
		mark=at position 0.5 with {\arrow{>}}}]{
		\draw [thick, black!60!green] (0.4,0)  arc (360:0:0.4);
		\node at (0, 0) {$\mathbf{A}$};
		\draw [thick, black!60!green] (3.4,0)  arc (360:0:0.4);
		\node at (3, 0) {$\mathbf{B}$};
		\draw [thick, black!60!green] (6.4,0)  arc (360:0:0.4);
		\node at (6, 0) {$\mathbf{C}$};
		\draw [thick, postaction={decorate}] (0.34, 0.2) .. controls (1,0.5) and (2,0.5) .. (2.66,0.2);
		\node at (1.5, 0.7) {$f$};
		\draw [thick, postaction={decorate}] (3.34, 0.2) .. controls (4,0.5) and (5,0.5) .. (5.66,0.2);
		\node at (4.5, 0.7) {$g$};
		\draw [thick, postaction={decorate}] (0.22, -0.32) .. controls (2, -1.5) and (4, -1.5) .. (5.78, -0.32);
		\node at (3, -1.6) {$g\circ f$};
	}
\end{figure}
    \item This composition is associative which implies that for three morphisms, $f$, $g$, $h$, if $g\circ f$ and $h\circ g$ exist, then $h\circ g\circ f := h\circ(g\circ f) = (h\circ g)\circ f$.
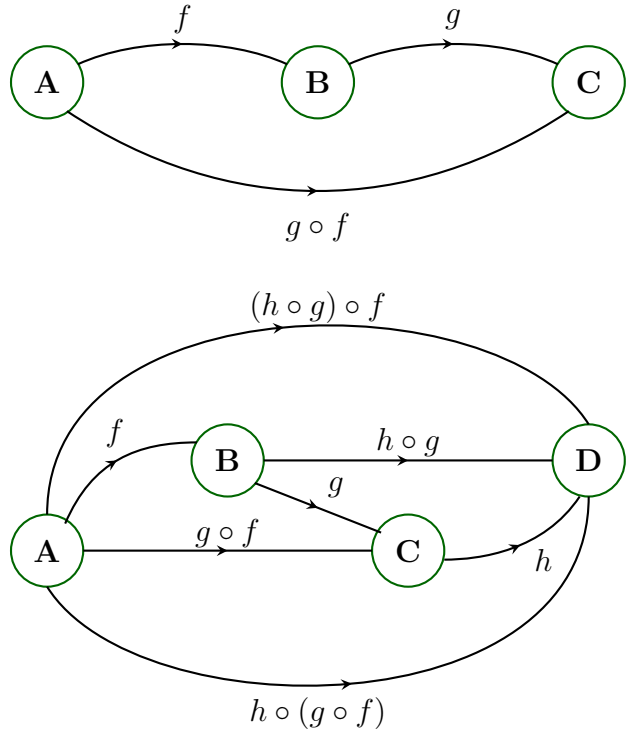
    \begin{figure}[h!]
	\centering
	\tikz[scale = 1.2, baseline = 2.5ex, decoration={
		markings,
		mark=at position 0.5 with {\arrow{>}}}]{
		\draw [thick, black!60!green] (0.4,0)  arc (360:0:0.4);
		\node at (0, 0) {$\mathbf{A}$};
		\draw [thick, black!60!green] (2.4,1)  arc (360:0:0.4);
		\node at (2, 1) {$\mathbf{B}$};
		\draw [thick, black!60!green] (4.4,0)  arc (360:0:0.4);
		\node at (4, 0) {$\mathbf{C}$};
		\draw [thick, black!60!green] (6.4,1)  arc (360:0:0.4);
		\node at (6, 1) {$\mathbf{D}$};
		\draw [thick, postaction={decorate}] (0.2, 0.3) .. controls (0.5, 1) and (1,1.2) .. (1.66, 1.2);
		\node at (0.75, 1.3) {$f$};
		\draw [thick, postaction={decorate}] (0.4, 0) -- (3.6, 0);
		\node at (2, 0.2) {$g\circ f$};
		\draw [thick, postaction={decorate}] (4.4, -0.1) .. controls (4.8, -0.1) and (5.5, 0) .. (5.9, 0.6);
		\node at (5.5, -0.1) {$h$};
		\draw [thick, postaction={decorate}] (2.3, 0.75) -- (3.7, 0.2);
		\node at (3.2, 0.7) {$g$};
		\draw [thick, postaction={decorate}] (2.4, 1) -- (5.6, 1);
		\node at (4, 1.2) {$h\circ g$};
		\draw [thick, postaction={decorate}] (0, -0.4) .. controls (1, -2) and (6, -2) .. (6, 0.6);
		\node at (3, -1.8) {$h\circ (g\circ f)$};
		\draw [thick, postaction={decorate}] (0, 0.4) .. controls (0, 3) and (5, 3) .. (6, 1.4);
		\node at (3, 2.7) {$(h\circ g)\circ f$};
	}
    \end{figure}

    \item For any object $X\in OB(\mathcal{C})$ there exists an identity morphism, $1_X\in\rm{Hom}(X, X)$ such that for $f\in\rm{Hom}(A, B)$, $1_B\circ f=f=f\circ 1_A$.  
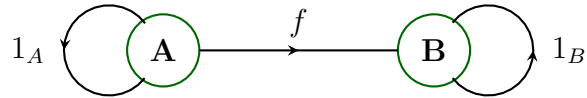
    \begin{figure}[h!]
	\centering
	\tikz[scale = 1.2, baseline = 2.5ex, decoration={
		markings,
		mark=at position 0.5 with {\arrow{>}}}]{
		\draw [thick, black!60!green] (0.4,0)  arc (360:0:0.4);
		\node at (0, 0) {$\mathbf{A}$};
		\draw [thick, black!60!green] (3.4,0)  arc (360:0:0.4);
		\node at (3, 0) {$\mathbf{B}$};
		\draw [thick,domain=37:322, postaction={decorate}] plot ({-0.6+0.5*cos(\x)}, {0.5*sin(\x)});
		\node at (-1.5, 0) {$1_{A}$};
		\draw [thick,domain=-142:142, postaction={decorate}] plot ({3.6+0.5*cos(\x)}, {0.5*sin(\x)});
		\node at (4.5, 0) {$1_{B}$};
		\draw [thick, postaction={decorate}] (0.4, 0) -- (2.6, 0);
		\node at (1.5, 0.3) {$f$};
	}
    \end{figure}
\end{enumerate}
For every morphism $f\in\rm{Hom}(A,B)$ we define the {\it source} $s$ and {\it target} $t$ maps as $s(f)=A$ and $t(f)=B$, which essentially mean that the source of a morphism is the tail of the arrow while the target is the head of the arrow. Not all morphisms can be composed and this is formalized with the ideas of the sources and targets of morphisms, as $f\circ g$ only makes sense if $t(g)=s(f)$.

\subsection*{Groupoids}
We can now define the groupoid, $G_{\mathcal{C}}$ as a category where all morphisms are invertible. The composition rule in the category can be taken as the product of elements in the groupoid, $fg$ is defined if $t(g)=s(f)$. For each morphism we have a left and a right identity that satisfy
\begin{equation}\nonumber
    1_{s(g)}g=g=g1_{t(g)}.
\end{equation}
As each morphism has an inverse, there exists $g^{-1}\in G_{\mathcal{C}}$ for every $g\in G_{\mathcal{C}}$ such that
\begin{equation}\nonumber
    gg^{-1} = 1_{t(g)}~~g^{-1}g = 1_{s(g)}.
\end{equation}
Thus the groupoid structure is quite similar to that of a group barring the facts that not all elements of a groupoid can be multiplied with each other and there is no unique identity element in the groupoid.

\subsection*{Groupoid algebra $\mathbb{C}G_{\mathcal{C}}$}
The groupoid $G_{\mathcal{C}}$ can be provided with a vector space or linear structure to convert it to a groupoid algebra, $\mathbb{C}G_{\mathcal{C}}$ in a manner similar to the construction of the group algebra. The set $\{\phi_g : g\in G_{\mathcal{C}}\}$ form the basis elements with the dual basis given by $\{\Psi^g : g\in G_{\mathcal{C}}\}$ such that $\Psi^g(\phi_h) = \delta(g,h)$.
The product map in this algebra is given by the map $m:\mathbb{C}G_{\mathcal{C}}\otimes \mathbb{C}G_{\mathcal{C}}\rightarrow \mathbb{C}G_{\mathcal{C}}$, as 
\begin{equation}\nonumber 
    m(\phi_g\otimes \phi_h) := \phi_{gh},
\end{equation}
if $t(h)=s(g)$ or 0 otherwise. This algebra has a unique unit $\eta$ given by
\begin{equation}\nonumber 
    \eta = \sum\limits_{X\in Ob(\mathcal{C}}~\phi_{1_X}.
\end{equation}
The dual algebra structure is the same as in the group algebra case, namely the co-product is given by the map, $\Delta : \mathbb{C}G_{\mathcal{C}}^*\otimes \mathbb{C}G_{\mathcal{C}}^*\rightarrow \mathbb{C}G_{\mathcal{C}}^*$,
\begin{equation}\nonumber
    \Delta\left(\Psi^g\otimes\Psi\right) := \Psi^g*\Psi^h = \delta(g,h)\Psi^g,
\end{equation}
and the co-unit $\epsilon$ is given by 
\begin{equation}\nonumber 
    \epsilon = \sum\limits_{g\in G_{\mathcal{C}}}~\Psi^g.
\end{equation}
The antipode map $S:\mathbb{C}G_{\mathcal{C}}\rightarrow \mathbb{C}G_{\mathcal{C}}$ is defined as
\begin{equation}\nonumber 
    S(\phi_g) = \phi_{g^{-1}}.
\end{equation}
The set $\langle m, \eta, \Delta, \epsilon, S\rangle$ defines the groupoid algebra as an example of a {\it weak Hopf algebra}.  

\section{General groupoid toric code}
\label{app:gengroupoid}
Kitaev's quantum double model can be defined for an arbitrary groupoid. We will show this by proving that $\left[A_v, B_f\right]=0$ for a general groupoid. We will work on the square lattice by choice, the proof carries over for arbitrary lattices as well. We begin with the general definition of the vertex and the face operator on the square lattice.
\begin{align}\label{eq:faceopgengroupoid}
	&B_{f}^{h}\;\tikz[scale = 2, baseline = 4ex, decoration={
		markings,
		mark=at position 0.5 with {\arrow{>}}}]{
		\draw[postaction={decorate}, thick] (0,0) -- (1, 0);
		\draw[postaction={decorate}, thick] (1,0) -- (1, 1);
		\draw[postaction={decorate}, thick] (0,0) -- (0, 1);
		\draw[postaction={decorate}, thick] (0,1) -- (1, 1);
		\node at (0.5, 0.5) {$f$};
		\node at (-0.2, 0.5) {$g_{1}$};
		\node at (0.5, 1.2) {$g_{2}$};
		\node at (1.2, 0.5) {$g_{3}$};
		\node at (0.5, -0.2) {$g_{4}$};
		\draw [thick,domain=180:-90, ->] plot ({0.5+0.3*cos(\x)}, {0.5+0.3*sin(\x)});
	}\nonumber\\
	=&\delta(g_{1}g_{2}g_{3}^{-1}g_{4}^{-1}, h)\delta(t(g_{1}), s(g_{2}))\delta(t(g_{2}), s(g_{3}^{-1}))\nonumber\\
	&\times\delta(t(g_{3}^{-1}), s(g_{4}^{-1}))\delta(s(g_{1}), s(h))\delta(t(g_{4}^{-1}), t(h))
	\tikz[scale = 2, baseline = 4ex, decoration={
		markings,
		mark=at position 0.5 with {\arrow{>}}}]{
		\draw[postaction={decorate}, thick] (0,0) -- (1, 0);
		\draw[postaction={decorate}, thick] (1,0) -- (1, 1);
		\draw[postaction={decorate}, thick] (0,0) -- (0, 1);
		\draw[postaction={decorate}, thick] (0,1) -- (1, 1);
		\node at (0.5, 0.5) {$f$};
		\node at (-0.2, 0.5) {$g_{1}$};
		\node at (0.5, 1.2) {$g_{2}$};
		\node at (1.2, 0.5) {$g_{3}$};
		\node at (0.5, -0.2) {$g_{4}$};
		\draw [thick,domain=180:-90, ->] plot ({0.5+0.3*cos(\x)}, {0.5+0.3*sin(\x)});
	}
\end{align}

\begin{align}\label{eq:vertexopgengroupoid}
	&A_{v}^{g}\;\tikz[baseline=-0.5ex, scale=1.5, decoration={
		markings,
		mark=at position 0.5 with {\arrow{>}}}]{
		\draw[thick, postaction={decorate}] (-1, 0) -- (0, 0);
		\node[anchor=north] at (-0.5, 0) {$g_{1}$};
		\draw[thick, postaction={decorate}] (0, 0) -- (1, 0);
		\node[anchor=south] at (0.5, 0) {$g_{3}$};
		\draw[thick, postaction={decorate}] (0, -1) -- (0, 0);
		\node[anchor=west] at (0, -0.5) {$g_{2}$};
		\draw[thick, postaction={decorate}] (0, 0) -- (0, 1);
		\node[anchor=east] at (0, 0.5) {$g_{4}$};
		\node[anchor=north west] {$v$};
	}\nonumber\\
	=&\delta(t(g_{1}), s(g^{-1}))\delta(t(g_{2}), s(g^{-1}))\nonumber\\
	&\times \delta(t(g), s(g_{3}))\delta(t(g), s(g_{4}))\;
	\tikz[baseline=-0.5ex, scale=1.5, decoration={
		markings,
		mark=at position 0.5 with {\arrow{>}}}]{
		\draw[thick, postaction={decorate}] (-1, 0) -- (0, 0);
		\node[anchor=north] at (-0.5, 0) {$g_{1}g^{-1}$};
		\draw[thick, postaction={decorate}] (0, 0) -- (1, 0);
		\node[anchor=south] at (0.5, 0) {$gg_{3}$};
		\draw[thick, postaction={decorate}] (0, -1) -- (0, 0);
		\node[anchor=west] at (0, -0.5) {$g_{2}g^{-1}$};
		\draw[thick, postaction={decorate}] (0, 0) -- (0, 1);
		\node[anchor=east] at (0, 0.5) {$gg_{4}$};
		\node[anchor=north west] {$v$};
	}
\end{align}

The right hand side of Eq. \ref{eq:vertexopgengroupoid} can be further simplified by noting that $s(g^{-1}) = t(g)$. Thus the $\delta$ expressions in Eq. \ref{eq:vertexopgengroupoid} simplify to $\delta(t(g_1),t(g_2))\delta(s(g_3),s(g_4))\delta(t(g_1),s(g_3))$. When working in the space of object indices this would imply that all the indices around the vertex would have to be the same for a non-zero action of the vertex operator. We will use this simplified action of the vertex operator while proving the commutation relations between the vertex and face operators.

We will need to prove that $\left[A_v, B_f\right]=0$ for each corner of the square face. This distinction matters especially for the SW corner of the face or the NE corner of the vertex as the holonomy across the face is measured starting from  vertex at the SW corner. 

\subsection*{$\left[A_v, B_f\right]=0$ for $v$ at the NW(SE) face(vertex) corner - }
We will use the state configuration
\begin{align}\label{eq:genstateforavbf0proofNW}
	|g_{1}, g_{2}, g_{3}, g_{4}\rangle = \tikz[scale = 2, baseline = 4ex, decoration={
		markings,
		mark=at position 0.5 with {\arrow{>}}}]{
		\node[anchor=south east] at (0, 1) {$v$};
		\draw[postaction={decorate}, thick] (0,0) -- (1, 0);
		\draw[postaction={decorate}, thick] (1,0) -- (1, 1);
		\draw[postaction={decorate}, thick] (0,0) -- (0, 1);
		\draw[postaction={decorate}, thick] (0,1) -- (1, 1);
		\node at (0.5, 0.5) {$f$};
		\node at (-0.2, 0.5) {$g_{1}$};
		\node at (0.5, 1.2) {$g_{2}$};
		\node at (1.2, 0.5) {$g_{3}$};
		\node at (0.5, -0.2) {$g_{4}$};
		\draw [thick,domain=180:-90, ->] plot ({0.5+0.3*cos(\x)}, {0.5+0.3*sin(\x)});
	}
\end{align}
Note that we have omitted the other two edges at the NW vertex as they are not shared with the face, $f$. 
We first evaluate $A_v^gB_f^h$, for arbitrary morphisms, $g$, $h$, on this state, 
\begin{eqnarray}
    A_v^gB_f^h~\ket{g_1,g_2,g_3,g_4} & = & \delta\left(g_1g_2g_3^{-1}g_4^{-1}, h\right)\delta\left(t(g_1),s(g_2)\right)\delta\left(t(g_2),s(g_3^{-1})\right)\delta\left(t(g_3^{-1}),s(g_4^{-1})\right) \nonumber \\
    & \times & \delta\left(s(g_1),s(h)\right)\delta\left(t(g_4^{-1}),t(h)\right)~A_v^g~\ket{g_1,g_2,g_3,g_4} \nonumber \\
    & = & \delta\left(g_1g_2g_3^{-1}g_4^{-1}, h\right)\delta\left(t(g_1),s(g_2)\right)\delta\left(t(g_2),s(g_3^{-1})\right)\delta\left(t(g_3^{-1}),s(g_4^{-1})\right) \nonumber \\
    & \times & \delta\left(s(g_1),s(h)\right)\delta\left(t(g_4^{-1}),t(h)\right)~\delta\left(t(g_1),s(g_2)\right)\ket{g_1g^{-1},gg_2,g_3,g_4} \nonumber \\
    & = & \delta\left(g_1g_2g_3^{-1}g_4^{-1}, h\right)\delta\left(t(g_1),s(g_2)\right)\delta\left(t(g_2),s(g_3^{-1})\right)\delta\left(t(g_3^{-1}),s(g_4^{-1})\right) \nonumber \\
    & \times & \delta\left(s(g_1),s(h)\right)\delta\left(t(g_4^{-1}),t(h)\right)~\ket{g_1g^{-1},gg_2,g_3,g_4}. 
\end{eqnarray}
Now evaluating $B_f^hA_v^g$ on the state in Eq. \ref{eq:genstateforavbf0proofNW} we obtain,
\begin{eqnarray}
    B_f^hA_v^g~\ket{g_1,g_2,g_3,g_4} & = & \delta\left(t(g_1),s(g_2)\right)B_f^h~\ket{g_1g^{-1},gg_2,g_3,g_4} \nonumber \\
    & = & \delta\left(g_1g^{-1}gg_2g_3^{-1}g_4^{-1}, h\right)\delta\left(t(g_1g^{-1}),s(gg_2)\right)\delta\left(t(gg_2),s(g_3^{-1})\right)\delta\left(t(g_3^{-1}),s(g_4^{-1})\right) \nonumber \\
    & \times & \delta\left(s(g_1g^{1}),s(h)\right)\delta\left(t(g_4^{-1}),t(h)\right)~\delta\left(t(g_1),s(g_2)\right)~\ket{g_1g^{-1},gg_2,g_3,g_4} \nonumber \\
    & = & \delta\left(g_11_{s(g^{-1})}g_2g_3^{-1}g_4^{-1}, h\right)\delta\left(t(g^{-1}),s(g)\right)\delta\left(t(g_2),s(g_3^{-1})\right)\delta\left(t(g_3^{-1}),s(g_4^{-1})\right) \nonumber \\
    & \times & \delta\left(s(g_1),s(h)\right)\delta\left(t(g_4^{-1}),t(h)\right)~\delta\left(t(g_1),s(g_2)\right)~\ket{g_1g^{-1},gg_2,g_3,g_4} \nonumber \\
    & = & \delta\left(g_1g_2g_3^{-1}g_4^{-1}, h\right)\delta\left(t(g_1),s(g_2)\right)\delta\left(t(g_2),s(g_3^{-1})\right)\delta\left(t(g_3^{-1}),s(g_4^{-1})\right) \nonumber \\
    & \times & \delta\left(s(g_1),s(h)\right)\delta\left(t(g_4^{-1}),t(h)\right)~\ket{g_1g^{-1},gg_2,g_3,g_4}.
\end{eqnarray}
In this computation we have used the fact that $t(g)=s(g^{-1})$, $s(h_1h_2) = s(h_1)$ and $t(h_1h_2) = t(h_2)$. We have also used $t(g_1)=s(g^{-1})$ for non-zero action of the vertex operator, $A_v^{g}$ on the NW corner. Thus we find that 
\begin{equation}
    \left[A_v^g, B_f^h\right]=0,~~\forall~g,h\in\mathbb{C}G_{\mathcal{C}},
\end{equation}
and hence $\left[A_v, B_f\right]=0$ for the NW corner of the face. Note that the vertex operator, $A_v$ in this case commutes with the face operators measuring an arbitrary holonomy for this corner. Thus there are no restrictions on $h$ in this case. This can be verified in the examples considered in this paper, namely the case of the $\mathcal{S}^n_1$-toric codes.  

The above computations follow a similar pattern for the NE(SW) and SE(NW) corners of the face(vertex) operator and hence we omit those details here. We will proceed to the non-trivial case of the SW(NE) corner of the face(vertex) next. 

\subsection*{$\left[A_v, B_f\right]=0$ for $v$ at the SW(NE) face(vertex) corner - }
Once again we will evaluate on the state
\begin{align}\label{eq:genstateforavbf0proofSW}
	|g_{1}, g_{2}, g_{3}, g_{4}\rangle = \tikz[scale = 2, baseline = 4ex, decoration={
		markings,
		mark=at position 0.5 with {\arrow{>}}}]{
		\node[anchor=north east] at (0, 0) {$v$};
		\draw[postaction={decorate}, thick] (0,0) -- (1, 0);
		\draw[postaction={decorate}, thick] (1,0) -- (1, 1);
		\draw[postaction={decorate}, thick] (0,0) -- (0, 1);
		\draw[postaction={decorate}, thick] (0,1) -- (1, 1);
		\node at (0.5, 0.5) {$f$};
		\node at (-0.2, 0.5) {$g_{1}$};
		\node at (0.5, 1.2) {$g_{2}$};
		\node at (1.2, 0.5) {$g_{3}$};
		\node at (0.5, -0.2) {$g_{4}$};
		\draw [thick,domain=180:-90, ->] plot ({0.5+0.3*cos(\x)}, {0.5+0.3*sin(\x)});
	}
\end{align}
The action of $A_v^gB_f^h$ on the state in Eq. \ref{eq:genstateforavbf0proofSW} is 
\begin{eqnarray}
 A_v^gB_f^h~\ket{g_1,g_2,g_3,g_4} & = & \delta\left(g_1g_2g_3^{-1}g_4^{-1}, h\right)\delta\left(t(g_1),s(g_2)\right)\delta\left(t(g_2),s(g_3^{-1})\right)\delta\left(t(g_3^{-1}),s(g_4^{-1})\right) \nonumber \\
    & \times & \delta\left(s(g_1),s(h)\right)\delta\left(t(g_4^{-1}),t(h)\right)~A_v^g~\ket{g_1,g_2,g_3,g_4} \nonumber \\
    & = & \delta\left(g_1g_2g_3^{-1}g_4^{-1}, h\right)\delta\left(t(g_1),s(g_2)\right)\delta\left(t(g_2),s(g_3^{-1})\right)\delta\left(t(g_3^{-1}),s(g_4^{-1})\right) \nonumber \\
    & \times & \delta\left(s(g_1),s(h)\right)\delta\left(t(g_4^{-1}),t(h)\right)~\delta\left(s(g_1),s(g_4)\right)\ket{gg_1,g_2,g_3,gg_4} \nonumber \\
    & = & \delta\left(g_1g_2g_3^{-1}g_4^{-1}, h\right)\delta\left(t(g_1),s(g_2)\right)\delta\left(t(g_2),s(g_3^{-1})\right)\delta\left(t(g_3^{-1}),s(g_4^{-1})\right) \nonumber \\
    & \times & \delta\left(s(h),t(h)\right)~\ket{gg_1,g_2,g_3,gg_4},
    \end{eqnarray}
and the action of $B_f^hA_v^g$ on the state in Eq. \ref{eq:genstateforavbf0proofSW} is 
\begin{eqnarray}
  B_f^hA_v^g~\ket{g_1,g_2,g_3,g_4} & = &  \delta\left(s(g_1),s(g_4)\right)B_f^h~\ket{gg_1,g_2,g_3,gg_4} \nonumber \\
  & = & \delta\left(gg_1g_2g_3^{-1}g_4^{-1}g^{-1}, h\right)\delta\left(t(gg_1),s(g_2)\right)\delta\left(t(g_2),s(g_3^{-1})\right)\delta\left(t(g_3^{-1}),s(g_4^{-1}g)\right) \nonumber \\
  & \times & \delta\left(s(gg_1),s(h)\right)\delta\left(t(g_4^{-1}g^{-1}),t(h)\right)~\delta\left(s(g_1),s(g_4)\right)\ket{gg_1,g_2,g_3,gg_4} \nonumber \\
  & = & \delta\left(1_{t(g)}g_1g_2g_3^{-1}g_4^{-1}1_{t(g)}, g^{-1}hg\right)\delta\left(t(g_1),s(g_2)\right)\delta\left(t(g_2),s(g_3^{-1})\right) \nonumber \\
  & \times & \delta\left(t(g_3^{-1}),s(g_4^{-1})\right)\delta\left(s(g),s(h)\right)\delta\left(t(g^{-1}),t(h)\right) \nonumber \\
  & \times & \delta\left(s(g_1),t(g_4^{-1})\right)\ket{gg_1,g_2,g_3,gg_4} \nonumber \\
  & = & \delta\left(g_1g_2g_3^{-1}g_4^{-1}, h\right)\delta\left(t(g_1),s(g_2)\right)\delta\left(t(g_2),s(g_3^{-1})\right)\delta\left(t(g_3^{-1}),s(g_4^{-1})\right) \nonumber \\
  & \times & \delta\left(s(h),t(h)\right)~\ket{gg_1,g_2,g_3,gg_4}. \nonumber \\
  \end{eqnarray}
For the last equality to hold we require that $g^{-1}hg=h$ or $h$ has to be central in $\mathbb{C}G_{\mathcal{C}}$ for this condition to hold for an arbitrary $g$. We can verify this with the $\mathcal{S}^n_1$-toric codes considered here. It is easy to see that $\sum\limits_{j=1}^n~x_{jj}$ is central in the $\mathcal{S}^n_1$-groupoid algebra. Thus $\sum\limits_{j=1}^n~B_f^{x_{jj}}$ will commute with the vertex operators.

Thus we can define an exactly solvable model using $A_v=\sum\limits_{g\in G_{\mathcal{C}}}~A_v^g$ and the face operator $\sum\limits_{j=1}^n~B_f^{x_{jj}}$. For the $\mathcal{S}^2_1$-toric codes this would correspond to the Hamiltonian in Eq. \ref{eq:hmodel4}, the model with a unique ground state. To obtain more interesting models in the general groupoid case we will need to consider the complete set of orthogonal vertex operators. We leave this more general case to a future work.

\section{Models on manifolds with boundary}
\label{app:finitegdtc}
We will now briefly look at the GSD of the model in Eq. \ref{eq:hmodel1} on a manifold with smooth boundaries,
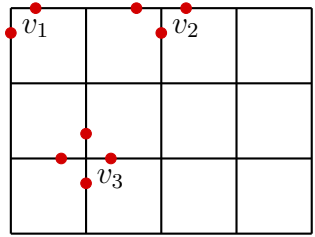
\begin{align*}
	\tikz[scale=1]{
		\foreach \a in {0, 1, 2, 3, 4}{
			\draw[thick] (\a, 0) -- (\a, 3);
		}
		\foreach \b in {0, 1, 2, 3}{
			\draw[thick] (0, \b) -- (4, \b);
		}
	\filldraw[red] (0, 2.67) circle (2pt);
	\filldraw[red] (0.33, 3) circle (2pt);
	\node[anchor=north west] at (0, 3) {$v_{1}$};
	\filldraw[red] (2, 2.67) circle (2pt);
	\filldraw[red] (1.67, 3) circle (2pt);
	\filldraw[red] (2.33, 3) circle (2pt);
	\node[anchor=north west] at (2, 3) {$v_{2}$};
	\filldraw[red] (1, 1.33) circle (2pt);
	\filldraw[red] (1, 0.67) circle (2pt);
	\filldraw[red] (0.67, 1) circle (2pt);
	\filldraw[red] (1.33, 1) circle (2pt);
	\node[anchor=north west] at (1, 1) {$v_{3}$};
	}
\end{align*}
\footnote{There are other choices for manifolds with boundary where we could have incomplete faces on any side of the lattice. We will not consider these possibilities here.} 
where $v_1$, $v_2$ and $v_3$ are examples of corner, boundary and bulk vertices respectively and this lattice is a 3 by 4 lattice or it has 12 faces. In general we will consider manifolds with smooth boundaries 
or a $m$ by $n$ lattice with $mn$ faces. Such lattices have a total of $(m+1)(n+1)$ vertices with 4 of them being corner vertices, $2(m+n)-4$ boundary vertices and $(m-1)(n-1)$ bulk vertices. The Hamiltonian for this model is written as 
\begin{equation}
    H = H_{\textrm{bulk}} + H_{\textrm{boundary}} + H_{\textrm{corner}},
\end{equation}
where 
\begin{eqnarray}
  H_{\textrm{bulk}} & = & -\sum\limits_{v=1}^{(m-1)(n-1)}~A_v -\sum\limits_{f=1}^{mn}~B_f, \nonumber \\
  H_{\textrm{boundary}} & = & -\sum\limits_{v=1}^{2(m+n)-4}~A_v^{\textrm{boundary}}, \nonumber \\
  H_{\textrm{corner}} & = & -\sum\limits_{v=1}^{4}~A_v^{\textrm{corner}}
\end{eqnarray}
with $A_v$ and $B_f$ given by eqs. \ref{eq:avmodel1} and \ref{eq:bfmodel1} respectively, and the vertex operators at the boundaries and corners have support on three and two qubit spaces respectively. There are four types of boundary vertex operators depending on whether they are acting on the top, bottom or left, right boundaries. However the choice of these operators is not unique and several choices lead to exactly solvable models. Here we make one such choice and analyze the consequences. Note that this model is not consistent with the bulk as if we `close' this surface by enforcing appropriate boundary conditions we will not obtain the vertex and face operators of Eqs. \ref{eq:avmodel1} and \ref{eq:bfmodel1}. Nevertheless we pick this model as it has interesting properties on the manifold with boundary.

We choose the four types of boundary operators as
\begin{align}
	A_{v}^{\text{boundary, top}} &= \frac{1}{2}\left[1+\tikz[baseline=-3.5ex]{
		\draw[thick] (-1, 0) -- (1, 0);
		\draw[thick] (0, 0) -- (0, -1);
		\filldraw[red] (-0.5, 0) circle (2pt) node[anchor=south, black] {$\scriptstyle X$};
		\filldraw[red] (0, -0.5) circle (2pt) node[anchor=west, black] {$\scriptstyle X$};
		\filldraw[red] (0.5, 0) circle (2pt) node[anchor=south, black] {$\scriptstyle X$};
	}\right]
	\frac{1}{2}\left[1-\tikz[baseline=-3.5ex]{
		\draw[thick] (-1, 0) -- (1, 0);
		\draw[thick] (0, 0) -- (0, -1);
		\filldraw[red] (-0.5, 0) circle (2pt) node[anchor=south, black] {$\scriptstyle Z$};
		\filldraw[red] (0, -0.5) circle (2pt) node[anchor=west, black] {$\scriptstyle $};
		\filldraw[red] (0.5, 0) circle (2pt) node[anchor=south, black] {$\scriptstyle Z$};
	}\right]\nonumber\\
	A_{v}^{\text{boundary, left}} &= \frac{1}{2}\left[1+\tikz[baseline=-0.5ex]{
		\draw[thick] (0, -1) -- (0, 1);
		\draw[thick] (0, 0) -- (1, 0);
		\filldraw[red] (0, -0.5) circle (2pt) node[anchor=east, black] {$\scriptstyle X$};
		\filldraw[red] (0.5, 0) circle (2pt) node[anchor=south, black] {$\scriptstyle X$};
		\filldraw[red] (0, 0.5) circle (2pt) node[anchor=east, black] {$\scriptstyle X$};
	}\right]
	\frac{1}{2}\left[1-\tikz[baseline=-0.5ex]{
		\draw[thick] (0, -1) -- (0, 1);
		\draw[thick] (0, 0) -- (1, 0);
		\filldraw[red] (0, -0.5) circle (2pt) node[anchor=east, black] {$\scriptstyle Z$};
		\filldraw[red] (0.5, 0) circle (2pt) node[anchor=south, black] {$\scriptstyle $};
		\filldraw[red] (0, 0.5) circle (2pt) node[anchor=east, black] {$\scriptstyle Z$};
	}\right]\nonumber\\
	A_{v}^{\text{boundary, bottom}} &= \frac{1}{2}\left[1+\tikz[baseline=2.5ex]{
	\draw[thick] (-1, 0) -- (1, 0);
	\draw[thick] (0, 0) -- (0, 1);
	\filldraw[red] (-0.5, 0) circle (2pt) node[anchor=north, black] {$\scriptstyle X$};
	\filldraw[red] (0, 0.5) circle (2pt) node[anchor=west, black] {$\scriptstyle X$};
	\filldraw[red] (0.5, 0) circle (2pt) node[anchor=north, black] {$\scriptstyle X$};
	}\right]
	\frac{1}{2}\left[1-\tikz[baseline=2.5ex]{
		\draw[thick] (-1, 0) -- (1, 0);
		\draw[thick] (0, 0) -- (0, 1);
		\filldraw[red] (-0.5, 0) circle (2pt) node[anchor=north, black] {$\scriptstyle Z$};
		\filldraw[red] (0, 0.5) circle (2pt) node[anchor=west, black] {$\scriptstyle $};
		\filldraw[red] (0.5, 0) circle (2pt) node[anchor=north, black] {$\scriptstyle Z$};
	}\right]\nonumber\\
	A_{v}^{\text{boundary, right}} &= \frac{1}{2}\left[1+\tikz[baseline=-0.5ex]{
	\draw[thick] (0, -1) -- (0, 1);
	\draw[thick] (0, 0) -- (-1, 0);
	\filldraw[red] (0, -0.5) circle (2pt) node[anchor=west, black] {$\scriptstyle X$};
	\filldraw[red] (-0.5, 0) circle (2pt) node[anchor=south, black] {$\scriptstyle X$};
	\filldraw[red] (0, 0.5) circle (2pt) node[anchor=west, black] {$\scriptstyle X$};
	}\right]
	\frac{1}{2}\left[1-\tikz[baseline=-0.5ex]{
		\draw[thick] (0, -1) -- (0, 1);
		\draw[thick] (0, 0) -- (-1, 0);
		\filldraw[red] (0, -0.5) circle (2pt) node[anchor=west, black] {$\scriptstyle Z$};
		\filldraw[red] (-0.5, 0) circle (2pt) node[anchor=south, black] {$\scriptstyle $};
		\filldraw[red] (0, 0.5) circle (2pt) node[anchor=west, black] {$\scriptstyle Z$};
	}\right]
\end{align}
The corner vertex operators take a similar form albeit the support on two qubit spaces. The operator on the $NW$ corner of the lattice is given by 
\begin{align}
	A_{v}^{NW} = \frac{1}{2} \left[1+\tikz[baseline=2.5ex]{
		\draw[thick] (0, 0) -- (0, 1) -- (1, 1);
		\filldraw[red] (0, 0.5) circle (2pt) node[anchor=east, black] {$\scriptstyle X$};
		\filldraw[red] (0.5, 1) circle (2pt) node[anchor=south, black] {$\scriptstyle X$};
	}\right]
	\frac{1}{2} \left[1-\tikz[baseline=2.5ex]{
		\draw[thick] (0, 0) -- (0, 1) -- (1, 1);
		\filldraw[red] (0, 0.5) circle (2pt) node[anchor=east, black] {$\scriptstyle Z$};
		\filldraw[red] (0.5, 1) circle (2pt) node[anchor=south, black] {$\scriptstyle Z$};
	}\right]
\end{align}
with similar expressions for the other three corners of the lattice. The GSD for this model can once again be computed using the techniques of Sec. \ref{subsec:gsd} and we find that to be
\begin{equation}
    GSD = 2^{mn-2},
\end{equation}
for a $m$ by $n$ lattice. Accounting for this degeneracy gets much harder and we will not attempt to do that here. Nevertheless we make some interesting observations to vindicate the choice of this model among others on the manifold with boundary. There are several symmetries supported on loops enclosing bulk vertices that partially account for the GSD. The interesting point is that none of these loops are non-contractible, in fact any symmetry operator supported on a non-contractible loop is equivalent to a symmetry operator supported on a contractible loop. To deduce this we first note the types of non-contractible loop symmetries present in this model. Unlike the surface code, which is based on the $\mathbb{Z}_2$-toric code, there are no non-contractible loop symmetries that connect two opposite boundaries of the manifold. In fact we can only make non-contractible loop symmetries that connect two qubit spaces of the top and right boundaries of the lattice,
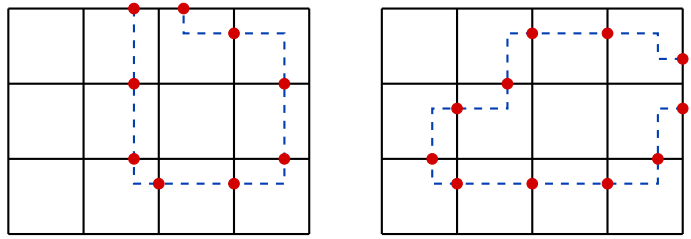
\begin{align*}
	\tikz[scale=1]{
		\foreach \a in {0, 1, 2, 3, 4}{
			\draw[thick] (\a, 0) -- (\a, 3);
		}
		\foreach \b in {0, 1, 2, 3}{
			\draw[thick] (0, \b) -- (4, \b);
		}
	\draw[thick, blue, dashed] (1.67, 3) -- (1.67, 0.67) -- (3.67, 0.67) -- (3.67, 2.67) -- (2.33, 2.67) -- (2.33, 3);
	\filldraw[red] (1.67, 3) circle (2pt);
	\filldraw[red] (1.67, 2) circle (2pt);
	\filldraw[red] (1.67, 1) circle (2pt);
	\filldraw[red] (2, 0.67) circle (2pt);
	\filldraw[red] (3, 0.67) circle (2pt);
	\filldraw[red] (3.67, 1) circle (2pt);
	\filldraw[red] (3.67, 2) circle (2pt);
	\filldraw[red] (3, 2.67) circle (2pt);
	\filldraw[red] (2.33, 3) circle (2pt);
	}\;\;\;\;\;\;\;\;
	\tikz[scale=1]{
	\foreach \a in {0, 1, 2, 3, 4}{
		\draw[thick] (\a, 0) -- (\a, 3);
	}
	\foreach \b in {0, 1, 2, 3}{
		\draw[thick] (0, \b) -- (4, \b);
	}
	\draw[thick, blue, dashed] (4, 2.33) -- (3.67, 2.33) -- (3.67, 2.67) -- (1.67, 2.67) -- (1.67, 1.67) -- (0.67, 1.67) -- (0.67, 0.67) -- (3.67, 0.67) -- (3.67, 1.67) -- (4, 1.67);
	\filldraw[red] (4, 2.33) circle (2pt);
	\filldraw[red] (1.67, 2) circle (2pt);
	\filldraw[red] (2, 2.67) circle (2pt);
	\filldraw[red] (2, 0.67) circle (2pt);
	\filldraw[red] (1, 1.67) circle (2pt);
	\filldraw[red] (0.67, 1) circle (2pt);
	\filldraw[red] (1, 0.67) circle (2pt);
	\filldraw[red] (3, 0.67) circle (2pt);
	\filldraw[red] (3.67, 1) circle (2pt);
	\filldraw[red] (3, 2.67) circle (2pt);
	\filldraw[red] (4, 1.67) circle (2pt);
	}
\end{align*}

which are equivalent to the contractible loop symmetries, 
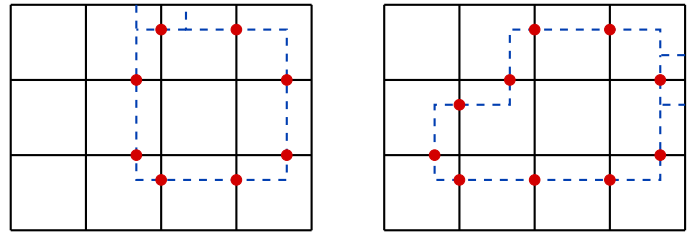
\begin{align*}
	\tikz[scale=1]{
		\foreach \a in {0, 1, 2, 3, 4}{
			\draw[thick] (\a, 0) -- (\a, 3);
		}
		\foreach \b in {0, 1, 2, 3}{
			\draw[thick] (0, \b) -- (4, \b);
		}
		\draw[thick, blue, dashed] (1.67, 3) -- (1.67, 0.67) -- (3.67, 0.67) -- (3.67, 2.67) -- (2.33, 2.67) -- (2.33, 3);
		\draw[thick, blue, dashed] (2.33, 2.67) -- (1.67, 2.67);
		\filldraw[red] (2, 2.67) circle (2pt);
		\filldraw[red] (1.67, 2) circle (2pt);
		\filldraw[red] (1.67, 1) circle (2pt);
		\filldraw[red] (2, 0.67) circle (2pt);
		\filldraw[red] (3, 0.67) circle (2pt);
		\filldraw[red] (3.67, 1) circle (2pt);
		\filldraw[red] (3.67, 2) circle (2pt);
		\filldraw[red] (3, 2.67) circle (2pt);
	}\;\;\;\;\;\;\;\;
	\tikz[scale=1]{
		\foreach \a in {0, 1, 2, 3, 4}{
			\draw[thick] (\a, 0) -- (\a, 3);
		}
		\foreach \b in {0, 1, 2, 3}{
			\draw[thick] (0, \b) -- (4, \b);
		}
		\draw[thick, blue, dashed] (4, 2.33) -- (3.67, 2.33) -- (3.67, 2.67) -- (1.67, 2.67) -- (1.67, 1.67) -- (0.67, 1.67) -- (0.67, 0.67) -- (3.67, 0.67) -- (3.67, 1.67) -- (4, 1.67);
		\draw[blue, thick, dashed] (3.67, 1.67) -- (3.67, 2.33);
		\filldraw[red] (1.67, 2) circle (2pt);
		\filldraw[red] (2, 2.67) circle (2pt);
		\filldraw[red] (2, 0.67) circle (2pt);
		\filldraw[red] (1, 1.67) circle (2pt);
		\filldraw[red] (0.67, 1) circle (2pt);
		\filldraw[red] (1, 0.67) circle (2pt);
		\filldraw[red] (3, 0.67) circle (2pt);
		\filldraw[red] (3.67, 1) circle (2pt);
		\filldraw[red] (3, 2.67) circle (2pt);
		\filldraw[red] (3.67, 2) circle (2pt);
	}
\end{align*}

by the vertex operators indicated by the $\times$ symbol.
On the other hand there are no non-contractible loop symmetries which originate and end on the left and bottom boundaries,
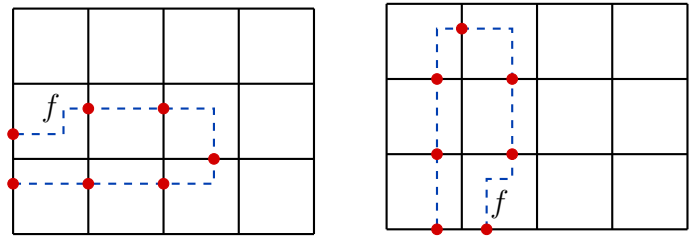
\begin{align*}
	\tikz[scale=1]{
		\foreach \a in {0, 1, 2, 3, 4}{
			\draw[thick] (\a, 0) -- (\a, 3);
		}
		\foreach \b in {0, 1, 2, 3}{
			\draw[thick] (0, \b) -- (4, \b);
		}
	\draw[thick, blue, dashed] (0, 0.67) -- (2.67, 0.67) -- (2.67, 1.67) -- (0.67, 1.67) -- (0.67, 1.33) -- (0, 1.33);
	\filldraw[red] (0, 0.67) circle (2pt);
	\filldraw[red] (1, 0.67) circle (2pt);
	\filldraw[red] (2, 0.67) circle (2pt);
	\filldraw[red] (2.67, 1) circle (2pt);
	\filldraw[red] (2, 1.67) circle (2pt);
	\filldraw[red] (1, 1.67) circle (2pt);
	\filldraw[red] (0, 1.33) circle (2pt);
	\node at (0.5, 1.67) {$f$};
	}\;\;\;\;\;\;\;\;
	\tikz[scale=1]{
	\foreach \a in {0, 1, 2, 3, 4}{
		\draw[thick] (\a, 0) -- (\a, 3);
	}
	\foreach \b in {0, 1, 2, 3}{
		\draw[thick] (0, \b) -- (4, \b);
	}
	\draw[thick, blue, dashed] (1.33, 0) -- (1.33, 0.67) -- (1.67, 0.67) -- (1.67, 2.67) -- (0.67, 2.67) -- (0.67, 0);
	\filldraw[red] (1.33, 0) circle (2pt);
	\filldraw[red] (1.67, 1) circle (2pt);
	\filldraw[red] (1.67, 2) circle (2pt);
	\filldraw[red] (1, 2.67) circle (2pt);
	\filldraw[red] (0.67, 2) circle (2pt);
	\filldraw[red] (0.67, 1) circle (2pt);
	\filldraw[red] (0.67, 0) circle (2pt);
	\node at (1.5, 0.33) {$f$};
	}
\end{align*}
as the face operators, indicated by $f$, are excited while forming the non-contractible loops. It is easy to convince oneself that it is impossible to make non-contractible loops lying at the left and bottom boundaries.  

The model on the manifold with boundary which is consistent with the bulk vertex and face operators of Eqs. \ref{eq:avmodel1} and \ref{eq:bfmodel1} is given by
\begin{align*}
	\tikz[scale=1]{
		\foreach \a in {0, 1, 2, 3, 4}{
			\draw[thick] (\a, 0) -- (\a, 3);
		}
		\foreach \b in {0, 1, 2, 3}{
			\draw[thick] (0, \b) -- (4, \b);
		}
		\node at (4.5, 2) {$\to$};
		\def\a{5.5}
		\def\b{2}
		\draw[thick] (-0.5+\a, \b) -- (\a, \b);
		\draw[thick] (\a, -0.5+\b) -- (\a, 0.5+\b);
		\filldraw[red] (-0.5+\a, \b) circle (2pt);
		\filldraw[red] (\a, -0.5+\b) circle (2pt);
		\filldraw[red] (\a, 0.5+\b) circle (2pt) node[anchor = south west, black] {$\scriptstyle Z$};
		\node at (-0.5, 2) {$\leftarrow$};
		\def\a{-1}
		\def\b{2}
		\draw[thick] (-0.5+\a, \b) -- (\a, \b);
		\draw[thick] (\a-0.5, -0.5+\b) -- (\a-0.5, 0.5+\b);
		\filldraw[red] (\a, \b) circle (2pt) node[anchor = south, black] {$\scriptstyle Z$};
		\filldraw[red] (\a - 0.5, -0.5+\b) circle (2pt);
		\filldraw[red] (\a-0.5, 0.5+\b) circle (2pt) node[anchor = south west, black] {$\scriptstyle Z$};
		\node at (2, -0.5) {$\downarrow$};
		\def\a{2}
		\def\b{-1.5}
		\draw[thick] (-0.5+\a, \b) -- (0.5+\a, \b);
		\draw[thick] (\a, \b) -- (\a, 0.5+\b);
		\filldraw[red] (\a, 0.5+\b) circle (2pt) node[anchor = south west, black] {$\scriptstyle Z$};
		\filldraw[red] (\a - 0.5, \b) circle (2pt);
		\filldraw[red] (\a+0.5, \b) circle (2pt) node[anchor = south west, black] {$\scriptstyle Z$};
		\node at (2, 3.5) {$\uparrow$};
		\def\a{2}
		\def\b{4.5}
		\draw[thick] (-0.5+\a, \b) -- (0.5+\a, \b);
		\draw[thick] (\a, \b) -- (\a, -0.5+\b);
		\filldraw[red] (\a, -0.5+\b) circle (2pt) node[anchor = south west, black] {$\scriptstyle $};
		\filldraw[red] (\a - 0.5, \b) circle (2pt);
		\filldraw[red] (\a+0.5, \b) circle (2pt) node[anchor = west, black] {$\scriptstyle Z$};
		\def\a{-1}
		\def\b{-1}
		\draw[thick] (\a, \b+0.5) -- (\a, \b) -- (\a+0.5, \b);
		\filldraw[red] (\a, \b+0.5) circle (2pt) node[anchor=south east, black] {$\scriptstyle Z$};
		\filldraw[red] (\a+0.5, \b) circle (2pt) node[anchor=north west, black] {$\scriptstyle Z$};
		\def\a{5}
		\def\b{-1}
		\draw[thick] (\a, \b) -- (\a+0.5, \b) -- (\a+0.5, \b+0.5);
		\filldraw[red] (\a, \b) circle (2pt) node[anchor=south east, black] {$\scriptstyle $};
		\filldraw[red] (\a+0.5, \b+0.5) circle (2pt) node[anchor=south west, black] {$\scriptstyle Z$};
		\def\a{5}
		\def\b{4}
		\draw[thick] (\a, \b) -- (\a+0.5, \b) -- (\a+0.5, \b-0.5);
		\filldraw[red] (\a, \b) circle (2pt) node[anchor=south east, black] {$\scriptstyle $};
		\filldraw[red] (\a+0.5, \b-0.5) circle (2pt) node[anchor=south west, black] {$\scriptstyle $};
		\def\a{-1}
		\def\b{4}
		\draw[thick] (\a, \b - 0.5) -- (\a, \b) -- (\a+0.5, \b);
		\filldraw[red] (\a, \b - 0.5) circle (2pt) node[anchor=south east, black] {$\scriptstyle $};
		\filldraw[red] (\a+0.5, \b) circle (2pt) node[anchor=south west, black] {$\scriptstyle Z$};
	}
\end{align*}
where we have not shown the $X$ part of the vertex operators. This model has a GSD of $2^{mn-1}$ and is in fact a non-Abelian model due to the boundary and corner vertex operators. We will analyze such models in the future.

\bibliographystyle{unsrt}
\normalem
\bibliography{refs}

\end{document}